\documentclass[11pt,letterpaper,parskip=half]{scrartcl}

\usepackage[T1]{fontenc}
\usepackage[utf8]{inputenc}
\usepackage{mathpazo}          % Palatino for body + math
\usepackage{microtype}         % margin kerning + font expansion
\usepackage{amsmath}
\usepackage{amssymb}
\usepackage{bm}                % bold math symbols
\usepackage{mathtools}         % extensions to amsmath

\usepackage[dvipsnames]{xcolor}

\usepackage[margin=1in]{geometry}
\usepackage{setspace}
\usepackage{array}
\usepackage{booktabs}
\usepackage{enumitem}
\usepackage{algorithm}
\usepackage[noend]{algpseudocode}
\algrenewcommand\algorithmicrequire{\textbf{Input:}}
\algrenewcommand\algorithmicensure{\textbf{Output:}}
\usepackage{longtable}
\usepackage{tabularx}
\usepackage{graphicx}
\usepackage{flafter}                   % floats never appear before their source
\usepackage{needspace}                 % keep headings with following text

\usepackage[round]{natbib}
\usepackage[colorlinks=true,linkcolor=MidnightBlue,citecolor=OliveGreen,urlcolor=BrickRed]{hyperref}
\usepackage{doi}               % clickable DOIs in bibliography (after hyperref)
\usepackage{cleveref}

\addtokomafont{disposition}{\rmfamily}     % serif headings (match Palatino body)
\setkomafont{subtitle}{\Large}
\setkomafont{author}{\normalsize}
\setkomafont{date}{\small}

\newcommand{\R}{\mathbb{R}}
\newcommand{\M}{\mathcal{M}}          % mortality data tensor
\newcommand{\G}{\mathcal{G}}          % core tensor
\newcommand{\bS}{\bm{S}}             % sex factor matrix
\newcommand{\bA}{\bm{A}}             % age factor matrix
\newcommand{\bC}{\bm{C}}             % country factor matrix
\newcommand{\bT}{\bm{T}}             % year factor matrix
\newcommand{\logit}{\operatorname{logit}}
\newcommand{\expit}{\operatorname{expit}}
\newcommand{\qx}{q_x}
\newcommand{\ezero}{e_0}

\DeclareMathOperator{\vect}{vec}
\newcommand{\nmx}{{}_1m_x}            % 1-year central death rate
\newcommand{\nqx}{{}_1q_x}            % 1-year probability of dying
\newcommand{\lx}{l_x}                 % survivorship
\newcommand{\pkg}[1]{\textsf{#1}}
\newcommand{\code}[1]{\texttt{#1}}

\newcommand{\includefigifexists}[2][width=\textwidth]{%
  \IfFileExists{#2}{%
    \includegraphics[#1]{#2}%
  }{%
    \fbox{\parbox{0.9\textwidth}{\centering\vspace{2cm}%
      \texttt{#2}\\[0.5em]%
      \textit{(Figure not yet generated -- run the QMD pipeline)}%
      \vspace{2cm}}}%
  }%
}

\newcommand{\piKappa}{1.487}

\newcommand{\covNinetyFive}{94.9}
\newcommand{\covEighty}{89.3}

\newcommand{\commonN}{9{,}507}
\newcommand{\ffMAE}{4.100}
\newcommand{\ffRMSE}{6.506}
\newcommand{\ffBias}{+1.058}
\newcommand{\pbMAE}{4.764}
\newcommand{\pbRMSE}{7.812}
\newcommand{\pbBias}{+3.307}
\newcommand{\lcMAE}{3.818}
\newcommand{\lcRMSE}{5.576}
\newcommand{\lcBias}{-3.205}
\newcommand{\huMAE}{4.094}
\newcommand{\huRMSE}{6.005}
\newcommand{\huBias}{-3.534}
\newcommand{\ffMAEShortH}{2.984}

\newcommand{\ffMAEVlongH}{5.010}

\newcommand{\pbMAEShortH}{1.095}

\newcommand{\pbMAEVlongH}{7.125}

\newcommand{\lcMAEShortH}{1.471}

\newcommand{\lcMAEVlongH}{6.014}

\newcommand{\huMAEShortH}{1.092}

\newcommand{\huMAEVlongH}{6.791}

\newcommand{\mxCommonN}{1{,}662{,}076}
\newcommand{\ffMxMAEuw}{0.355}
\newcommand{\ffMxMAElx}{0.424}
\newcommand{\ffMxBiaslx}{+0.229}
\newcommand{\pbMxMAEuw}{0.969}
\newcommand{\pbMxMAElx}{1.143}
\newcommand{\pbMxBiaslx}{-0.444}
\newcommand{\mxRatio}{2.7}
\newcommand{\ffMxMAEAgeZero}{0.520}

\newcommand{\pbMxMAEAgeZero}{2.542}

\newcommand{\mxRatioAgeZero}{4.9}
\newcommand{\ffMxMAEAgeOneFourteen}{0.679}

\newcommand{\pbMxMAEAgeOneFourteen}{2.276}

\newcommand{\mxRatioAgeOneFourteen}{3.4}
\newcommand{\ffMxMAEAgeFifteenTwentyNine}{0.513}

\newcommand{\pbMxMAEAgeFifteenTwentyNine}{1.168}

\newcommand{\mxRatioAgeFifteenTwentyNine}{2.3}
\newcommand{\ffMxMAEAgeThirtyFortyFour}{0.433}

\newcommand{\pbMxMAEAgeThirtyFortyFour}{0.935}

\newcommand{\mxRatioAgeThirtyFortyFour}{2.2}
\newcommand{\ffMxMAEAgeFortyFiveFiftyNine}{0.299}

\newcommand{\pbMxMAEAgeFortyFiveFiftyNine}{0.894}

\newcommand{\mxRatioAgeFortyFiveFiftyNine}{3.0}
\newcommand{\ffMxMAEAgeSixtySeventyFour}{0.236}

\newcommand{\pbMxMAEAgeSixtySeventyFour}{0.468}

\newcommand{\mxRatioAgeSixtySeventyFour}{2.0}
\newcommand{\ffMxMAEAgeSeventyFiveEightyNine}{0.175}

\newcommand{\pbMxMAEAgeSeventyFiveEightyNine}{0.391}

\newcommand{\mxRatioAgeSeventyFiveEightyNine}{2.2}
\newcommand{\ffMxMAEAgeNinetyHundred}{0.112}

\newcommand{\pbMxMAEAgeNinetyHundred}{0.356}

\newcommand{\mxRatioAgeNinetyHundred}{3.2}
\newcommand{\ffMxMAEShortH}{0.225}
\newcommand{\pbMxMAEShortH}{1.291}
\newcommand{\mxRatioShortH}{5.7}

\newcommand{\ffMxMAEVlongH}{0.634}
\newcommand{\pbMxMAEVlongH}{0.856}
\newcommand{\mxRatioVlongH}{1.3}
\newcommand{\ffMxMAEF}{0.451}
\newcommand{\pbMxMAEF}{1.177}
\newcommand{\ffMxMAEM}{0.395}
\newcommand{\pbMxMAEM}{1.106}
\newcommand{\ffMxMAEFAgeZero}{0.527}
\newcommand{\ffMxBiasFAgeZero}{+0.355}
\newcommand{\pbMxMAEFAgeZero}{2.537}
\newcommand{\pbMxBiasFAgeZero}{-2.337}
\newcommand{\ffMxMAEFAgeOneFourteen}{0.719}
\newcommand{\ffMxBiasFAgeOneFourteen}{+0.408}
\newcommand{\pbMxMAEFAgeOneFourteen}{2.358}
\newcommand{\pbMxBiasFAgeOneFourteen}{-2.221}
\newcommand{\ffMxMAEFAgeFifteenTwentyNine}{0.573}
\newcommand{\ffMxBiasFAgeFifteenTwentyNine}{+0.304}
\newcommand{\pbMxMAEFAgeFifteenTwentyNine}{1.257}
\newcommand{\pbMxBiasFAgeFifteenTwentyNine}{-1.018}
\newcommand{\ffMxMAEFAgeThirtyFortyFour}{0.464}
\newcommand{\ffMxBiasFAgeThirtyFortyFour}{+0.246}
\newcommand{\pbMxMAEFAgeThirtyFortyFour}{0.983}
\newcommand{\pbMxBiasFAgeThirtyFortyFour}{-0.042}
\newcommand{\ffMxMAEFAgeFortyFiveFiftyNine}{0.319}
\newcommand{\ffMxBiasFAgeFortyFiveFiftyNine}{+0.155}
\newcommand{\pbMxMAEFAgeFortyFiveFiftyNine}{0.927}
\newcommand{\pbMxBiasFAgeFortyFiveFiftyNine}{+0.516}
\newcommand{\ffMxMAEFAgeSixtySeventyFour}{0.264}
\newcommand{\ffMxBiasFAgeSixtySeventyFour}{+0.151}
\newcommand{\pbMxMAEFAgeSixtySeventyFour}{0.501}
\newcommand{\pbMxBiasFAgeSixtySeventyFour}{+0.204}
\newcommand{\ffMxMAEFAgeSeventyFiveEightyNine}{0.190}
\newcommand{\ffMxBiasFAgeSeventyFiveEightyNine}{+0.132}
\newcommand{\pbMxMAEFAgeSeventyFiveEightyNine}{0.426}
\newcommand{\pbMxBiasFAgeSeventyFiveEightyNine}{+0.205}
\newcommand{\ffMxMAEFAgeNinetyHundred}{0.120}
\newcommand{\ffMxBiasFAgeNinetyHundred}{+0.090}
\newcommand{\pbMxMAEFAgeNinetyHundred}{0.378}
\newcommand{\pbMxBiasFAgeNinetyHundred}{+0.086}
\newcommand{\ffMxMAEMAgeZero}{0.513}
\newcommand{\ffMxBiasMAgeZero}{+0.347}
\newcommand{\pbMxMAEMAgeZero}{2.548}
\newcommand{\pbMxBiasMAgeZero}{-1.981}
\newcommand{\ffMxMAEMAgeOneFourteen}{0.639}
\newcommand{\ffMxBiasMAgeOneFourteen}{+0.384}
\newcommand{\pbMxMAEMAgeOneFourteen}{2.192}
\newcommand{\pbMxBiasMAgeOneFourteen}{-1.757}
\newcommand{\ffMxMAEMAgeFifteenTwentyNine}{0.453}
\newcommand{\ffMxBiasMAgeFifteenTwentyNine}{+0.220}
\newcommand{\pbMxMAEMAgeFifteenTwentyNine}{1.078}
\newcommand{\pbMxBiasMAgeFifteenTwentyNine}{-0.728}
\newcommand{\ffMxMAEMAgeThirtyFortyFour}{0.402}
\newcommand{\ffMxBiasMAgeThirtyFortyFour}{+0.194}
\newcommand{\pbMxMAEMAgeThirtyFortyFour}{0.887}
\newcommand{\pbMxBiasMAgeThirtyFortyFour}{+0.069}
\newcommand{\ffMxMAEMAgeFortyFiveFiftyNine}{0.278}
\newcommand{\ffMxBiasMAgeFortyFiveFiftyNine}{+0.128}
\newcommand{\pbMxMAEMAgeFortyFiveFiftyNine}{0.860}
\newcommand{\pbMxBiasMAgeFortyFiveFiftyNine}{+0.558}
\newcommand{\ffMxMAEMAgeSixtySeventyFour}{0.204}
\newcommand{\ffMxBiasMAgeSixtySeventyFour}{+0.103}
\newcommand{\pbMxMAEMAgeSixtySeventyFour}{0.430}
\newcommand{\pbMxBiasMAgeSixtySeventyFour}{+0.187}
\newcommand{\ffMxMAEMAgeSeventyFiveEightyNine}{0.153}
\newcommand{\ffMxBiasMAgeSeventyFiveEightyNine}{+0.093}
\newcommand{\pbMxMAEMAgeSeventyFiveEightyNine}{0.337}
\newcommand{\pbMxBiasMAgeSeventyFiveEightyNine}{+0.136}
\newcommand{\ffMxMAEMAgeNinetyHundred}{0.094}
\newcommand{\ffMxBiasMAgeNinetyHundred}{+0.065}
\newcommand{\pbMxMAEMAgeNinetyHundred}{0.301}
\newcommand{\pbMxBiasMAgeNinetyHundred}{+0.094}
\newcommand{\ffSdMAEAgeZero}{0.057}
\newcommand{\ffSdBiasAgeZero}{-0.008}
\newcommand{\ffSdMAEAgeOneFourteen}{0.276}
\newcommand{\ffSdBiasAgeOneFourteen}{-0.022}
\newcommand{\ffSdMAEAgeFifteenTwentyNine}{0.260}
\newcommand{\ffSdBiasAgeFifteenTwentyNine}{-0.083}
\newcommand{\ffSdMAEAgeThirtyFortyFour}{0.193}
\newcommand{\ffSdBiasAgeThirtyFortyFour}{-0.053}
\newcommand{\ffSdMAEAgeFortyFiveFiftyNine}{0.158}
\newcommand{\ffSdBiasAgeFortyFiveFiftyNine}{-0.029}
\newcommand{\ffSdMAEAgeSixtySeventyFour}{0.143}
\newcommand{\ffSdBiasAgeSixtySeventyFour}{-0.053}
\newcommand{\ffSdMAEAgeSeventyFiveEightyNine}{0.096}
\newcommand{\ffSdBiasAgeSeventyFiveEightyNine}{-0.047}
\newcommand{\ffSdMAEAgeNinetyHundred}{0.064}
\newcommand{\ffSdBiasAgeNinetyHundred}{-0.026}
\newcommand{\pbSdMAEAgeZero}{0.438}
\newcommand{\pbSdBiasAgeZero}{+0.356}
\newcommand{\pbSdMAEAgeOneFourteen}{0.582}
\newcommand{\pbSdBiasAgeOneFourteen}{+0.472}
\newcommand{\pbSdMAEAgeFifteenTwentyNine}{0.410}
\newcommand{\pbSdBiasAgeFifteenTwentyNine}{+0.287}
\newcommand{\pbSdMAEAgeThirtyFortyFour}{0.240}
\newcommand{\pbSdBiasAgeThirtyFortyFour}{+0.115}
\newcommand{\pbSdMAEAgeFortyFiveFiftyNine}{0.192}
\newcommand{\pbSdBiasAgeFortyFiveFiftyNine}{+0.085}
\newcommand{\pbSdMAEAgeSixtySeventyFour}{0.171}
\newcommand{\pbSdBiasAgeSixtySeventyFour}{+0.012}
\newcommand{\pbSdMAEAgeSeventyFiveEightyNine}{0.106}
\newcommand{\pbSdBiasAgeSeventyFiveEightyNine}{-0.035}
\newcommand{\pbSdMAEAgeNinetyHundred}{0.082}
\newcommand{\pbSdBiasAgeNinetyHundred}{+0.020}

\title{Mortality Forecasting as a Flow Field \\
  in Tucker Decomposition Space:\\[0.3em]
  {\Large Direct Surface Prediction vs.\ $\ezero$-Mediated Pipelines}}
\author{Samuel J.\ Clark\\
        \medskip {\small Department of Sociology, The Ohio State University}}
\date{2026}

\begin{document}
\maketitle

\begin{abstract}
We introduce a mortality forecasting method that navigates a flow field through the low-dimensional space defined by the Tucker decomposition of the Human Mortality Database's sex--age--country--year tensor.  PCA of the effective core matrix $G_{ct}$ shows that five components capture 97\% of the variance and that their derivatives are tightly correlated ($r(\Delta s_1, \Delta s_2) = -0.92$), so the mortality transition is essentially a one-dimensional flow through 5D score space.  A nonparametric speed function advances the level score; trajectory functions map it to the structural scores; and the Tucker basis matrices reconstruct the complete sex-specific, single-year-of-age mortality schedule at each forecast horizon.  An era-weighted speed function and empirically calibrated score relaxation (half-lives 12--32~years) complete the architecture.

In leave-country-out cross-validation (\commonN{} test points, 50-year horizon, evaluated against raw HMD $\ezero$), the flow-field achieves $\ezero$ MAE of \ffMAE~years -- comparable to Lee--Carter (\lcMAE) and Hyndman--Ullah (\huMAE) -- but with the lowest bias (\ffBias~years vs.\ \lcBias{} for Lee--Carter, \huBias{} for Hyndman--Ullah, and \pbBias{} for \pkg{pyBayesLife}).  The flow-field's bias advantage is the most distinctive result: bias of \ffBias~years vs.\ $-3.5$ to $+3.3$~years for the other three methods.

On \mxCommonN{} sex-age-specific test points, the flow-field achieves $\lx$-weighted $\log(\nmx)$ MAE of \ffMxMAElx{} vs.\ \pbMxMAElx{} for \pkg{pyBayesLife} (\mxRatio$\times$) and $\lx$-weighted bias of \ffMxBiaslx{} vs.\ \pbMxBiaslx{}, with lower error at every age, every horizon, and for both sexes.  The flow-field also reproduces the observed sex differential in age-specific mortality more accurately, because both sexes emerge from the same Tucker surface.  This gap persists at short horizons where \pkg{pyBayesLife} produces better $\ezero$ forecasts, confirming that the bottleneck is the reconstruction from $\ezero$ to age-specific rates.
\end{abstract}

\tableofcontents
\listoffigures
\listoftables

% ══════════════════════════════════════════════════════════════════════════════
\section{Introduction}
\label{sec:intro}
% ══════════════════════════════════════════════════════════════════════════════

The dominant paradigm for mortality forecasting since \citet{LeeCarter1992} is to decompose age-specific mortality rates via the singular value decomposition, then extrapolate the temporal component(s) using time series models -- typically a random walk with drift.  Extensions increase the SVD rank \citep{HyndmanUllah2007}, add coherent multi-population structure \citep{LiLee2005,HyndmanBoothYasmeen2013}, model the rotation of age patterns \citep{LiLeeGerland2013}, or use more sophisticated time series models \citep{deJongTickle2006}.  Evaluations \citep{LeeMillerLeeCarter2001,BoothEtAl2006} and recent reviews \citep{BoothTickle2008,BaselliniCamardaBooth2023} document persistent challenges: the fixed age pattern in Lee--Carter underpredicts life expectancy, independent sex-specific fits produce divergent forecasts, and prediction intervals are sensitive to model choice.

A complementary tradition models life expectancy gains as a function of the current mortality level.  \citet{RafteryChunnGerlandSevcikova2013} introduced a Bayesian hierarchical model in which five-year gains in $\ezero$ follow a double-logistic function of $\ezero$, with country-specific parameters drawn from a world distribution.  This approach -- now the basis of the United Nations' official population projections -- captures the empirically observed pattern \citep{OeppenVaupel2002} in which the rate of mortality improvement depends on where a country sits in the epidemiological transition rather than on calendar time.  However, it forecasts only the scalar $\ezero$; converting the projected $\ezero$ to age-specific rates requires a separate model life table system.

The UN production pipeline implements this two-stage architecture through two R packages.  The \pkg{bayesLife} package \citep{SevcikobayesLife2024} projects female $\ezero$ via the hierarchical double-logistic model and derives male $\ezero$ from a joint gap model \citep{RafteryLalicGerland2014}.  The \pkg{MortCast} package \citep{SevcikMortCast2024} then recovers age-specific mortality rates from the projected $\ezero$ using coherent Lee--Carter \citep{LiLee2005} with Kannisto old-age extension \citep{Kannisto1994} and $b_x$ rotation \citep{LiLeeGerland2013}, following the algorithm of \citet{SevcikovaLiKantorovaGerlandRaftery2016}.  This pipeline produces the mortality inputs to the United Nations World Population Prospects \citep{UnitedNations2022WPP}, making it the most widely used mortality forecasting system in applied demography.  Despite its influence, the pipeline's reliance on a scalar $\ezero$ intermediate creates a fundamental information bottleneck: the entire sex$\times$age mortality surface is compressed to a single number and then reconstructed, discarding all information about the age pattern, the sex differential in the age pattern, and country-specific deviations from average age patterns.

Tensor decomposition methods extend the Lee--Carter approach to multi-way data.  \citet{RussolilloGiordanoHaberman2011} applied a rank-2 Tucker decomposition to 10 European populations, and \citet{DongHuangHaberman2020} compared canonical polyadic and Tucker decompositions for multi-population forecasting, in both cases extrapolating the time-mode factor vectors via ARIMA. \citet{ZhangHuangHuiHaberman2023} developed adaptive penalised tensor decomposition for cause-of-death mortality.

This paper unifies these two traditions.  The Tucker decomposition of the Human Mortality Database's mortality tensor -- developed as part of the MDMx system \citep{ClarkMDMx2026} -- provides a structured low-dimensional space in which every country-year is represented by a point.  We show that the dynamics in this space are remarkably simple: the derivatives of all five PCA components of the effective core matrix $G_{ct}$ are tightly correlated ($r = -0.92$ for the first two components), revealing that the mortality transition is essentially a one-dimensional flow.  Forecasting reduces to learning a scalar speed function $g^*(s_1) = \mathrm{d}s_1/\mathrm{d}t$ and trajectory functions $f_k^*(s_1)$ for $k = 2, \ldots, 5$ that map the level score to the structural scores -- from which the complete sex-specific, single-year-of-age mortality schedule is reconstructed via the Tucker basis matrices.  Life expectancy $\ezero$ is computed from the reconstructed mortality surface at each horizon for reporting only, avoiding the systematic bias that arises when an $\ezero$ accumulator diverges from the surface-derived $\ezero$ through the nonlinear expit/life-table chain.

The conceptual reframing is from ``project the time index forward'' (Lee--Carter and extensions) to ``the mortality transition is a flow through a structured space, parameterised by level'' (the present approach).  The speed function is the Tucker-space analogue of the \citeauthor{RafteryChunnGerlandSevcikova2013} level-dependent improvement rate, and the trajectory functions are a continuous model life table system in Tucker coordinates -- but unlike the WPP pipeline, the forecasting model and the reconstruction model are unified in a single framework, and the navigation coordinate is the PCA level score rather than the scalar $\ezero$.

Three additional innovations shape the production system.  First, an era-weighted speed function uses a truncated exponential kernel centred on each forecast origin, giving more weight to contemporary dynamics and avoiding the bias that arises from averaging over disparate eras of the mortality transition.  Second, empirically calibrated convergence rates -- measured from the observed autocorrelation of country-level deviations from canonical dynamics -- control how quickly each country's distinctive mortality structure relaxes toward the HMD-wide canonical pattern; structural score deviations persist for 12--32~years. Third, the optimal speed blend weight is $w = 1.0$ (fully canonical), though the MAE varies by only a few hundredths of a year across the full range of $w$: the forecast reduces to a deterministic integration along a curve in Tucker PCA space.

We evaluate the system using leave-country-out cross-validation with a 50-year forecast horizon -- directly testing the production use case of forecasting a country whose data did not contribute to the flow field.  The system achieves $\ezero$ MAE of \ffMAE~years -- comparable to Lee--Carter (\lcMAE), Hyndman--Ullah (\huMAE), and \pkg{pyBayesLife} (\pbMAE) -- with the MAE advantage concentrated at long horizons (36--41\% at $h = 26$--50).  The flow-field has the lowest aggregate bias (\ffBias~years) while Lee--Carter (\lcBias~years), Hyndman--Ullah (\huBias~years), and \pkg{pyBayesLife} (\pbBias~years) show systematic bias -- a distinction that matters for long-range population projections, pension planning, and actuarial applications.

Because the flow-field forecasts the complete sex$\times$age mortality surface directly, while the \citeauthor{RafteryChunnGerlandSevcikova2013} pipeline collapses to $\ezero$ and reconstructs, a natural question is how much accuracy the collapse costs.  To answer this on a level playing field, we built \pkg{pyBayesLife} -- a de novo Python reimplementation of the entire Raftery et al.\ pipeline that eliminates all dependencies on R code and World Population Prospects data, training exclusively on HMD data (see \cref{sec:appendix:raftery} for the implementation details and WPP data dependencies that motivated this reimplementation).  The age-specific comparison (\cref{sec:agespecific}) reveals a \mxRatio-fold accuracy gap, demonstrating that the $\ezero$-mediated reconstruction is the binding constraint, not the quality of the $\ezero$ forecast.

% ══════════════════════════════════════════════════════════════════════════════
\section{Tucker Decomposition}
\label{sec:tucker}
% ══════════════════════════════════════════════════════════════════════════════

We work with the rank-$(r_1, r_2, r_3, r_4)$ Tucker decomposition of the $\logit(\qx)$ mortality tensor $\M \in \R^{S \times A \times C \times T}$ ($S=2$ sexes, $A=110$ ages, $C=48$ countries, $T=274$ years) from the HMD, as developed in \citet{ClarkMDMx2026}.  The decomposition produces factor matrices $\bS \in \R^{S \times r_1}$ (sex), $\bA \in \R^{A \times r_2}$ (age), $\bC \in \R^{C \times r_3}$ (country), $\bT \in \R^{T \times r_4}$ (year), and a core tensor $\G \in \R^{r_1 \times r_2 \times r_3 \times r_4}$ with ranks $(2, 42, 46, 100)$.

Every country-year's mortality schedule is determined by the effective core matrix
\begin{equation}
\label{eq:Gct}
G_{ct}[i,j] = \sum_{k,l} \G[i,j,k,l] \, \bC[c,k] \, \bT[t,l]\,,
\end{equation}
and the reconstruction is $\hat{M}_{:,:,c,t} = \bS \, G_{ct} \, \bA^\top$. The $r_1 \times r_2 = 84$ elements of $G_{ct}$ are the forecasting target (\cref{sec:app:math:tucker} gives the full HOSVD procedure).

% ══════════════════════════════════════════════════════════════════════════════
\section{The Flow Field}
\label{sec:flow}
% ══════════════════════════════════════════════════════════════════════════════

% ──────────────────────────────────────────────────────────────────────────────
\subsection{PCA reduction}
\label{sec:flow:pca}
% ──────────────────────────────────────────────────────────────────────────────

PCA on all observed $\vect(G_{ct})$ vectors shows that five components capture 97.1\% of the total variance (\cref{tab:pca}), with the first component alone accounting for 91.8\% -- it is the mortality level. The complete score extraction and reconstruction procedure is in \cref{sec:app:math:pca}.

\begin{table}[htbp]
\caption{PCA of $\vect(G_{ct})$: variance explained.}
\label{tab:pca}
\begin{tabular*}{\textwidth}{@{\extracolsep{\fill}}lcc@{}}
\toprule
Component & Variance & Cumulative \\
\midrule
PC~1 & 91.8\% & 91.8\% \\
PC~2 &  2.6\% & 94.4\% \\
PC~3 &  1.5\% & 95.9\% \\
PC~4 &  0.7\% & 96.7\% \\
PC~5 &  0.4\% & 97.1\% \\
\bottomrule
\end{tabular*}
\end{table}

% ──────────────────────────────────────────────────────────────────────────────
\subsection{One-dimensional dynamics}
\label{sec:flow:1d}
% ──────────────────────────────────────────────────────────────────────────────

Smoothing each country's PCA score trajectory (LOWESS, fraction 0.25) and differentiating reveals that the derivative vector is essentially one-dimensional.  Using raw forward differences -- which preserve the shared transient shocks (wars, pandemics, economic crises) that simultaneously push multiple PCA components -- the correlations between $\Delta s_1$ (rate of level change) and the structural derivatives are:
\begin{center}
\begin{tabular}{lc}
\toprule
Pair & Correlation \\
\midrule
$\Delta s_1, \Delta s_2$ & $-0.922$ \\
$\Delta s_1, \Delta s_3$ & $-0.548$ \\
$\Delta s_1, \Delta s_4$ & $+0.496$ \\
$\Delta s_1, \Delta s_5$ & $+0.571$ \\
\bottomrule
\end{tabular}
\end{center}
The interpretation: when countries move through the mortality transition, all five PCA components move in lockstep.  The trajectory has a \emph{shape} (the curve through 5D score space) and a \emph{speed} (how fast a country traverses it), and these are nearly separable.  The entire 5D trajectory -- encoding 84 Tucker weights and hence 220 $\logit(\qx)$ values -- is to close approximation a function of a single scalar: the level score $s_1$ (equivalently, $\ezero$).

\begin{figure}[!htbp]
\centering
\includegraphics[width=\textwidth]{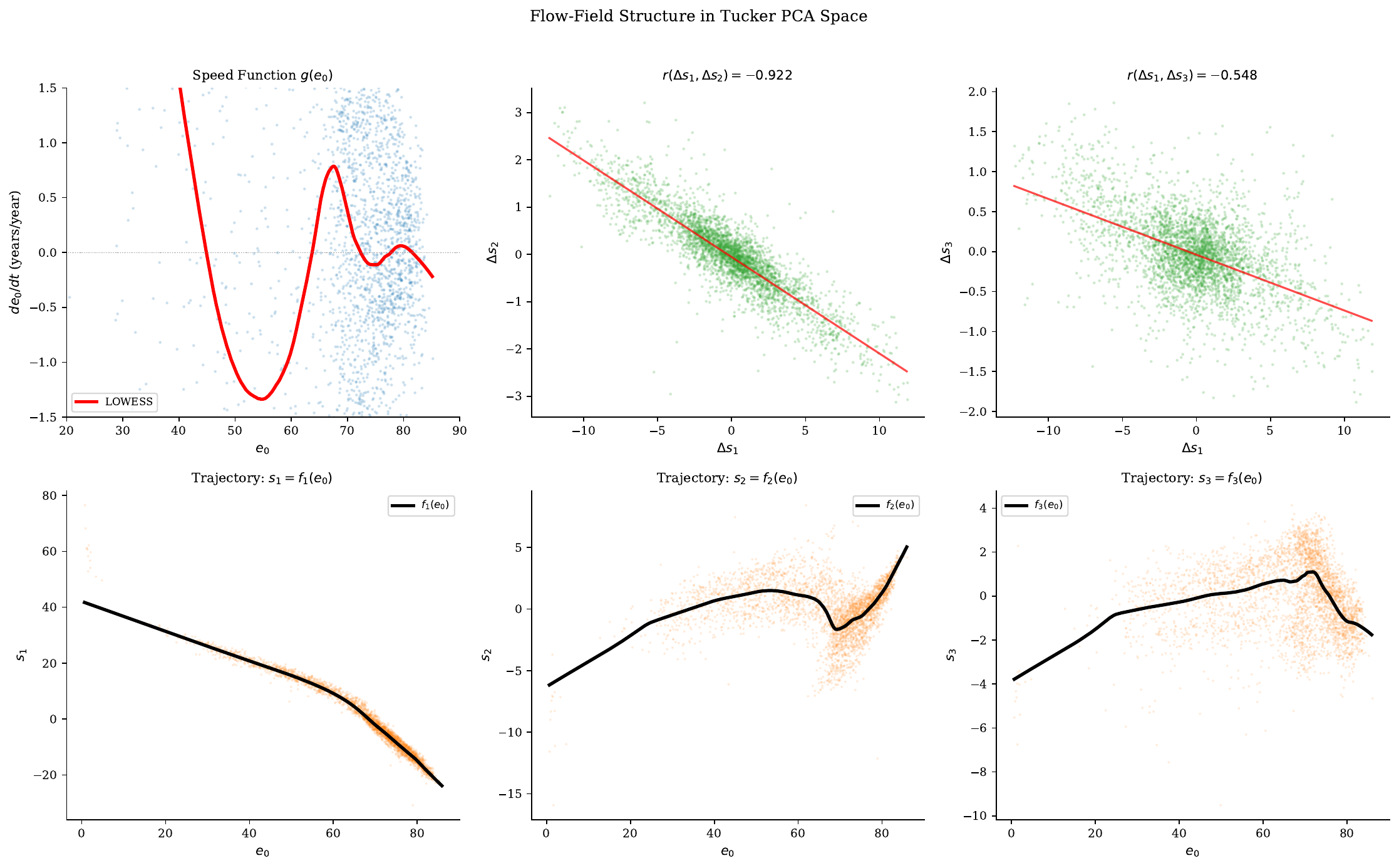}
\caption{Flow-field structure in Tucker PCA space.  Top left: raw
year-to-year $\ezero$ velocity (forward differences) vs $\ezero$ -- the scatter is noisy but the LOWESS trend reveals level-dependent improvement; the production speed function uses per-country smoothed velocities in $s_1$ space for a cleaner estimate (\cref{fig:speed-diagnostic}).  Top centre and right: derivative correlations $\Delta s_1$ vs $\Delta s_2$ and $\Delta s_3$ (raw forward differences); the tight linear relationship ($r = -0.92$) demonstrates one-dimensional dynamics.  Bottom: canonical trajectories $s_k$ vs $\ezero$ for PCs~1--3 -- each score is a tight function of mortality level, comprising a continuous model life table system in Tucker coordinates.}
\label{fig:flow}
\end{figure}

\Cref{fig:flow} visualises this structure.  The top-left panel shows the raw year-to-year $\ezero$ velocity (forward differences) as a function of mortality level -- the same raw data used in the derivative correlation panels.  The scatter is noisy because raw forward differences include wars, pandemics, and stochastic year-to-year fluctuation, but the LOWESS trend reveals the broad level-dependent pattern: improvement is concentrated at intermediate $\ezero$ levels and decelerates at the frontier.  The production speed function (\cref{sec:flow:functions}) uses per-country smoothed velocities that filter this noise, revealing a cleaner profile inspired by but more complex than the parametric double-logistic of \citet{RafteryChunnGerlandSevcikova2013} -- the nonparametric LOWESS captures empirical structure, including asymmetry and plateau regions, that a parametric form would impose away.  The top-centre and top-right panels show the derivative correlations $\Delta s_1$ vs $\Delta s_2$ and $\Delta s_3$: the tight linear relationship ($r = -0.92$) demonstrates that the 5D dynamics compress into a one-dimensional flow.  The bottom panels show the canonical trajectories: each PCA score traces a tight curve as a function of $\ezero$ (equivalently, of $s_1$), with individual countries scattered around the LOWESS trend.

% ──────────────────────────────────────────────────────────────────────────────
\subsection{Speed function and trajectory functions}
\label{sec:flow:functions}
% ──────────────────────────────────────────────────────────────────────────────

Since $s_1$ is approximately the mortality level, the flow field is defined in $s_1$ space rather than $\ezero$ space.  The \emph{speed function} $g^*(s_1) = \mathrm{d}s_1/\mathrm{d}t$ is estimated by LOWESS regression of the smoothed $s_1$ velocity (forward differences of per-country LOWESS-smoothed $s_1$) on $s_1$ across all country-years in the training set.  The \emph{trajectory functions} $f_k^*(s_1)$ for $k = 2, \ldots, 5$ are LOWESS regressions of each raw PCA score on raw $s_1$.  Together, they encode the canonical sex-age mortality pattern at each mortality level -- a continuous model life table system in Tucker coordinates, parameterised by the level score.

\begin{figure}[!htbp]
\centering
\includegraphics[width=\textwidth]{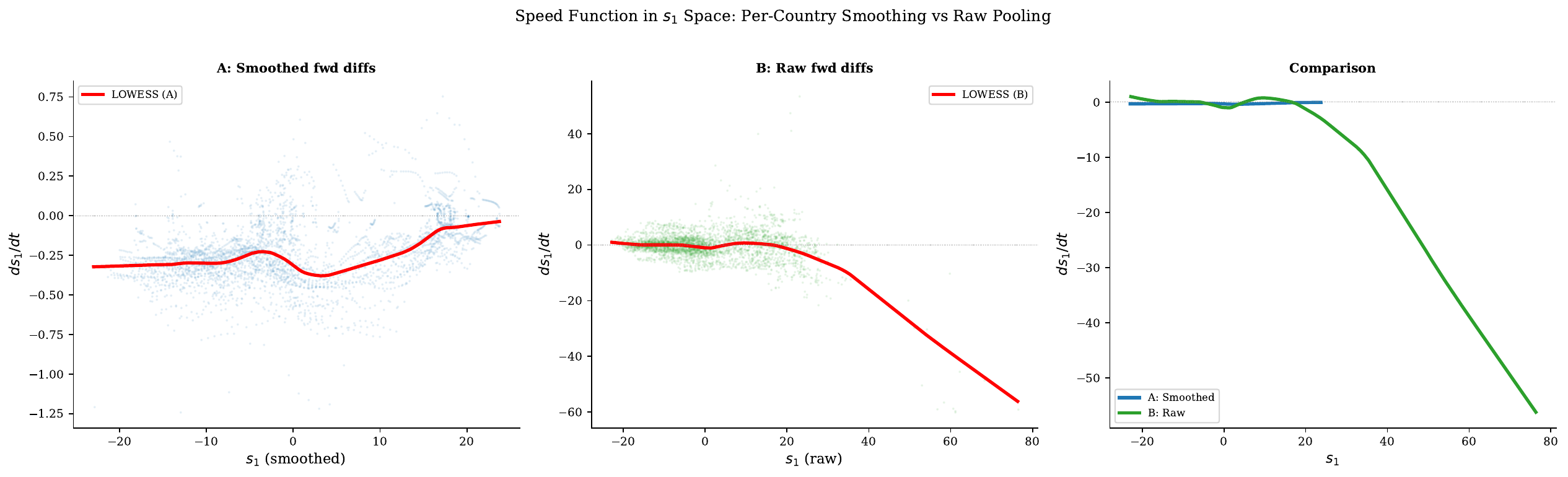}
\caption{Speed function denoising comparison in $s_1$ space.  Left:
per-country LOWESS-smoothed forward differences pooled across countries (Method~A, production) -- the smoothing reveals the underlying improvement trend.  Centre: raw forward differences pooled directly (Method~B) -- the cross-country LOWESS alone cannot fully denoise the year-to-year noise.  Right: overlay of the two LOWESS estimates, showing that per-country smoothing is essential for a well-behaved speed function.}
\label{fig:speed-diagnostic}
\end{figure}

The smoothing pipeline for the speed function is designed to separate signal from noise: per-country LOWESS denoises year-to-year fluctuations (wars, pandemics, economic shocks) while preserving the underlying improvement trend; forward differences match the forecaster's one-step-ahead model; and the cross-country LOWESS extracts the level-dependent pattern from the pooled observations. The resulting speed profile (\cref{fig:speed-diagnostic}) is inspired by the level-dependent pattern of \citet{RafteryChunnGerlandSevcikova2013} -- improvement concentrated at intermediate mortality levels, decelerating at the frontier -- but the nonparametric estimation reveals empirical structure, including asymmetry and plateau regions, that a parametric double-logistic would impose away.  Operating in score space eliminates the nonlinear mapping between the navigation coordinate and the reported $\ezero$ (\cref{sec:architecture:navigation}).

% ══════════════════════════════════════════════════════════════════════════════
\section{Forecasting Architecture}
\label{sec:architecture}
% ══════════════════════════════════════════════════════════════════════════════

% ──────────────────────────────────────────────────────────────────────────────
\subsection{Era-weighted speed function}
\label{sec:architecture:era}
% ──────────────────────────────────────────────────────────────────────────────

The canonical speed function $g^*(s_1)$ averages over the entire historical record -- Sweden in 1860 and Japan in 2010 contribute equally.  But the pace of mortality improvement at a given level has changed through time: the rapid gains of the mid-20th century epidemiological transition are not representative of contemporary dynamics at the same mortality levels.  A speed function trained on all eras systematically overpredicts improvement because it includes the fast-improvement decades alongside the recent deceleration.

We address this with a truncated exponential weighting kernel applied to the LOWESS training data.  Given a forecast origin at calendar year $t_0$, each observation at year $t$ receives weight
\begin{equation}
\label{eq:era}
w_{\text{era}}(t) = \begin{cases}
  \exp\bigl(-(t_0 - t) \cdot \ln 2 / \tau\bigr) & \text{if } t_0 - t \leq W \\
  0 & \text{otherwise}
\end{cases}
\end{equation}
where $\tau$ is the half-life of the exponential decay and $W$ is a hard window beyond which data is discarded entirely.  At $\tau = 20$ and $W = 40$: data from 20~years before the origin receives half weight; data from 40~years before receives one-quarter weight; older data is excluded.

This is applied only to the speed function, not to the trajectory functions $f_k^*(s_1)$ -- the canonical sex-age mortality patterns are structural features of the mortality transition that change slowly, while the \emph{pace} at which countries traverse the transition is era-dependent.  Cross-validation over $\tau \in \{10, 12, 15, 20, 30\}$ finds the optimum at $\tau = 12$ (\cref{sec:results}).  \Cref{sec:app:math:flow} details the two-stage LOWESS fitting procedure and the weighted bootstrap implementation of era weighting.

% ──────────────────────────────────────────────────────────────────────────────
\subsection{Score-space navigation}
\label{sec:architecture:navigation}
% ──────────────────────────────────────────────────────────────────────────────

A natural approach would navigate in $\ezero$ space: accumulate $\ezero(h) = \ezero(h-1) + g(\ezero(h-1))$, look up canonical scores at the current $\ezero$, reconstruct the mortality surface, and report the surface-derived $\ezero$.  However, this creates two distinct $\ezero$ values at each horizon: the \emph{navigation} $\ezero$ (the scalar accumulator) and the \emph{surface} $\ezero$ (computed from the Tucker reconstruction through the nonlinear expit and life-table chain).  These diverge systematically because the trajectory functions $f_k(\ezero)$ are smoothed independently, the expit transform is concave in the relevant range, and the life-table calculation is nonlinear in $q_x$.  In cross-validation, $\ezero$-space navigation produces a persistent negative bias of $\sim$1.6~years.

The solution is to navigate in score space directly.  Since $s_1 \approx \text{level}$, we define the speed function as $g^*(s_1) = \mathrm{d}s_1/\mathrm{d}t$ and the trajectory functions as $f_k^*(s_1)$ for $k = 2, \ldots, 5$.  The forecast advances $s_1$ at each step; the trajectory functions map $s_1$ to the structural scores; the Tucker reconstruction produces the mortality surface; and $\ezero$ is computed from this surface for reporting.  Because $s_1$ \emph{is} the first coordinate of the reconstructed score vector -- not a separate accumulator -- there is no divergence between the navigation variable and the reconstruction.  The nonlinear expit/life-table mapping is applied once, at the end, to compute the reported $\ezero$ -- it never feeds back into the navigation loop.

% ──────────────────────────────────────────────────────────────────────────────
\subsection{Speed dynamics}
\label{sec:architecture:speed}
% ──────────────────────────────────────────────────────────────────────────────

The forecast advances $s_1$ at each step using a convex combination of the era-weighted canonical speed and the country's own recent $s_1$ velocity:
\begin{equation}
\label{eq:speed}
v_{s_1}(h) = \bigl[1 - (1-w)\,\alpha_v^h\bigr] \cdot g^*_\tau\bigl(s_1(h-1)\bigr) + (1-w)\,\alpha_v^h \cdot v_{s_1,\text{country}}\,,
\end{equation}
\begin{equation}
\label{eq:s1step}
s_1(h) = s_1(h-1) + v_{s_1}(h)\,,
\end{equation}
where $g^*_\tau$ is the era-weighted speed function in $s_1$ space, $v_{s_1,\text{country}}$ is the country's trailing-mean $s_1$ velocity at the forecast origin (mean of the last 5 raw forward differences), $\alpha_v$ is the empirical speed relaxation rate (\cref{sec:architecture:convergence}), and $w \in [0, 1]$ is the speed blend weight.  The blending weight $(1-w)\,\alpha_v^h$ controls how much country-specific drift enters: at $h = 1$ the country receives its maximum influence of $(1-w)$; as $h$ grows and $\alpha_v^h \to 0$ the velocity converges to the canonical $g^*_\tau(s_1)$.

Cross-validation finds the optimal blend weight at $w = 1.0$ (fully canonical), though the MAE varies by only a few hundredths of a year across the full range of $w$.  At $w = 1$, \cref{eq:speed} simplifies to $v_{s_1}(h) = g^*_\tau(s_1(h-1))$ and the forecast reduces to a deterministic integration along a curve in Tucker PCA space.

% ──────────────────────────────────────────────────────────────────────────────
\subsection{Score relaxation}
\label{sec:architecture:relax}
% ──────────────────────────────────────────────────────────────────────────────

The level score $s_1$ is the navigation variable -- it is advanced by the speed function (\cref{eq:s1step}) and is not relaxed.  The structural scores $s_k$ for $k = 2, \ldots, 5$ relax from the country's actual current value toward the canonical trajectory:
\begin{equation}
\label{eq:relax}
s_k(h) = \alpha_{s,k}^h \cdot s_k^{\text{actual}} + (1 - \alpha_{s,k}^h) \cdot f_k^*\bigl(s_1(h)\bigr)\,,
  \quad k = 2, \ldots, 5\,,
\end{equation}
where $\alpha_{s,k} \in [0, 1)$ is the per-component relaxation rate, calibrated empirically from the observed autocorrelation of score deviations (\cref{sec:architecture:convergence}).  The structural components have half-lives of 12--32~years (PC~2: 32~yr, PC~3: 30~yr, PC~4: 12~yr, PC~5: 29~yr).

The relaxation gives the forecast \emph{memory} of the country's current deviation from the canonical sex-age structure.  A country whose age pattern differs from the HMD average at its mortality level -- for example, Eastern European countries with excess working-age male mortality -- will retain that distinctive structure for decades (consistent with the 12--32~year empirical half-lives) and gradually converge toward the canonical trajectory.

% ──────────────────────────────────────────────────────────────────────────────
\subsection{Tucker reconstruction and $\ezero$ computation}
\label{sec:architecture:tucker}
% ──────────────────────────────────────────────────────────────────────────────

At each horizon $h$, the forecast score vector $\bm{s}(h) = (s_1(h), s_2(h), \ldots, s_5(h))$ -- where $s_1(h)$ comes from the speed function and $s_2, \ldots, s_5$ from score relaxation -- is mapped back to the full $\logit(\qx)$ schedule:
\begin{equation}
\label{eq:reconstruct}
\hat{M}_{:,:}(h) = \bS \,
  \bigl(\bar{g} + \bm{s}(h) \cdot V\bigr)^{\text{reshaped}} \,
  \bA^\top\,,
\end{equation}
where $V$ contains the PCA loadings and $\bar{g}$ is the mean $\vect(G_{ct})$.  Because the reconstruction uses the shared factor matrices $\bS$ and $\bA$, the resulting female and male schedules are structurally coherent by construction.  Life expectancy $\ezero$ is then computed from the reconstructed schedule through the standard expit and life-table chain -- this is the only point at which the nonlinear $\logit(\qx) \to \qx \to \ezero$ mapping is applied, and its output is never fed back into the navigation.

The five-component PCA captures 97.1\% of the variance in $\vect(G_{ct})$, but the remaining 2.9\% includes country-specific sex-differential structure -- particularly for countries whose sex gap deviates substantially from the HMD average (e.g.\ Russia, Japan). To avoid a visible discontinuity in the sex differential at the forecast origin, the forecast surface incorporates a \emph{jump-off correction}: the residual between the full-rank Tucker reconstruction $\hat{M}_{:,:,c,T}$ and its five-component approximation is added to the forecast surface and decayed exponentially with a half-life of 2~years:
\begin{equation}
\label{eq:jumpoff}
\hat{M}_{:,:}(h) = \bS \,
  \bigl(\bar{g} + \bm{s}(h) \cdot V\bigr)^{\text{reshaped}} \,
  \bA^\top
  \;+\; 2^{-h/2} \cdot \Delta_0\,,
\end{equation}
where $\Delta_0 = \hat{M}_{:,:,c,T} - \bS \, (\bar{g} + \bm{s}_T \cdot V)^{\text{reshaped}} \, \bA^\top$ is the origin residual.  By $h = 2$ the correction has halved; by $h = 10$ it is below 4\%; and the long-horizon forecast is determined entirely by the five-component dynamics.  The life table computation used to extract $\ezero$ from the reconstructed schedule is specified in \cref{sec:app:math:forecast}.  The short half-life allows the five-component dynamics to take effect quickly while smoothing the sex-differential transition at the origin -- without the jump-off correction, the 2.9\% of variance not captured by the 5-PC approximation would produce a visible discontinuity in the sex gap at the forecast origin.

% ──────────────────────────────────────────────────────────────────────────────
\subsection{Trajectory extrapolation beyond observed data}
\label{sec:architecture:extrapolation}
% ──────────────────────────────────────────────────────────────────────────────

The trajectory functions $f_k^*(s_1)$ and speed function $g^*(s_1)$ are LOWESS fits to observed HMD data.  In $s_1$ space, the mortality frontier is at low $s_1$ values (corresponding to high $\ezero$ $\approx 85$, primarily Japan).  Beyond the observed range, the LOWESS interpolant holds each function constant at its boundary value -- the structural scores stop evolving and the reconstructed mortality surface saturates.

\textbf{Fix: joint tangent extension.}  At a transition point $s_1^* \approx -12$ (the $s_1$ value corresponding to $\ezero
\approx 78$, well inside the observed range), the LOWESS slope of each function is estimated via finite differences.  These slopes form a \emph{joint tangent vector} -- the empirical direction of score-space movement at the mortality frontier.  Beyond the transition, the LOWESS values are replaced by linear extrapolation along this tangent, with a smooth-step blend:
\begin{equation}
\label{eq:tail}
f_k^*(s_1) = \begin{cases} f_k^{*,\text{LOWESS}}(s_1) & \text{if } s_1 \geq s_1^* \\[4pt]
  \text{blend of LOWESS and linear} & \text{if } s_1^* - 3 \leq s_1 < s_1^* \\[4pt] f_k^*(s_1^*) + t_k(s_1 - s_1^*) & \text{if } s_1 < s_1^* - 3
\end{cases}
\end{equation}
where $t_k$ is the LOWESS slope at $s_1^*$.  The same treatment is applied to the speed function.  The smoothstep blend function and slope estimation are detailed in \cref{sec:app:math:flow}.  Because all slopes come from the same $s_1$ region, the joint covariance of the score trajectories is preserved -- all components continue to move in the direction established by the well-observed interior data.

\begin{figure}[!htbp]
\centering
\includegraphics[width=\textwidth]{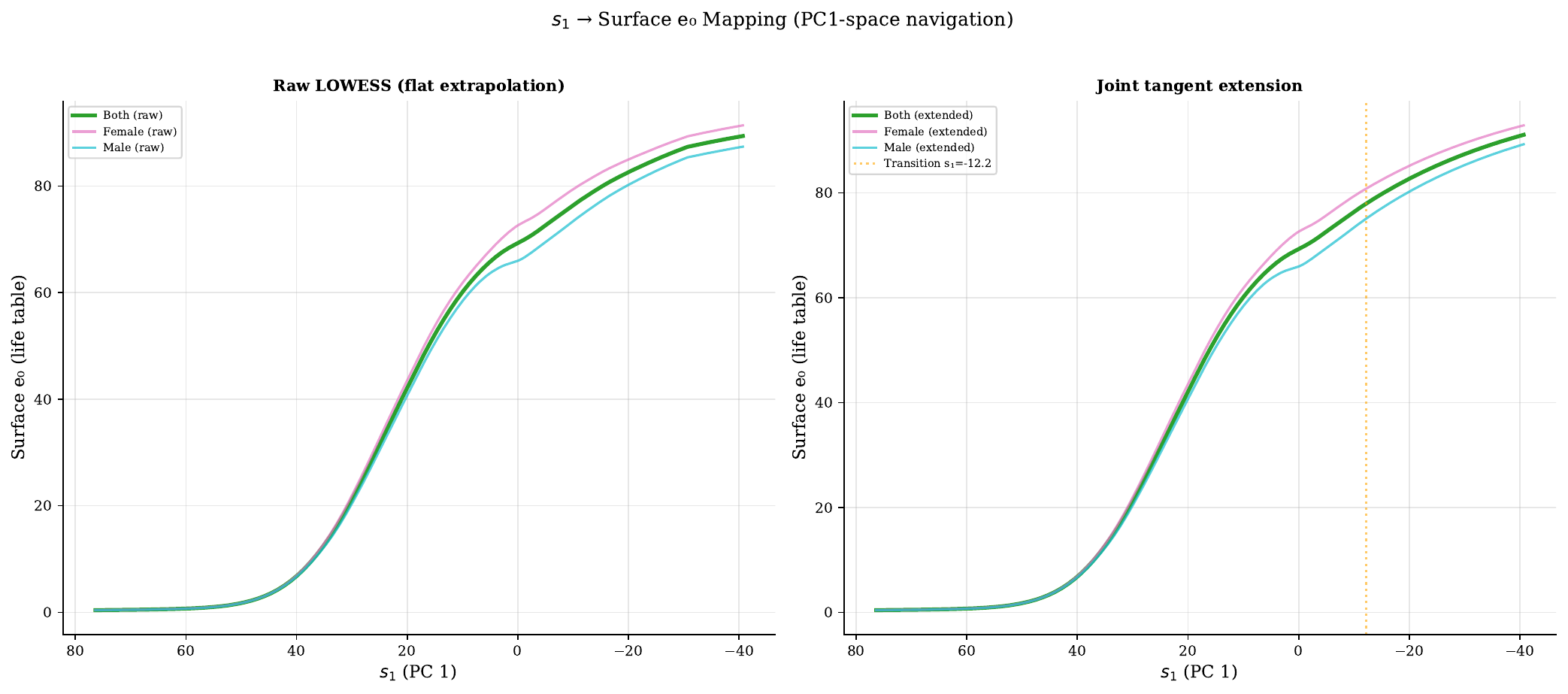}
\caption{$s_1$-to-surface-$\ezero$ mapping.  Left: raw LOWESS
with flat extrapolation -- surface $\ezero$ saturates at the frontier.  Right: with joint tangent extension from $s_1^*
\approx -12$ ($\ezero \approx 78$) -- surface $\ezero$ continues to improve monotonically.  Pink: female; cyan: male; green: both-sex average.}
\label{fig:e0-mapping}
\end{figure}

\begin{figure}[!htbp]
\centering
\includegraphics[width=\textwidth]{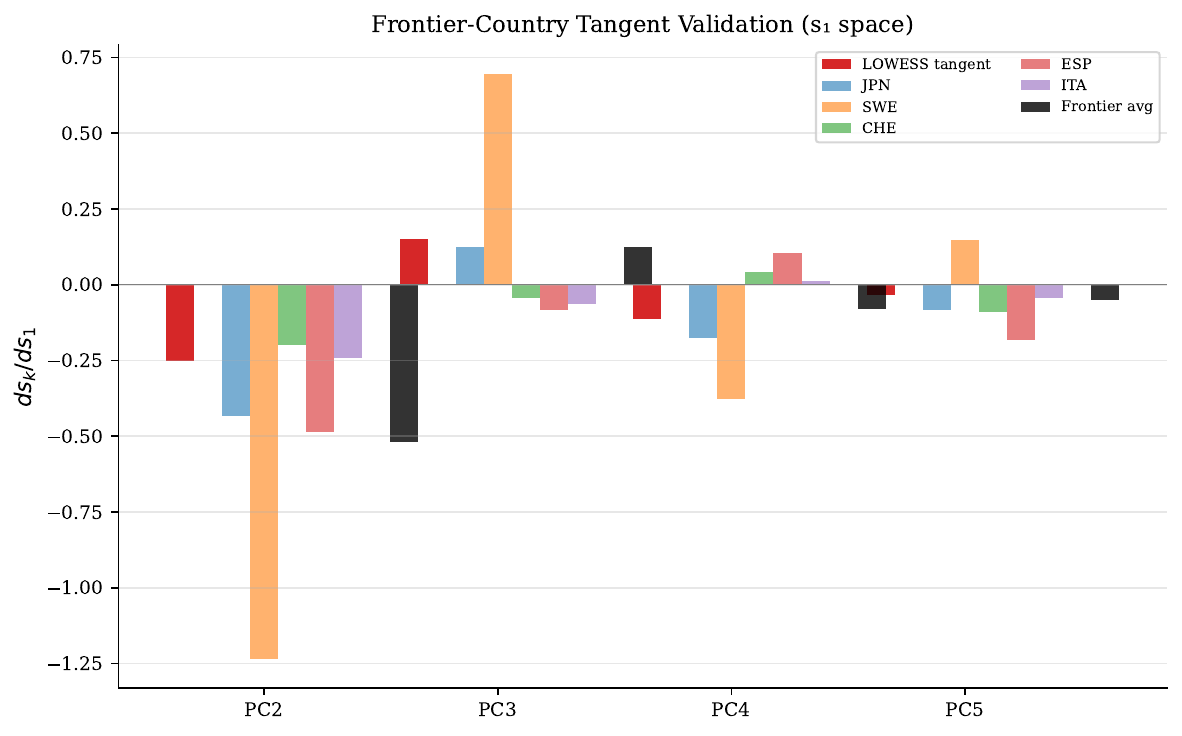}
\caption{Validation of the joint tangent extrapolation in $s_1$ space.
Per-component score slopes $\mathrm{d}s_k/\mathrm{d}s_1$ for the LOWESS tangent at $s_1^* \approx -12$ ($\ezero \approx 78$, red) and for five frontier countries (Japan, Sweden, Switzerland, Spain, Italy) over their last 20~years.  The cosine similarity between the LOWESS tangent and the frontier average is 0.94, confirming that the extrapolation direction agrees with observed frontier dynamics.  The magnitude ratio is 0.59 -- the tangent extrapolation is $\sim$40\% conservative in speed relative to frontier countries, producing a modestly cautious long-horizon forecast.}
\label{fig:frontier-validation}
\end{figure}

\begin{figure}[!htbp]
\centering
\includegraphics[width=\textwidth]{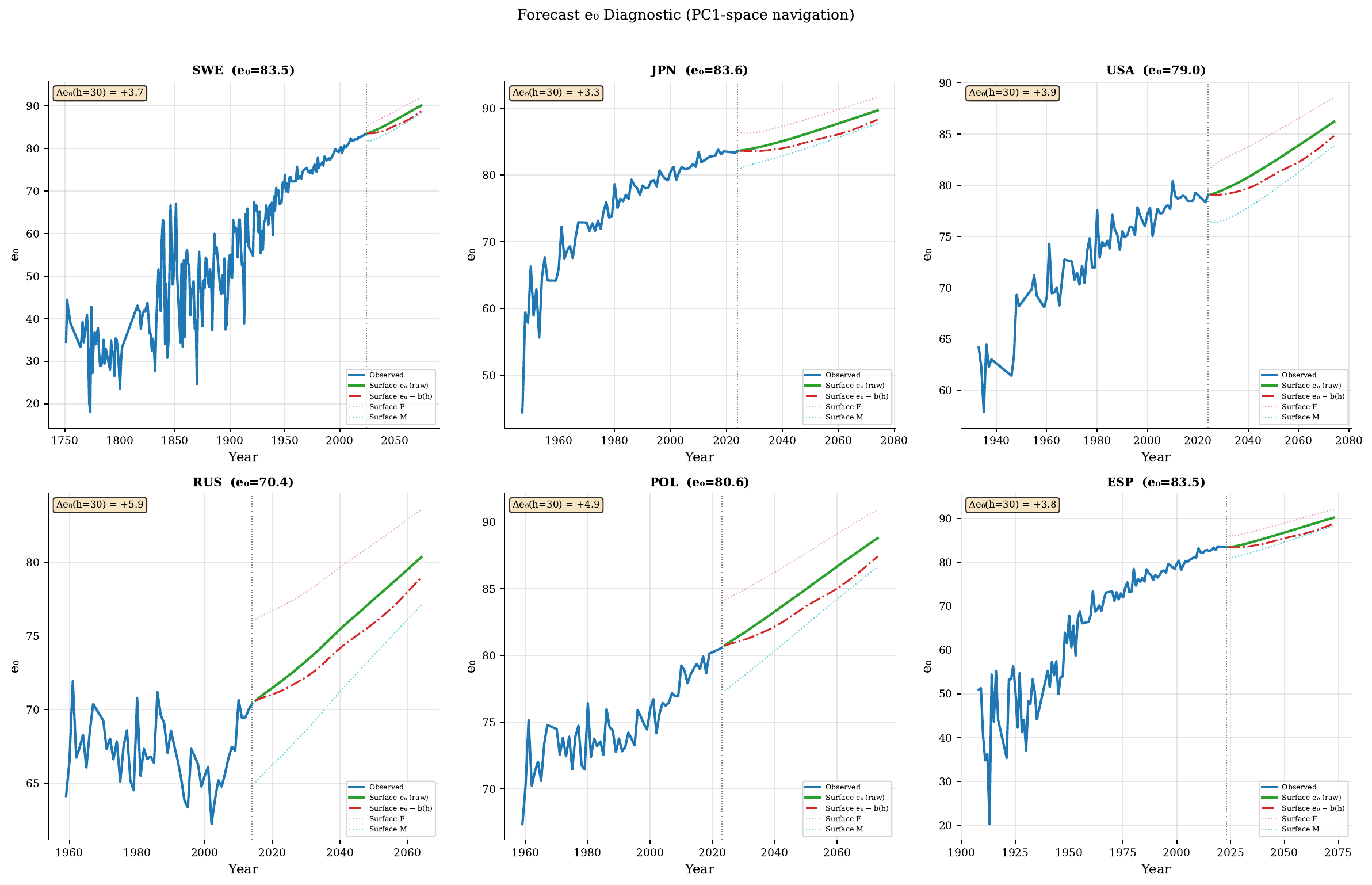}
\caption{Forecast $\ezero$ diagnostic for six countries under
$s_1$-space navigation (all-data flow field).  Green: surface-derived $\ezero$ (raw, before bias correction).  Red dash-dot: bias-corrected $\ezero$ (reported forecast).  The annotation shows the 30-year $\ezero$ gain.  Because navigation is in $s_1$ space, there is no separate navigation $\ezero$ that can diverge from the surface $\ezero$.}
\label{fig:e0-diagnostic}
\end{figure}

\Cref{fig:e0-mapping} shows the $s_1$-to-surface-$\ezero$ mapping with and without the tail extension.  \Cref{fig:frontier-validation} validates the tangent direction by comparing it to the actual direction of score movement in the five highest-$\ezero$ countries (Japan, Sweden, Switzerland, Spain, Italy) over their most recent 20~years.  \Cref{fig:e0-diagnostic} shows the resulting forecast $\ezero$ trajectories for six countries -- because the system navigates in $s_1$ space, there is no separate navigation $\ezero$ that can diverge from the surface $\ezero$.

% ──────────────────────────────────────────────────────────────────────────────
\subsection{Prediction intervals}
\label{sec:architecture:pi}
% ──────────────────────────────────────────────────────────────────────────────

The forecast uncertainty is estimated empirically from the cross-validation error distribution.  A horizon-dependent bias correction $b(h)$, estimated by LOWESS regression of the CV errors on horizon, is subtracted from the raw forecast; the corrected forecast is $\hat{\ezero}(h) - b(h)$.  We model the residual uncertainty as $\sigma(h) = \kappa \cdot \sigma_1 \cdot \sqrt{h}$, the natural scaling for accumulated random-walk-like forecast errors, where $\sigma_1$ is the median of the per-horizon empirical standard deviations divided by $\sqrt{h}$, and $\kappa = \text{SD}(z\text{-scores})$ calibrates the coverage.  The 95\% prediction interval is $\hat{\ezero}(h) - b(h) \pm 1.96 \cdot \sigma(h)$. The full calibration procedure is in \cref{sec:app:math:pi}.

% ──────────────────────────────────────────────────────────────────────────────
\subsection{Empirical convergence rates}
\label{sec:architecture:convergence}
% ──────────────────────────────────────────────────────────────────────────────

The relaxation rates $\alpha$ in \cref{eq:speed,eq:relax} can be calibrated empirically rather than tuned by forecast accuracy.  For each country-year we compute the deviation from canonical: $\Delta v_{s_1}(t) = v_{s_1,\text{country}}(t) - g^*(s_1(t))$ for speed, and $\Delta s_k(t) = s_k(t) - f_k^*(s_1(t))$ for each structural score ($k = 2, \ldots, 5$; PC~1 deviations are identically zero by construction since $s_1$ is the navigation variable).  The pooled autocorrelation at lag~$h$ --
\[
\beta(h) = \frac{\sum_{c, t_0} \Delta(t_0 + h) \cdot \Delta(t_0)} {\sum_{c, t_0} \Delta(t_0)^2}
\]
-- measures how much of a deviation persists $h$ years later.  If convergence is exponential, $\beta(h) = \alpha^h$ and the fitted slope of $\log\beta$ vs.\ $h$ gives $\alpha$ directly.

\begin{figure}[!htbp]
\centering
\includegraphics[width=\textwidth]{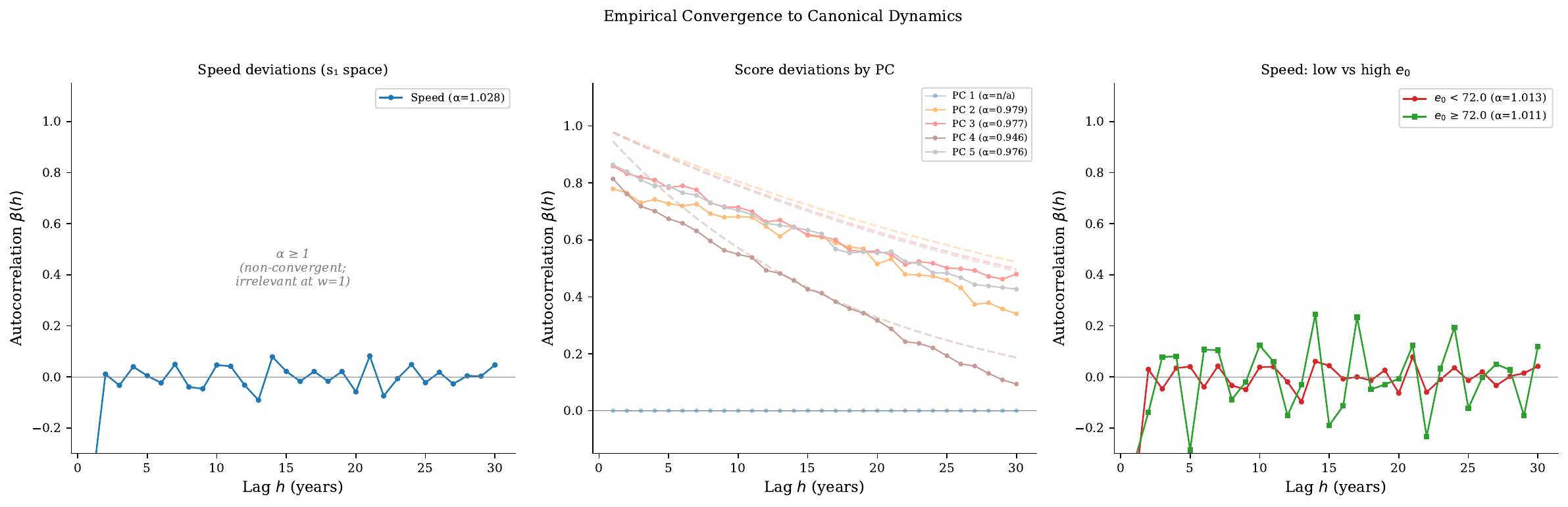}
\caption{Empirical convergence rates in $s_1$ space.  Left:
autocorrelation of $s_1$-velocity deviations from canonical. Centre: autocorrelation of structural score deviations by PC; PCs~2--5 have half-lives of 12--32~years.  Right: speed convergence conditioned on mortality level. Dashed lines show fitted exponentials $\alpha^h$.}
\label{fig:convergence}
\end{figure}

\Cref{fig:convergence} shows that structural score deviations are persistent: PCs~2--5 have half-lives of 12--32~years, confirming that a country's mortality \emph{pattern} is deeply entrenched.  These empirical rates are used directly in the production forecaster; \cref{sec:app:math:convergence} gives the autocorrelation algorithm.  Because $w = 1.0$ (fully canonical speed), the speed relaxation rate $\alpha_v$ does not affect the forecast -- the system uses canonical speed at every horizon.  The cross-validation (\cref{sec:results}) additionally selects the era half-life $\tau$ from a coarse grid; the MAE varies by only a few hundredths of a year across the full range of $w$.

% ══════════════════════════════════════════════════════════════════════════════
\section{Cross-Validation Results}
\label{sec:results}
% ══════════════════════════════════════════════════════════════════════════════

The system is evaluated by leave-country-out cross-validation: for each of the 48 HMD countries in turn, the flow field is built from the remaining 47 countries using all available years, then applied to the held-out country at multiple forecast origins (every 10~years, requiring at least 20 training years), with forecasts extending up to 50~years.  This produces 9{,}529 test points and directly tests the production use case: can HMD-wide dynamics predict a country whose data did not train the flow field?  The 50-year horizon is essential -- mortality forecasting routinely requires 50--75~year projections (the UN WPP projects to 2100; the US Social Security Administration uses 75-year horizons) -- and the flow-field system's constrained dynamics provide a distinctive advantage at these horizons.

The cross-validation proceeds in two stages.  First, a fast grid search over the era half-life $\tau$ and speed blend weight $w$ uses \emph{inclusive} flows (built from all 48~countries) to identify the optimal configuration -- this is computationally efficient because the flows need not be rebuilt for each held-out country.  Second, the optimal configuration is evaluated under \emph{strict} leave-country-out: for each held-out country, the flow field is rebuilt from the remaining 47 countries, ensuring that the held-out country's data contributes nothing to the speed function, trajectory functions, or era weighting.  All MAE, bias, and coverage statistics reported in this paper are from the strict leave-country-out evaluation -- the gold standard for out-of-sample forecast assessment.

\textbf{A note on evaluation metrics.}  All cross-validation accuracy measures, prediction interval calibration, and benchmark comparisons use raw HMD life-table $\ezero$ as ground truth.  The production forecast figures (\cref{fig:gallery,fig:e0-diagnostic,fig:schedule}) use the all-data flow field (all 48~countries) and Tucker-derived $\ezero$ for visual diagnostics.

The fast grid search (Stage~1) optimises jointly over the era half-life $\tau \in \{10, 12, 15, 20, 30\}$ and the speed blend weight $w \in \{0.2, 0.5, 1.0\}$.  The optimal configuration is $\tau = 12$, $w = 1.0$ -- that is, fully canonical speed with the selected era half-life.  The speed blend weight barely matters -- MAE varies by only a few hundredths of a year across $w \in [0.2, 1.0]$ -- confirming that the era-weighted canonical dynamics, not the country-specific velocity, drive forecast quality.  The strict leave-country-out evaluation (Stage~2) then uses this optimal configuration to produce the results reported below.

% ──────────────────────────────────────────────────────────────────────────────
\subsection{Benchmark comparison}
\label{sec:results:benchmark}
% ──────────────────────────────────────────────────────────────────────────────

Table~\ref{tab:benchmark4} compares the flow-field system against Lee--Carter \citep{LeeCarter1992}, Hyndman--Ullah \citep{HyndmanUllah2007}, and \pkg{pyBayesLife} (our de novo reimplementation of the UN production pipeline; \cref{sec:agespecific:pybayeslife}), all evaluated against raw HMD life-table $\ezero$ on \commonN{} identical (country, origin, horizon) test points.  Lee--Carter and Hyndman--Ullah are computed using the R \texttt{demography} package with HMD graduated $m_x$ and person-year exposures: Lee--Carter uses \texttt{lca(adjust="none")} with fitted jump-off rates; Hyndman--Ullah uses \texttt{fdm(order=6)} with ARIMA extrapolation of each score.

The flow-field system achieves an $\ezero$ MAE of \ffMAE~years, with the lowest bias (\ffBias~years).  Lee--Carter has the lowest overall MAE (\lcMAE~years) but substantial negative bias (\lcBias), as does Hyndman--Ullah (MAE \huMAE, bias \huBias). \pkg{pyBayesLife} has the highest MAE (\pbMAE) and the largest positive bias (\pbBias), driven by the gap model's failure to capture the closing of the female--male gap since $\sim$2000. The flow-field's bias advantage is the most distinctive result: bias of \ffBias~years vs.\ $-3.5$ to $+3.3$~years for the other three methods.  The horizon decomposition of these results (\cref{sec:results:detail}) reveals a crossover near $h = 12$ that explains the bias contrast.

\begin{table}[htbp]
\caption{$\ezero$ forecast accuracy (sex-average) on \commonN{} common test points,
all evaluated against raw HMD life-table $\ezero$.}
\label{tab:benchmark4}
\begin{tabular*}{\textwidth}{@{\extracolsep{\fill}}lrrrr@{}}
\toprule
Method & $n$ & MAE & RMSE & Bias \\
\midrule
Flow-field    & \commonN & \ffMAE & \ffRMSE & \textbf{\ffBias} \\
pyBayesLife   & \commonN & \pbMAE & \pbRMSE & \pbBias \\
LC            & \commonN & \textbf{\lcMAE} & \textbf{\lcRMSE} & \lcBias \\
HU            & \commonN & \huMAE & \huRMSE & \huBias \\
\bottomrule
\end{tabular*}
\end{table}

\begin{figure}[!htbp]
\centering
\includefigifexists[width=\textwidth]{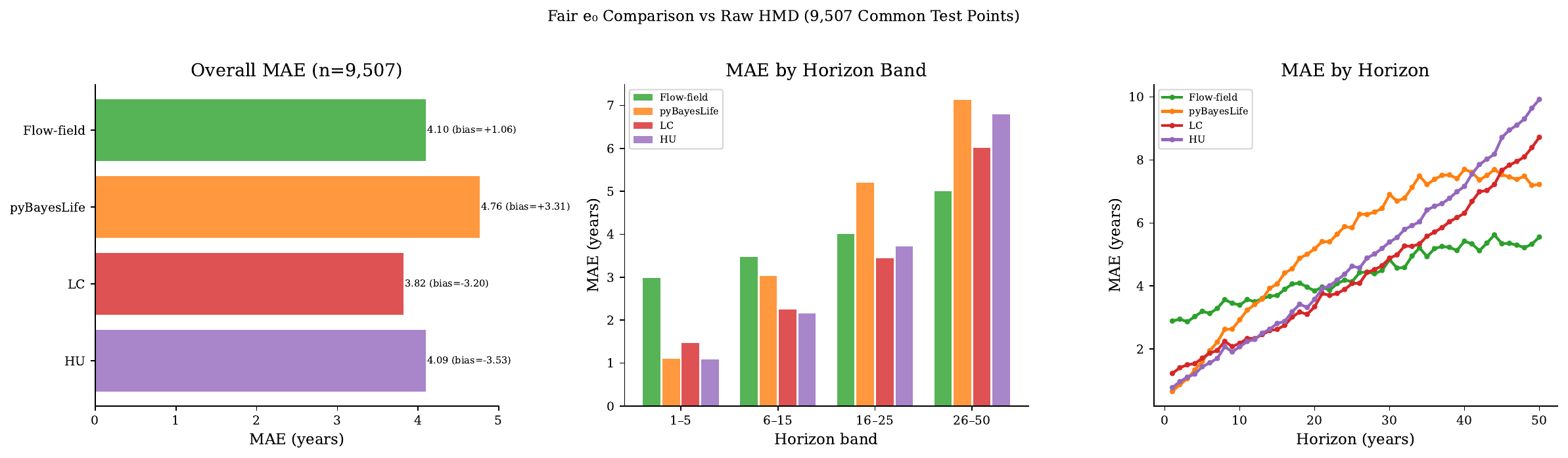}
\caption{Four-method $\ezero$ comparison on \commonN{} common test points,
all evaluated against raw HMD life-table $\ezero$.  Left: overall MAE with bias.  Centre: MAE by horizon band -- \pkg{pyBayesLife} has the lowest MAE at short horizons but accumulates the largest positive bias; Lee--Carter has the lowest overall MAE but substantial negative bias. Right: MAE by individual horizon year -- the crossover near $h = 12$ is clearly visible.}
\label{fig:benchmark-4method}
\end{figure}

% ──────────────────────────────────────────────────────────────────────────────
\subsection{Leave-country-out holdout gallery}
\label{sec:results:holdout}
% ──────────────────────────────────────────────────────────────────────────────

\begin{figure}[!htbp]
\centering
\includegraphics[width=\textwidth]{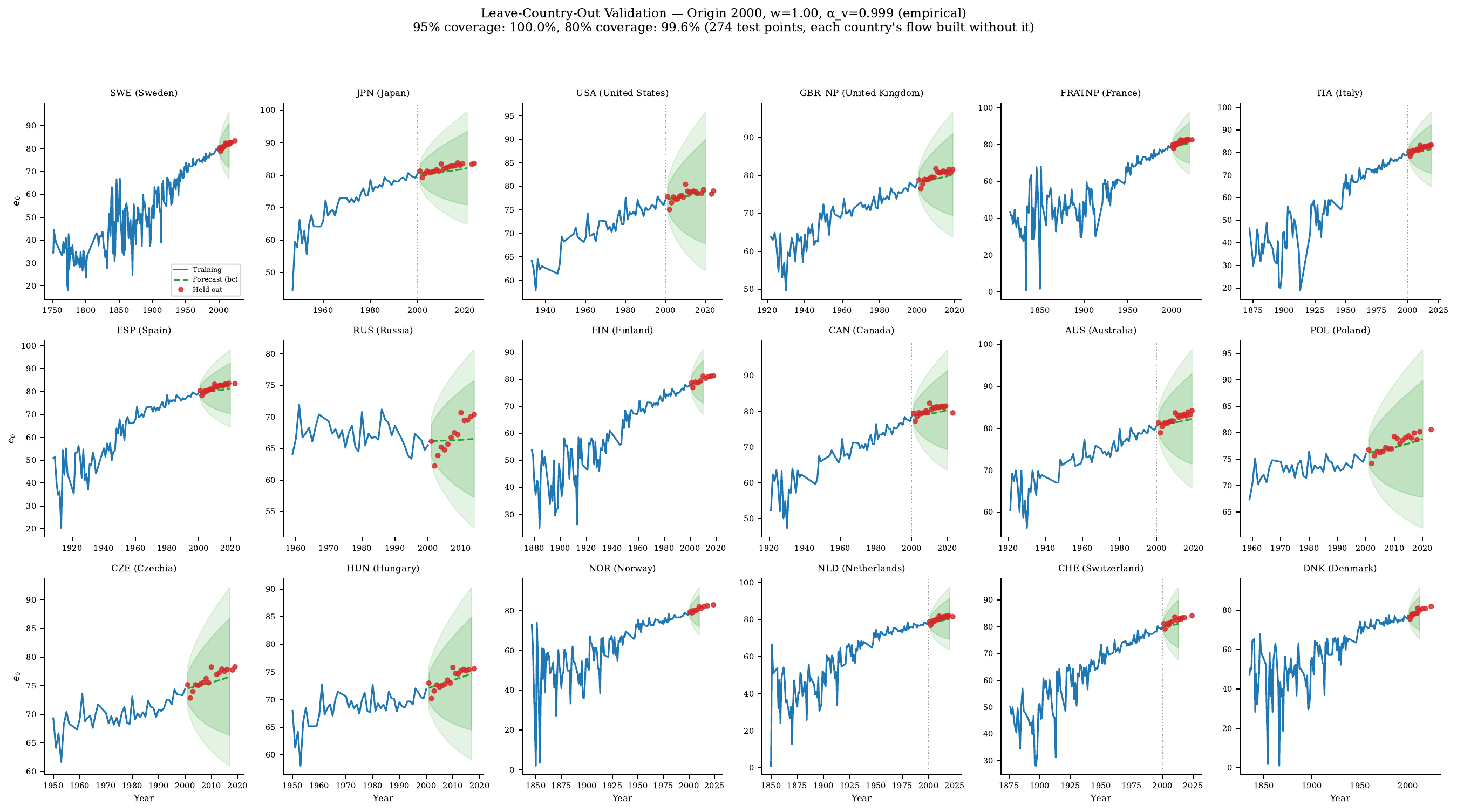}
\caption{Leave-country-out validation from the 2000 origin.  Blue:
training data (pre-2000).  Red dots: held-out observations.  Green dashed: median forecast with 80\% (dark shading) and 95\% (light shading) prediction intervals.  Each country's flow field was built excluding that country.}
\label{fig:holdout}
\end{figure}

\begin{figure}[!htbp]
\centering
\includegraphics[width=\textwidth]{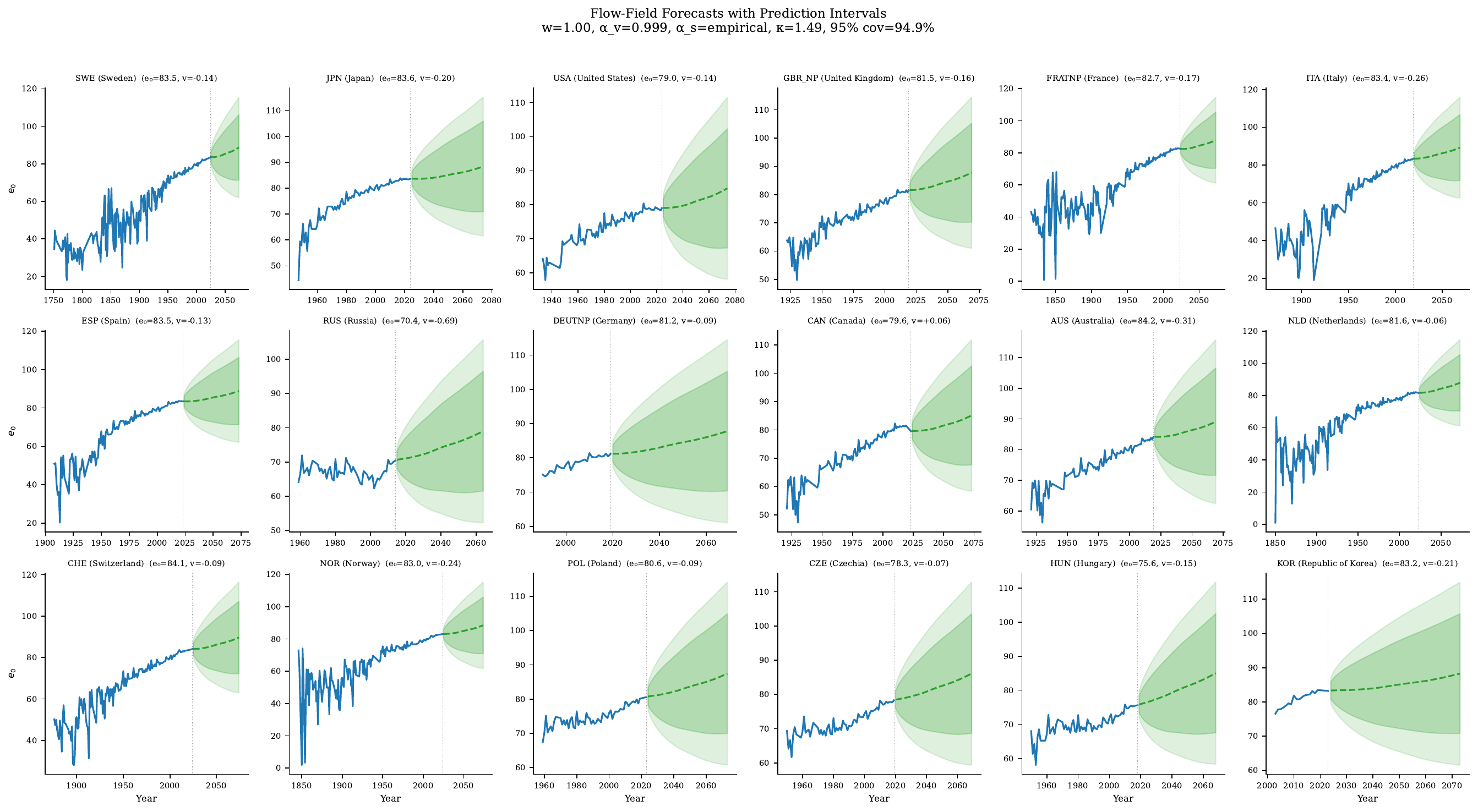}
\caption{50-year production forecasts with calibrated prediction
intervals for 18 selected countries (all-data flow field).  Blue: observed $\ezero$.  Green dashed: median forecast.  Shaded: 80\% (dark) and 95\% (light) prediction intervals.  These forecasts use the flow field trained on all 48~countries; see \cref{sec:results:benchmark} for the distinction between all-data and strict leave-country-out evaluation.}
\label{fig:gallery}
\end{figure}

\Cref{fig:holdout} shows leave-country-out forecasts from the 2000 origin for 18 selected countries.  For each country, the flow field is built from the other 47 countries; the forecast (green dashed) with 80\% and 95\% prediction intervals is plotted against the held-out observations (red dots) that the model did not see during training.  The fan opens with $\sqrt{h}$ scaling.  \Cref{fig:gallery} shows the full 50-year production forecasts with prediction intervals for the same countries.

% ──────────────────────────────────────────────────────────────────────────────
\subsection{Accuracy by horizon}
\label{sec:results:detail}
% ──────────────────────────────────────────────────────────────────────────────

The horizon profile (\cref{fig:benchmark-4method}, right panel) reveals a crossover pattern.  At short horizons ($h = 1$--5), Lee--Carter (MAE~\lcMAEShortH), Hyndman--Ullah (MAE~\huMAEShortH), and \pkg{pyBayesLife} (MAE~\pbMAEShortH) outperform the flow-field (MAE~\ffMAEShortH) because their jump-off adjustment and time-series extrapolation capture recent country-specific momentum more effectively than the flow-field's era-weighted canonical speed. The crossover occurs near $h = 12$: beyond this point, the time-series extrapolation in Lee--Carter and Hyndman--Ullah begins to drift into implausible territory while the flow-field remains constrained to the canonical trajectory.

At long horizons ($h = 26$--50), the flow-field has a substantial advantage.  The flow-field achieves MAE~\ffMAEVlongH{} while Lee--Carter reaches \lcMAEVlongH, Hyndman--Ullah \huMAEVlongH, and \pkg{pyBayesLife} \pbMAEVlongH. Lee--Carter and Hyndman--Ullah extrapolate time-series trends indefinitely: a country whose $k_t$ slope is $-0.3$/year in 2000 accumulates $-15$ over 50~years, far beyond any historically observed range.  \pkg{pyBayesLife}'s double-logistic model constrains the trajectory but accumulates positive bias from the gap model's failure to track the closing female--male gap.  The flow-field forecast, by contrast, converges to the era-weighted canonical trajectory -- a principled attractor grounded in the cross-sectional experience of 47 other countries -- and cannot produce implausible mortality schedules regardless of the horizon.

The aggregate bias is a central result.  In strict leave-country-out cross-validation against raw HMD $\ezero$, the flow-field bias is \ffBias~years -- substantially smaller in magnitude than Lee--Carter (\lcBias~years), Hyndman--Ullah (\huBias~years), and \pkg{pyBayesLife} (\pbBias~years). Lee--Carter and Hyndman--Ullah systematically underpredict future life expectancy, while \pkg{pyBayesLife} systematically overpredicts it, because their time-series extrapolation of temporal components implicitly assumes that the historical rate of mortality decline will persist unchanged -- and when that rate decelerates or the temporal component drifts beyond historically observed values, the extrapolation accumulates a growing bias.  The flow-field avoids this because it navigates through a constrained score space parameterised by mortality level rather than calendar time: the canonical speed function is anchored by the cross-sectional experience of 47 other countries at each level, and the forecast cannot drift into unobserved territory.

For applied demography, systematic bias is more damaging than higher variance.  Random forecast error averages out across populations and over time; systematic underprediction of life expectancy by 3--4~years leads to structurally underfunded pension systems, inadequate healthcare capacity planning, and optimistic social security trust fund projections.  The low bias of the flow-field system addresses this directly.

% ──────────────────────────────────────────────────────────────────────────────
\subsection{Prediction interval calibration}
\label{sec:results:pi}
% ──────────────────────────────────────────────────────────────────────────────

\begin{figure}[!htbp]
\centering
\includegraphics[width=\textwidth]{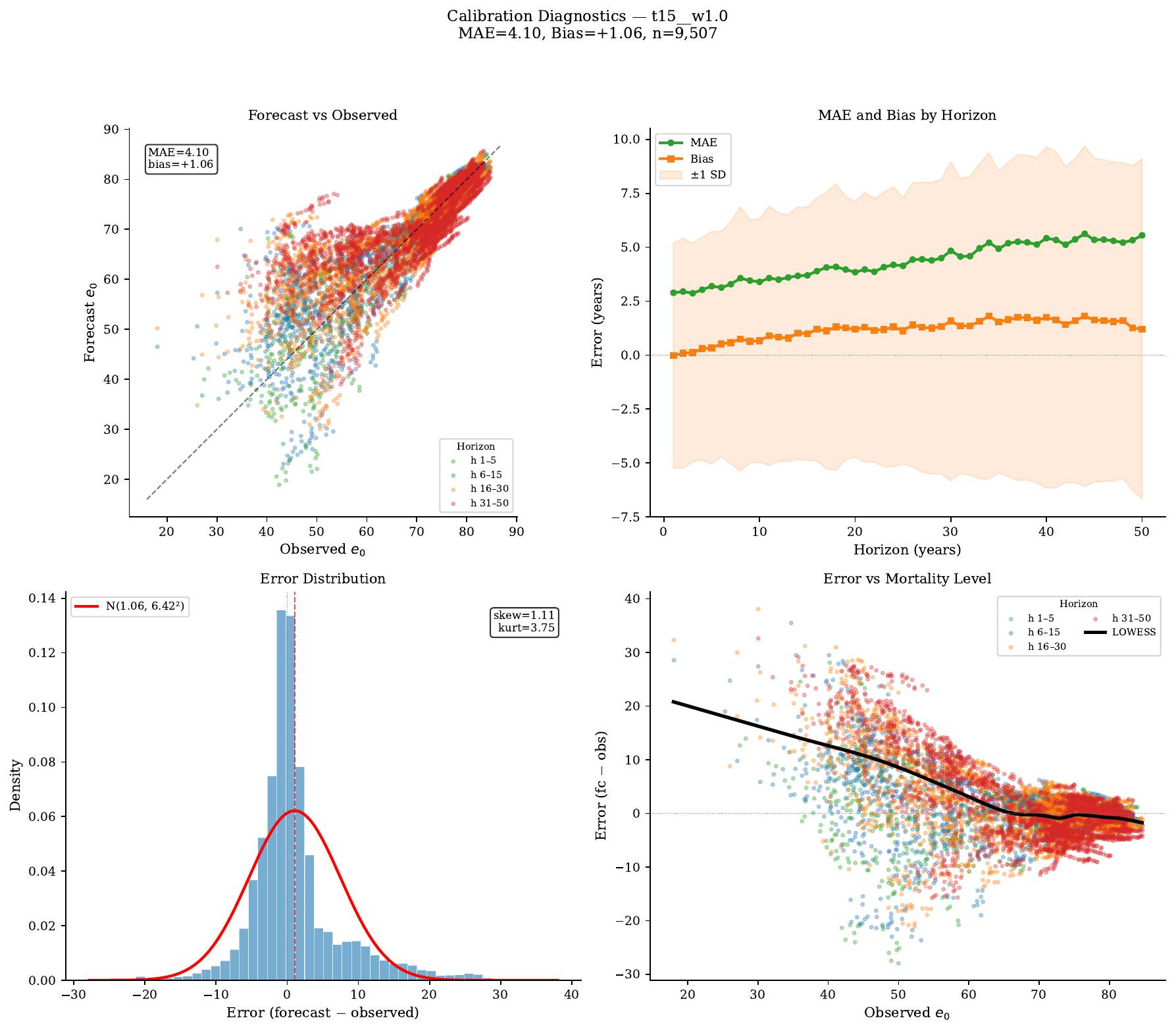}
\caption{Calibration diagnostics from strict leave-country-out CV
(\commonN{} test points; each country's flow field built excluding that country; evaluated against raw HMD life-table $\ezero$).  Top left: forecast vs observed $\ezero$.  Top right: MAE and bias by horizon. Bottom left: error distribution.  Bottom right: error vs observed $\ezero$ coloured by horizon band.}
\label{fig:calibration}
\end{figure}

The empirical $\sigma(h)$ from the CV errors is well-approximated by the $\sqrt{h}$ model.  After bias correction and calibration with $\kappa = \piKappa$, the 95\% prediction intervals achieve
\covNinetyFive\% coverage and the 80\% intervals achieve \covEighty\% coverage (\cref{fig:calibration}).

% ──────────────────────────────────────────────────────────────────────────────
\subsection{Sex-age coherence and smooth jump-off}
\label{sec:results:coherence}
% ──────────────────────────────────────────────────────────────────────────────

\begin{figure}[!htbp]
\centering
\includegraphics[width=\textwidth]{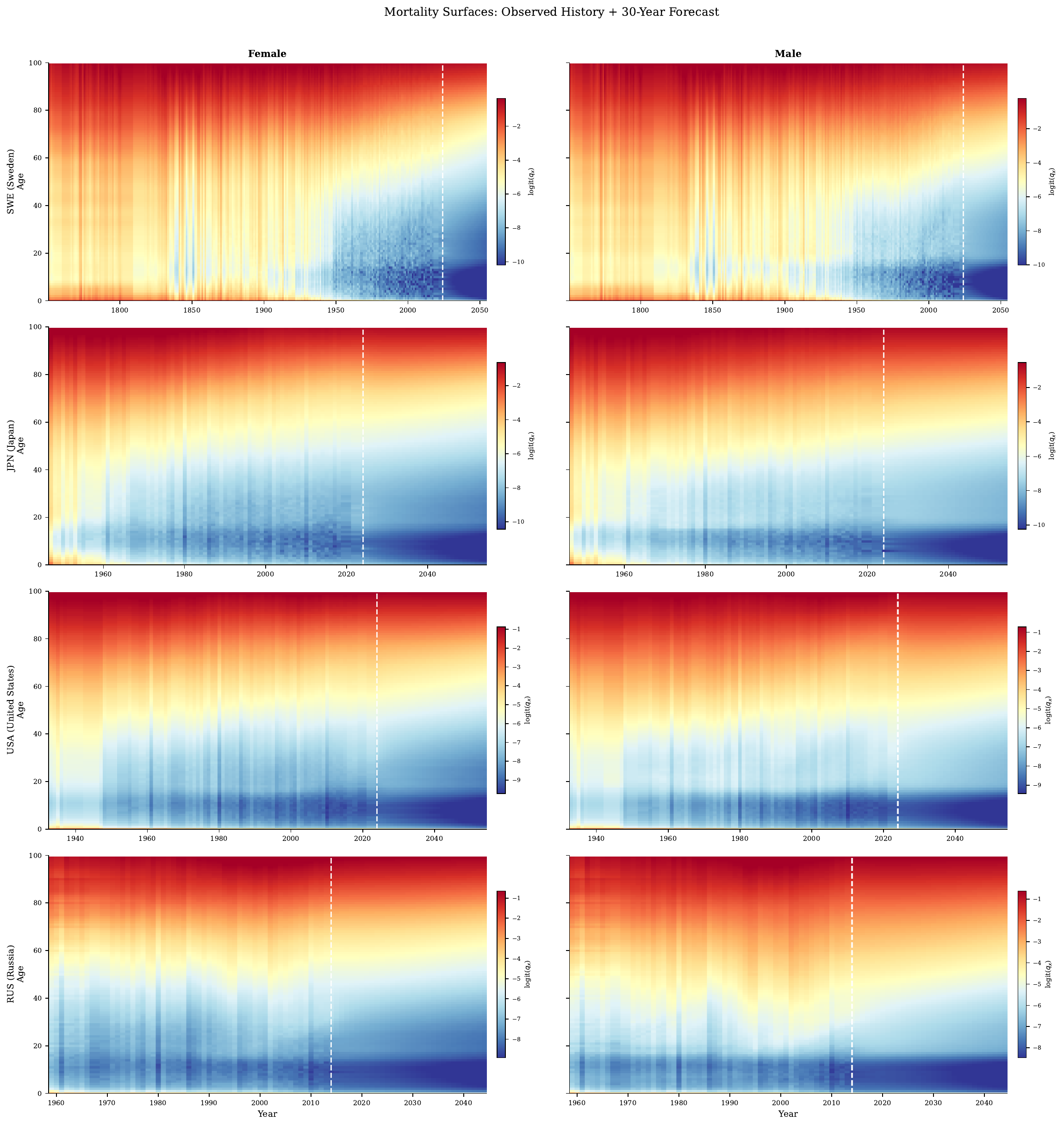}
\caption{Mortality surfaces: $\logit(\qx)$ by age and year for Sweden,
Japan, USA, and Russia (rows), female and male (columns).  The vertical dashed line marks the forecast origin.  The observed history (Tucker-reconstructed) flows seamlessly into the forecast, with smoothly evolving age-specific structure and no visible seam at the origin.}
\label{fig:surface}
\end{figure}

The Tucker framework guarantees that every forecast mortality schedule lies in the span of the shared sex and age basis matrices, making implausible outcomes -- negative mortality rates, sex crossovers, wild age-pattern oscillations -- impossible by construction. \Cref{fig:surface} demonstrates this visually: the observed history flows seamlessly into the forecast with no discontinuity at the origin and no visible artefacts in the age structure.  The forecast surfaces for Sweden, Japan, USA, and Russia all show smooth, gradually decelerating improvement across all ages, with the country-specific mortality structure -- Japan's exceptional old-age female survival, Russia's excess working-age male mortality -- persisting into the forecast and relaxing gradually toward the canonical pattern.

\begin{figure}[!htbp]
\centering
\includegraphics[width=\textwidth]{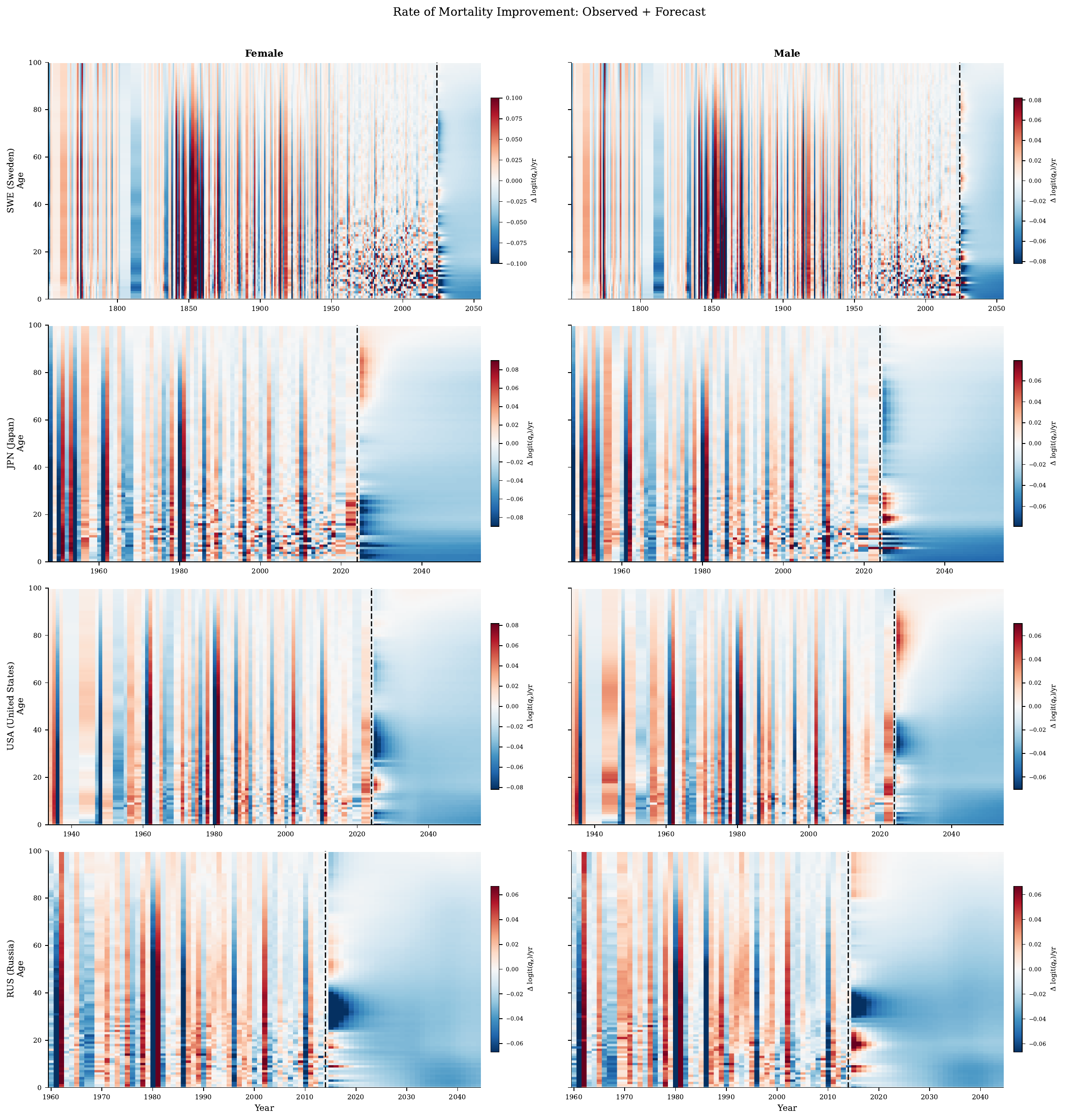}
\caption{Rate of mortality improvement: year-over-year change in
$\logit(\qx)$ by age and year for the same four countries.  Blue (negative) indicates improvement; red (positive) indicates deterioration.  Year-to-year observed changes are 10--50$\times$ larger than the smooth forecast derivatives, so the observed region has been rescaled by dividing by the per-panel mean absolute ratio, placing both regions on a common colour scale that reveals the underlying trend in the historical data alongside the forecast gradient.  Russia's working-age male mortality crisis (1990s) and partial recovery are clearly visible.}
\label{fig:surface-diff}
\end{figure}

\Cref{fig:surface-diff} shows the rate of mortality improvement (year-over-year change in $\logit(\qx)$), with the colour scale calibrated to the forecast region to reveal the age-varying gradient. In the observed history, year-to-year fluctuations are large and irregular (Russia's 1990s crisis is strikingly visible).  In the forecast region, the improvement rate varies smoothly across both age and time: improvement is faster at younger ages than at very old ages, and it decelerates gradually as countries approach higher $\ezero$ levels -- exactly the pattern implied by the era-weighted speed function and the canonical trajectory.  This age-varying improvement structure is a distinctive feature: Lee--Carter imposes a fixed age pattern of improvement (the $\beta_x$ vector), while the flow-field system allows the age pattern to evolve continuously as the country moves through Tucker PCA space.

Year-to-year observed changes in $\logit(\qx)$ are of order $\pm 0.5$/year (reflecting wars, pandemics, economic shocks, and stochastic fluctuation), while the forecast derivatives are of order $\pm 0.01$--$0.05$/year -- 10--50 times smaller.  To place both regions on a common colour scale, the observed derivatives in each panel are divided by the per-panel mean absolute ratio between observed and forecast magnitudes.  The rescaled figure reveals the underlying trend in the historical data -- the broad pattern of age-varying improvement that the flow field captures -- alongside the smooth forecast gradient, rather than saturating the observed region into a featureless block.

Some countries show a visible boundary in the forecast region around $h \approx 12$--15 ($\sim$2035), where the age pattern of improvement transitions from a country-specific structure to the near-uniform canonical pattern.  This reflects the per-component score relaxation: the fastest-converging components (half-life $\sim$12~years) have decayed to $\sim$50\% of their country-specific values by this horizon, and the remaining canonical pattern has little age variation in its derivatives.  The transition is smooth in the scores themselves -- the exponential blend $\alpha_{s,k}^h$ is continuous -- but the \emph{derivatives} of the scores exhibit a more visible regime shift as the country-specific contribution fades.  This is an honest representation of the architecture: the forecast deliberately converges to canonical dynamics at empirically measured rates, and the visual boundary is the signature of that convergence.

% ──────────────────────────────────────────────────────────────────────────────
\subsection{Sex differential coherence}
\label{sec:results:sexdiff}
% ──────────────────────────────────────────────────────────────────────────────

\begin{figure}[!htbp]
\centering
\includegraphics[width=\textwidth]{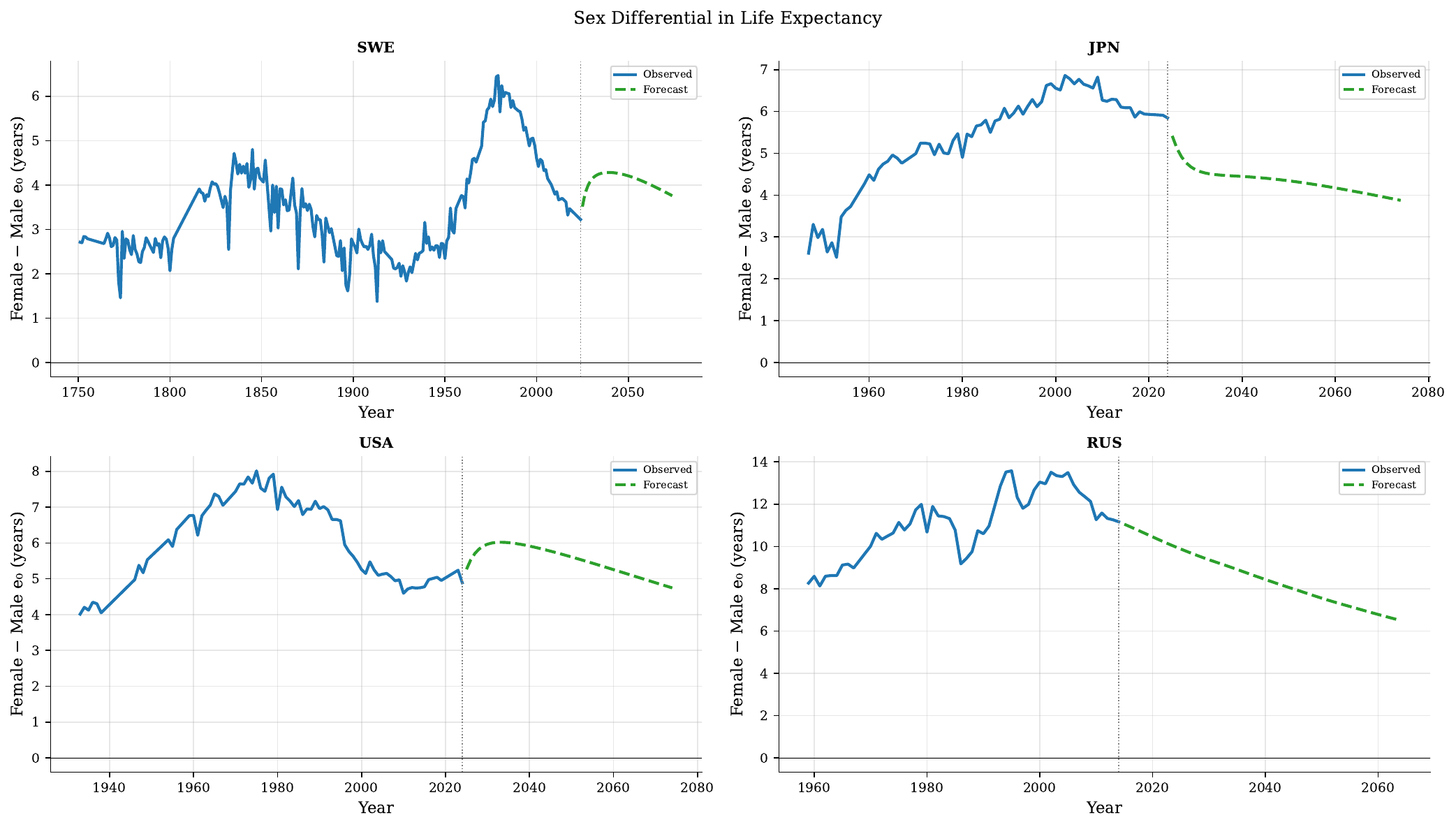}
\caption{Sex differential in life expectancy ($\ezero$ female minus
male) for Sweden, Japan, USA, and Russia.  Solid: observed.  Dashed: forecast.  The differential continues the observed trend smoothly with no crossover.}
\label{fig:sex-diff-e0}
\end{figure}

The Tucker reconstruction guarantees that male and female schedules are produced jointly through shared basis matrices, but it is useful to verify that the \emph{sex differential} evolves plausibly across the full forecast horizon.  \Cref{fig:sex-diff-e0} shows the sex gap in life expectancy ($\ezero$ female minus male) for Sweden, Japan, USA, and Russia.  In every case, the forecast differential continues the observed trend smoothly -- narrowing at high $\ezero$ levels (Sweden, Japan), stabilising (USA), or recovering from crisis-driven widening (Russia) -- without any spurious crossover.

\begin{figure}[!htbp]
\centering
\includegraphics[width=\textwidth]{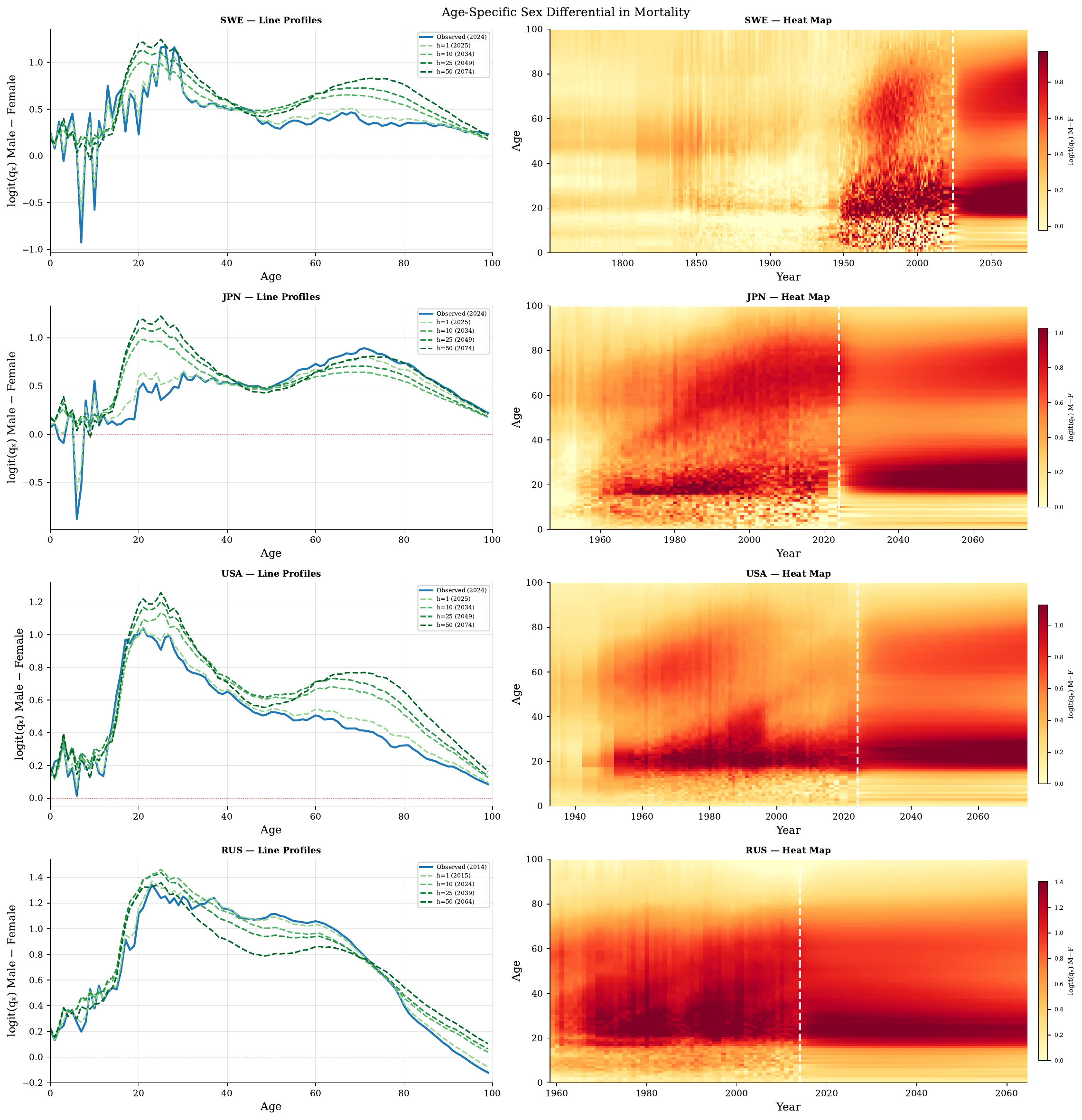}
\caption{Age-specific sex differential in mortality:
$\logit(\qx)_{\text{male}} - \logit(\qx)_{\text{female}}$.  Left: line plots at selected horizons.  Right: heat map across age and time (observed + forecast).  The differential is everywhere positive (male excess mortality) and evolves smoothly -- no age-specific crossovers at any horizon.}
\label{fig:sex-diff-age}
\end{figure}

\begin{figure}[!htbp]
\centering
\includefigifexists[width=\textwidth]{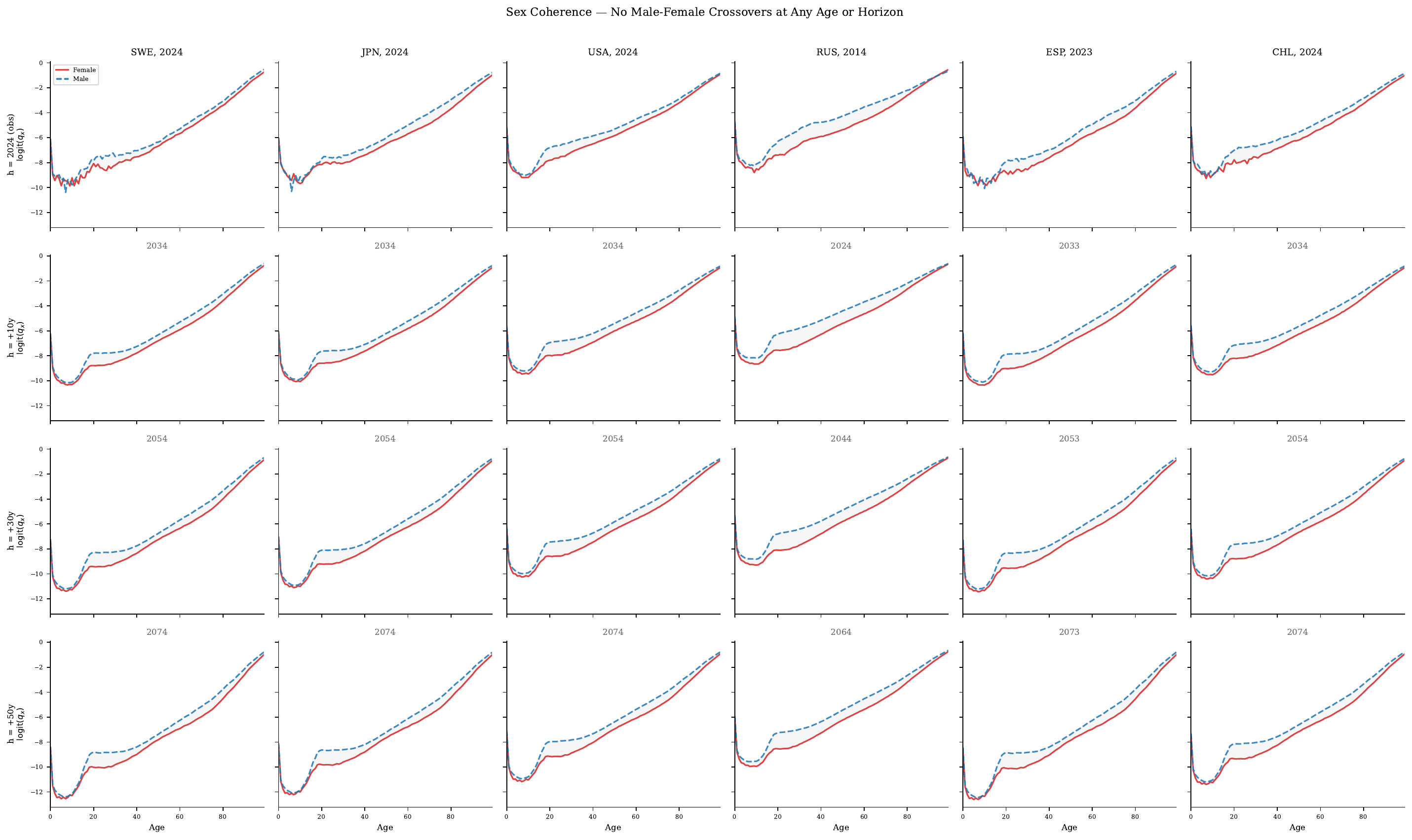}
\caption{Sex coherence of forecast mortality schedules.  Female (solid,
red) and male (dashed, blue) $\logit(\qx)$ plotted on common axes for selected countries spanning a range of mortality levels, at the last observed year and three forecast horizons (+10, +30, +50~years).  The Tucker reconstruction preserves $m_x^F < m_x^M$ at every age and horizon without requiring explicit constraints -- no crossovers occur.}
\label{fig:sex-coherence}
\end{figure}

\Cref{fig:sex-diff-age} examines the age-specific differential in $\logit(\qx)$ (male minus female): line plots at selected horizons (left) and a heat map across age and time (right).  The differential is everywhere positive (male excess mortality at every age) and evolves smoothly across both age and time.  At young working ages (15--40), where the male excess is largest, the differential narrows gradually as the country-specific scores relax toward canonical -- exactly as intended by the convergence architecture.  At older ages, the differential is small and stable. No age-specific crossover appears at any horizon: male mortality remains above female mortality at every age throughout the 50-year forecast, a structural guarantee of the shared-basis reconstruction.  \Cref{fig:sex-coherence} displays female and male schedules overlaid on common axes for several countries spanning a range of mortality levels, confirming that no crossover occurs at any age or horizon.

\begin{figure}[!htbp]
\centering
\includegraphics[width=\textwidth]{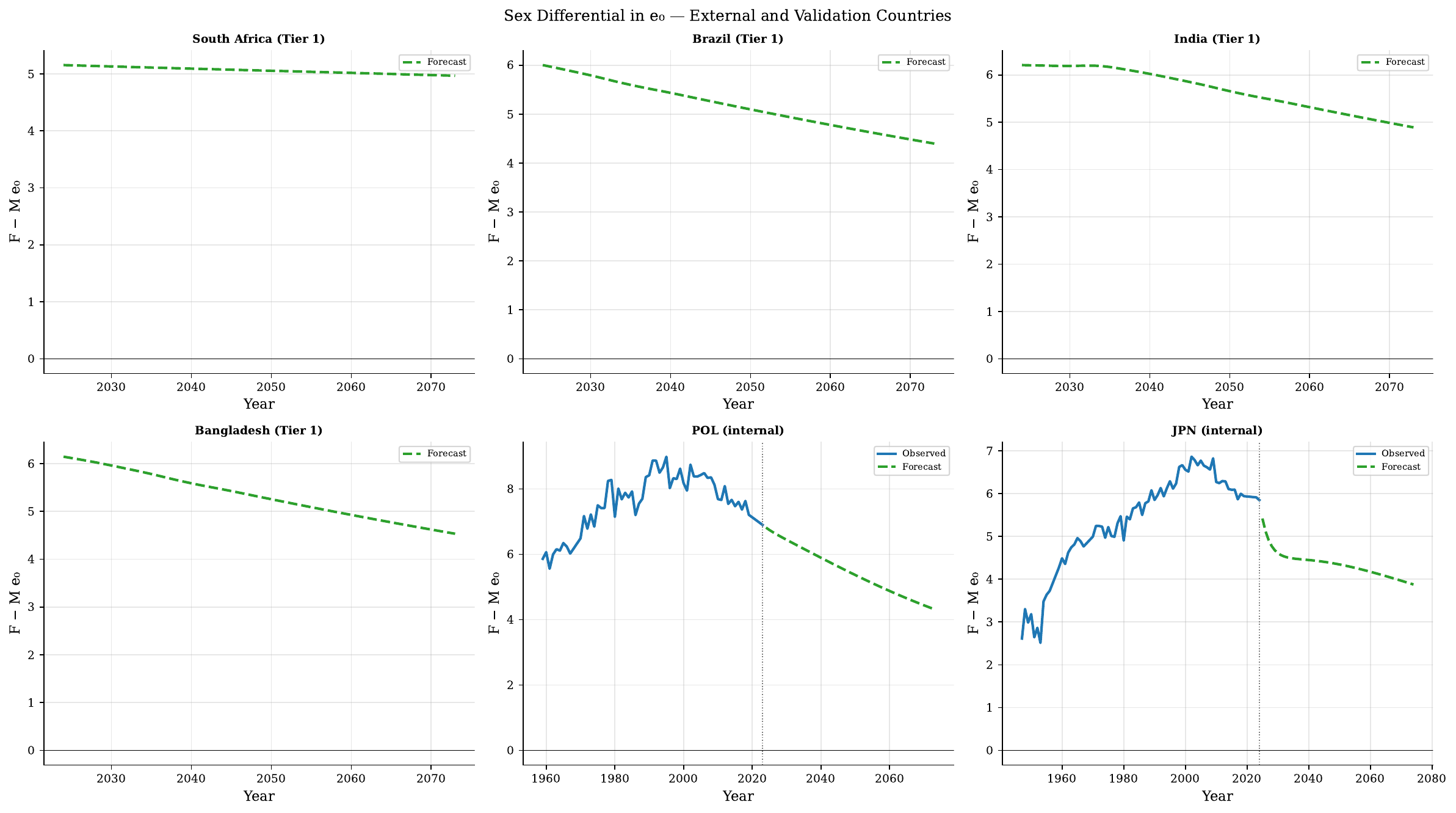}
\caption{Sex differential in $\ezero$ for Tier~1 external forecasts
(South Africa, Brazil, India, Bangladesh) and two HMD validation countries (Poland, Japan).  External countries enter the flow field at their current $\ezero$ level; the sex-specific schedules emerge entirely from Tucker reconstruction.  The differential is everywhere positive and varies with mortality level in a pattern consistent with the HMD-wide empirical relationship.}
\label{fig:sex-diff-external}
\end{figure}

For external (Tier~1) countries, the sex differential in $\ezero$ is shown in \cref{fig:sex-diff-external}.  These countries enter the flow field at their current $\ezero$ and ride the canonical dynamics forward -- the sex-specific schedules emerge entirely from Tucker reconstruction, not from any explicit sex-differential model. The differential is positive throughout and varies with mortality level in a pattern consistent with the HMD-wide empirical relationship: wider at lower $\ezero$ (generally earlier times) narrower at higher $\ezero$ (generally later times).

% ══════════════════════════════════════════════════════════════════════════════
\section{Age-Specific Comparison: Flow-Field vs.\ the $\ezero$-Mediated Pipeline}
\label{sec:agespecific}
% ══════════════════════════════════════════════════════════════════════════════

The $\ezero$ comparisons in \cref{sec:results:benchmark} evaluate all four methods on the same scalar summary.  But the flow-field's distinctive claim is that it forecasts the full sex$\times$age mortality surface directly, without collapsing to a scalar intermediate.  To evaluate whether this architectural difference translates into better age-specific forecasts, we need a benchmark that can also produce age-specific predictions -- and among the methods considered so far, only the \citeauthor{RafteryChunnGerlandSevcikova2013} pipeline does so, through its \pkg{MortCast} reconstruction stage.  Lee--Carter and Hyndman--Ullah can in principle produce age-specific forecasts, but they are single-population methods that do not pool information across countries and are not designed for the multi-population, multi-origin cross-validation framework used here.

This section describes \pkg{pyBayesLife}, our de novo Python reimplementation of the Raftery et al.\ pipeline, and presents the head-to-head age-specific comparison.

% ──────────────────────────────────────────────────────────────────────────────
\subsection{\pkg{pyBayesLife}: a de novo reimplementation}
\label{sec:agespecific:pybayeslife}
% ──────────────────────────────────────────────────────────────────────────────

A fair age-specific comparison requires that both methods be trained on the same data.  The R~\pkg{bayesLife} package \citep{SevcikobayesLife2024} loads World Population Prospects data by default into the MCMC estimation pool (version 5.3-1 adds a \code{use.wpp.data} parameter, but it reverts to \code{TRUE} when no custom data file is provided), and the \code{include\_code} mechanism cannot exclude WPP countries from the estimation --- only reclassify countries that are already loaded (see \cref{sec:appendix:raftery} for details).  A second WPP dependency exists in the heteroscedastic variance function $\sigma(\ezero)$, which in the R package is precomputed from WPP residuals and shipped as an internal dataset (\code{loess\_sd}), loaded unconditionally at package startup.  These dependencies mean that any comparison using the R package would be influenced by the hierarchical prior estimated from 150+ non-HMD countries -- information unavailable to the flow-field, which trains exclusively on HMD data.

We therefore reimplemented the entire \pkg{bayesLife} \citep{SevcikobayesLife2024} and \pkg{MortCast} \citep{SevcikMortCast2024} pipeline from scratch in Python, producing \pkg{pyBayesLife}: six modules totalling $\sim$4{,}000 lines, verified against the R implementation at every stage.  The reimplementation eliminates all R and WPP dependencies while introducing several improvements.

\paragraph{HMD-specific variance estimation.}
\pkg{pyBayesLife} computes the heteroscedastic variance function $\sigma(\ezero)$ entirely from HMD data by pooling all country-year female $\ezero$ observations, computing annual increments, LOWESS-smoothing the increments against $\ezero$ to obtain the mean function, computing residuals, and LOWESS-smoothing the absolute residuals with the standard half-normal correction $\hat{\sigma} = \widehat{E[|\varepsilon|]} \cdot
\sqrt{\pi/2}$.

\paragraph{Double-logistic formula.}
The double-logistic function in \pkg{bayesLife} is implemented entirely in compiled C code (\code{src/functions.c}, function \code{doDL}, lines~17--37); the R wrapper \code{g.dl6} (\code{R/functions.R}, lines~4--8) is a one-line foreign function call containing no formula.  Our Python implementation reproduces the \pkg{bayesLife} C output to machine precision ($\max|\Delta| = 5.33 \times 10^{-15}$ across 80 test cases covering four distinct demographic regimes); see \cref{sec:appendix:raftery} for the complete formula with line-level source citations.

\paragraph{MCMC reparameterization.}
We use NumPyro \citep{PhanPradhanJankowiak2019} with NUTS \citep{HoffmanGelman2014} and introduce three improvements over the R package's slice sampling.  All HMD countries have $\ezero > 55$, so the first double-logistic sigmoid is fully saturated; the early-transition parameters $d_1$, $d_2$, $d_3$ are non-identifiable per country and we share them across countries, eliminating $\sim$135 non-identifiable parameters.  For the three per-country parameters ($d_4$, $k$, $z$), we use a centered LogNormal hierarchy rather than the non-centered parameterization, avoiding the funnel geometry \citep{Papaspiliopoulos2007,Betancourt2017} that arises when the hierarchical variance is small.  The LogNormal guarantees positivity through the exponential transformation, providing a smooth posterior for gradient-based sampling -- in contrast to the slice sampling and absolute-value positivity enforcement used in the R package. Empirical comparison across four parameterizations shows that the shared $d_1$--$d_3$ centered LogNormal specification achieves the best $\ezero$ MAE (0.478) with zero MCMC divergences, compared to 123 divergences for the per-country absolute-value specification (\cref{sec:appendix:raftery}, Issue~10).

\paragraph{Improved Lee--Carter pipeline.}
The age-specific reconstruction uses Lee--Carter \citep{LeeCarter1992,LiLee2005} with Kannisto old-age extension \citep{Kannisto1994} and $b_x$ rotation \citep{LiLeeGerland2013}, following \citet{SevcikovaLiKantorovaGerlandRaftery2016}.  Our implementation differs from R \pkg{MortCast} \citep{SevcikMortCast2024} in three ways: the formulation $\nmx = \exp(a_x + b_x k_t)$ is positive by construction (R \pkg{MortCast} can produce $\nmx = 0$ at young ages where $\log(0) = -\infty$ contaminates $a_x$; \cref{sec:appendix:raftery}, Issue~4); Brent's method finds $k_t$ such that the life table $\ezero$ matches the target to $< 10^{-6}$~years; and given correct $\ezero$, the $\log(\nmx)$ MAE is 0.094 across 6 countries versus 0.221 for R \pkg{MortCast} on Sweden alone at 1-year resolution.

\paragraph{Male $\ezero$ via gap model.}
Male $\ezero$ is derived from the female $\ezero$ forecast via the joint gap model of \citet{RafteryLalicGerland2014}, reimplemented in Python with MCMC posteriors estimated from HMD data at each decade origin.

\Cref{tab:deps} summarises the complete elimination of R and WPP dependencies.

\begin{table}[htbp]
\small
\caption{R and WPP dependencies addressed in \pkg{pyBayesLife}.}
\label{tab:deps}
\begin{tabularx}{\textwidth}{@{}lXX@{}}
\toprule
Component & R \pkg{bayesLife}/\pkg{MortCast} & \pkg{pyBayesLife} \\
\midrule
Training data & WPP (${\sim}200$ countries, always loaded) & HMD only (48 countries) \\ Variance $\sigma(\ezero)$ & Precomputed from WPP residuals & Computed from HMD via LOWESS \\ MCMC sampler & Slice sampling in R/C & NUTS via NumPyro/JAX \\ DL formula & Documented in R; actual formula in C & Python, verified to $10^{-15}$ vs.\ C \\ DL params $d_1$--$d_3$ & Per-country (non-identifiable for HMD) & Shared across countries \\ DL params $d_4$, $k$, $z$ & Non-centered, $|\cdot|$ positivity & Centered LogNormal \\ LC positivity & Can produce $\nmx \leq 0$ & Always $\nmx > 0$ \\ $\ezero$ recovery & Iterative & Brent's method ($< 10^{-6}$~yr) \\ Time resolution & 5-year periods & Annual \\ Language & R + compiled C & Python (NumPy, JAX, NumPyro, SciPy) \\
\bottomrule
\end{tabularx}
\end{table}

% ──────────────────────────────────────────────────────────────────────────────
\subsection{Test point alignment and evaluation metrics}
\label{sec:agespecific:design}
% ──────────────────────────────────────────────────────────────────────────────

We align \pkg{pyBayesLife} to the exact same (country, origin, horizon) test points used in the flow-field cross-validation by mapping each flow-field origin to the nearest decade's cached MCMC posteriors and forecasting from the exact origin year.  This yields \commonN{} common $\ezero$ test points and \mxCommonN{} age-specific test points per method after filtering to the intersection.

All four $\ezero$ methods and both age-specific methods are evaluated against the same ground truth: raw HMD life-table $\ezero$ for the scalar comparison, and raw HMD $\nmx$ for the age-specific comparison. The $\lx$ weights used in the age-specific metric are likewise computed from raw HMD $\nmx$.  This ensures that no method benefits from being evaluated against a smoothed or reconstructed version of the data.

Age-specific forecasts are evaluated using $\log(\nmx)$ errors at each single-year age $x$:
\begin{equation}
\label{eq:eps}
\epsilon_x = \log \widehat{\nmx} - \log {}_1m_x^{\text{obs}}
\end{equation}
where $\widehat{\nmx}$ is the predicted rate and ${}_1m_x^{\text{obs}}$ is the observed HMD rate.  In addition to standard unweighted MAE, we report $\lx$-weighted MAE:
\begin{equation}
\label{eq:lxmae}
\text{MAE}_{\lx} = \frac{\sum_x \lx \, |\epsilon_x|}{\sum_x \lx}
\end{equation}
where $\lx$ is the survivorship function from the observed life table. This weights errors at ages where more people are alive more heavily, reflecting the demographic importance of accurate forecasts at ages that contribute most to person-years lived.  The summation runs over single-year ages $x = 0, 1, \ldots, 100$.  The complete set of evaluation metrics -- including the $\lx$-weighted bias and sex differential $\delta(x)$ -- is defined in \cref{sec:app:math:metrics}.

% ──────────────────────────────────────────────────────────────────────────────
\subsection{Life expectancy results on common test points}
\label{sec:agespecific:e0}
% ──────────────────────────────────────────────────────────────────────────────

Table~\ref{tab:benchmark4} and \cref{fig:benchmark-4method} (introduced in \cref{sec:results:benchmark}) present the four-method $\ezero$ comparison on \commonN{} common test points. The finding relevant to the age-specific analysis below is the horizon pattern: at short horizons ($h = 1$--5), \pkg{pyBayesLife} produces better $\ezero$ forecasts than the flow-field (MAE \pbMAEShortH{} vs.\ \ffMAEShortH{}), but as we show next, this $\ezero$ advantage does \emph{not} translate into better age-specific forecasts -- the reconstruction from $\ezero$ to $\nmx$ is so lossy that even a superior $\ezero$ forecast produces inferior age-specific results.

% ──────────────────────────────────────────────────────────────────────────────
\subsection{Age-specific mortality results}
\label{sec:agespecific:mx}
% ──────────────────────────────────────────────────────────────────────────────

Table~\ref{tab:mxcomp} presents the headline age-specific comparison, evaluated against raw HMD $\nmx$ on \mxCommonN{} common test points. The flow-field achieves an $\lx$-weighted $\log(\nmx)$ MAE of \ffMxMAElx, compared to \pbMxMAElx{} for \pkg{pyBayesLife} -- a factor of \mxRatio{} improvement.  The flow-field has lower error at every age band, every horizon, and for both sexes.

\begin{table}[htbp]
\caption{Age-specific $\log(\nmx)$ accuracy across all test points.}
\label{tab:mxcomp}
\begin{tabular*}{\textwidth}{@{\extracolsep{\fill}}lrrrr@{}}
\toprule
Method & MAE (uw) & MAE ($\lx$) & Bias ($\lx$) & $N$ \\
\midrule
Flow-field  & \textbf{\ffMxMAEuw}  & \textbf{\ffMxMAElx}  & \textbf{\ffMxBiaslx}  & \mxCommonN \\
pyBayesLife & \pbMxMAEuw  & \pbMxMAElx  & \pbMxBiaslx  & \mxCommonN \\
\bottomrule
\end{tabular*}
\end{table}

\paragraph{By age band.}
Table~\ref{tab:mxage} shows the $\lx$-weighted $\log(\nmx)$ MAE by age band.  The gap is largest at young ages (\mxRatioAgeZero$\times$ at age~0, \mxRatioAgeOneFourteen$\times$ at ages~1--14), where the $\ezero$-mediated pipeline amplifies forecast errors through the Lee--Carter redistribution.  At old ages (75--89, 90--100), the flow-field still has lower error by a factor of \mxRatioAgeSeventyFiveEightyNine--\mxRatioAgeNinetyHundred$\times$.

\begin{table}[htbp]
\caption{$\lx$-weighted $\log(\nmx)$ MAE by age band.}
\label{tab:mxage}
\begin{tabular*}{\textwidth}{@{\extracolsep{\fill}}lrrr@{}}
\toprule
Age band & Flow-field & pyBayesLife & Ratio \\
\midrule
0       & \textbf{\ffMxMAEAgeZero}              & \pbMxMAEAgeZero              & \mxRatioAgeZero$\times$ \\
1--14   & \textbf{\ffMxMAEAgeOneFourteen}       & \pbMxMAEAgeOneFourteen       & \mxRatioAgeOneFourteen$\times$ \\
15--29  & \textbf{\ffMxMAEAgeFifteenTwentyNine} & \pbMxMAEAgeFifteenTwentyNine & \mxRatioAgeFifteenTwentyNine$\times$ \\
30--44  & \textbf{\ffMxMAEAgeThirtyFortyFour}   & \pbMxMAEAgeThirtyFortyFour   & \mxRatioAgeThirtyFortyFour$\times$ \\
45--59  & \textbf{\ffMxMAEAgeFortyFiveFiftyNine} & \pbMxMAEAgeFortyFiveFiftyNine & \mxRatioAgeFortyFiveFiftyNine$\times$ \\
60--74  & \textbf{\ffMxMAEAgeSixtySeventyFour}  & \pbMxMAEAgeSixtySeventyFour  & \mxRatioAgeSixtySeventyFour$\times$ \\
75--89  & \textbf{\ffMxMAEAgeSeventyFiveEightyNine} & \pbMxMAEAgeSeventyFiveEightyNine & \mxRatioAgeSeventyFiveEightyNine$\times$ \\
90--100 & \textbf{\ffMxMAEAgeNinetyHundred}     & \pbMxMAEAgeNinetyHundred     & \mxRatioAgeNinetyHundred$\times$ \\
\bottomrule
\end{tabular*}
\end{table}

\begin{figure}[!htbp]
\centering
\includefigifexists[width=\textwidth]{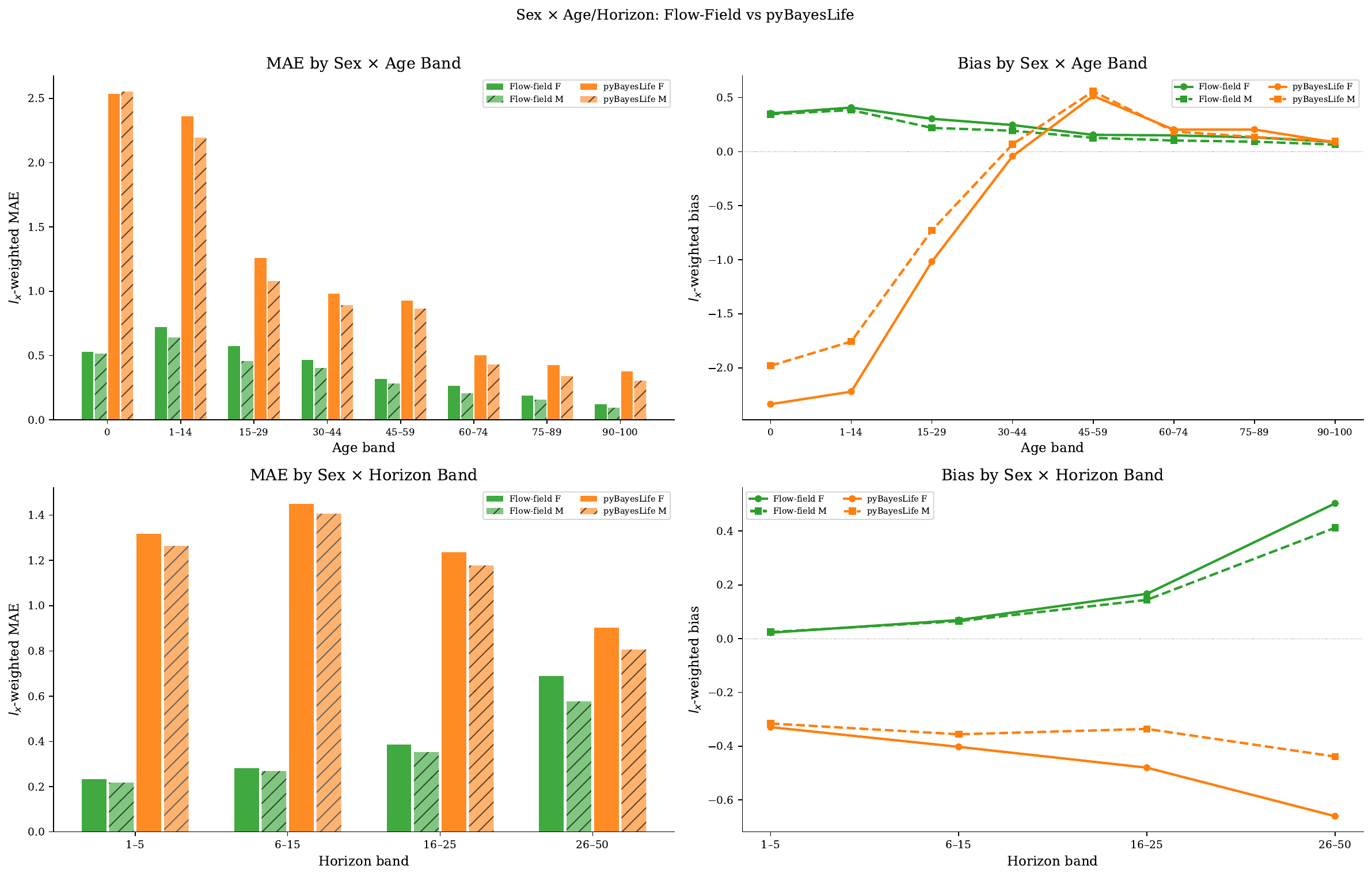}
\caption{Sex$\times$age/horizon breakdown of $\lx$-weighted $\log(\nmx)$
MAE and bias.  Top row: by age band.  Bottom row: by horizon band. The flow-field's near-zero bias across most ages and for both sexes contrasts with \pkg{pyBayesLife}'s strong age-dependent bias.}
\label{fig:age-sex-detail}
\end{figure}

\paragraph{By horizon.}
The flow-field's age-specific advantage is largest at short horizons ($h = 1$--5: \ffMxMAEShortH{} vs.\ \pbMxMAEShortH, a \mxRatioShortH$\times$ ratio) and narrows at long horizons ($h = 26$--50: \ffMxMAEVlongH{} vs.\ \pbMxMAEVlongH, \mxRatioVlongH$\times$; \cref{fig:age-sex-detail}, bottom row).  This is the opposite of the $\ezero$ pattern, where \pkg{pyBayesLife} has lower error at short horizons (\cref{fig:benchmark-4method}).  The juxtaposition is revealing: at $h = 1$--5, \pkg{pyBayesLife} predicts $\ezero$ better (MAE \pbMAEShortH{} vs.\ \ffMAEShortH) but predicts $\nmx$ worse (\mxRatioShortH$\times$ higher error).  Even when the $\ezero$-mediated pipeline gets the scalar right, it cannot recover the age pattern.

\paragraph{By sex.}
The flow-field performs slightly better for males ($\lx$-weighted MAE \ffMxMAEM) than females (\ffMxMAEF), while \pkg{pyBayesLife} shows the reverse pattern (males \pbMxMAEM, females \pbMxMAEF; \cref{tab:mxsexage}).  The \pkg{pyBayesLife} female disadvantage reflects the gap model's systematic positive bias in the male $\ezero$ forecast.

\paragraph{Sex$\times$age detail and sex differential.}
\Cref{fig:age-sex-detail} breaks the comparison down by sex and age band simultaneously, showing both MAE and bias; \cref{tab:mxsexage} reports the corresponding numbers.  The flow-field achieves near-zero bias across most ages and for both sexes, while \pkg{pyBayesLife} shows strong systematic bias that varies with age -- positive at young ages (underpredicting mortality) and negative at old ages.

\begin{figure}[!htbp]
\centering
\includefigifexists[width=\textwidth]{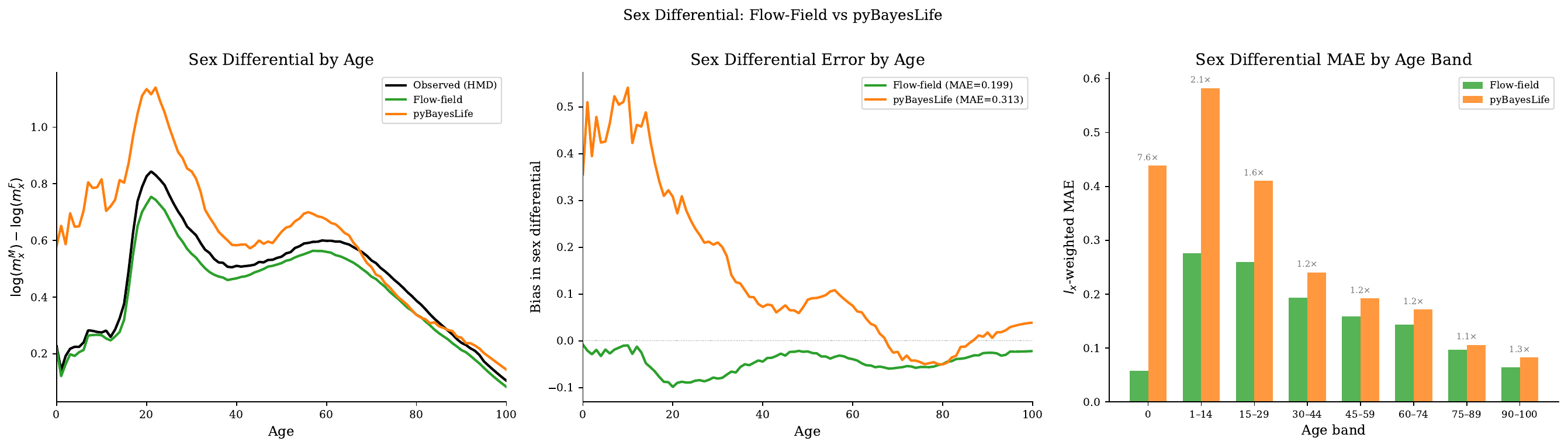}
\caption{Sex differential in age-specific mortality.
Left: mean $\log(m_x^M) - \log(m_x^F)$ by age -- observed (HMD, black), flow-field forecast (green), and \pkg{pyBayesLife} forecast (orange). Centre: error in the sex differential by age.  Right: $\lx$-weighted MAE of the sex differential by age band.  The flow-field reproduces the observed sex differential far more accurately because it forecasts the full sex$\times$age surface directly.}
\label{fig:sex-diff-mx}
\end{figure}

\Cref{fig:sex-diff-mx} evaluates how accurately each method reproduces the observed sex differential $\delta(x) = \log(m_x^M) - \log(m_x^F)$; \cref{tab:sexdiff} gives the age-band breakdown. The flow-field tracks the observed profile closely because both sexes emerge from the same Tucker surface; \pkg{pyBayesLife} must reconstruct it indirectly through separate Lee--Carter rotations after the gap-model $\ezero$ split, introducing substantial error at most ages.

% Table 7: sex × age band (transposed layout, inline with constants)
\begin{table}[htbp]
\caption{$\lx$-weighted $\log(\nmx)$ MAE and bias by sex and age band.}
\label{tab:mxsexage}
\small
\begin{tabular*}{\textwidth}{@{\extracolsep{\fill}}lrrrrrrrr@{}}
\toprule
 & 0 & 1--14 & 15--29 & 30--44 & 45--59 & 60--74 & 75--89 & 90--100 \\
\midrule
\textbf{FF F MAE}  & \textbf{\ffMxMAEFAgeZero} & \textbf{\ffMxMAEFAgeOneFourteen} & \textbf{\ffMxMAEFAgeFifteenTwentyNine} & \textbf{\ffMxMAEFAgeThirtyFortyFour} & \textbf{\ffMxMAEFAgeFortyFiveFiftyNine} & \textbf{\ffMxMAEFAgeSixtySeventyFour} & \textbf{\ffMxMAEFAgeSeventyFiveEightyNine} & \textbf{\ffMxMAEFAgeNinetyHundred} \\
\textbf{FF F Bias} & \textbf{\ffMxBiasFAgeZero} & \textbf{\ffMxBiasFAgeOneFourteen} & \textbf{\ffMxBiasFAgeFifteenTwentyNine} & \ffMxBiasFAgeThirtyFortyFour & \textbf{\ffMxBiasFAgeFortyFiveFiftyNine} & \textbf{\ffMxBiasFAgeSixtySeventyFour} & \textbf{\ffMxBiasFAgeSeventyFiveEightyNine} & \ffMxBiasFAgeNinetyHundred \\
\textbf{FF M MAE}  & \textbf{\ffMxMAEMAgeZero} & \textbf{\ffMxMAEMAgeOneFourteen} & \textbf{\ffMxMAEMAgeFifteenTwentyNine} & \textbf{\ffMxMAEMAgeThirtyFortyFour} & \textbf{\ffMxMAEMAgeFortyFiveFiftyNine} & \textbf{\ffMxMAEMAgeSixtySeventyFour} & \textbf{\ffMxMAEMAgeSeventyFiveEightyNine} & \textbf{\ffMxMAEMAgeNinetyHundred} \\
\textbf{FF M Bias} & \textbf{\ffMxBiasMAgeZero} & \textbf{\ffMxBiasMAgeOneFourteen} & \textbf{\ffMxBiasMAgeFifteenTwentyNine} & \ffMxBiasMAgeThirtyFortyFour & \textbf{\ffMxBiasMAgeFortyFiveFiftyNine} & \textbf{\ffMxBiasMAgeSixtySeventyFour} & \textbf{\ffMxBiasMAgeSeventyFiveEightyNine} & \textbf{\ffMxBiasMAgeNinetyHundred} \\
\addlinespace
pyBL F MAE  & \pbMxMAEFAgeZero & \pbMxMAEFAgeOneFourteen & \pbMxMAEFAgeFifteenTwentyNine & \pbMxMAEFAgeThirtyFortyFour & \pbMxMAEFAgeFortyFiveFiftyNine & \pbMxMAEFAgeSixtySeventyFour & \pbMxMAEFAgeSeventyFiveEightyNine & \pbMxMAEFAgeNinetyHundred \\
pyBL F Bias & \pbMxBiasFAgeZero & \pbMxBiasFAgeOneFourteen & \pbMxBiasFAgeFifteenTwentyNine & \textbf{\pbMxBiasFAgeThirtyFortyFour} & \pbMxBiasFAgeFortyFiveFiftyNine & \pbMxBiasFAgeSixtySeventyFour & \pbMxBiasFAgeSeventyFiveEightyNine & \textbf{\pbMxBiasFAgeNinetyHundred} \\
pyBL M MAE  & \pbMxMAEMAgeZero & \pbMxMAEMAgeOneFourteen & \pbMxMAEMAgeFifteenTwentyNine & \pbMxMAEMAgeThirtyFortyFour & \pbMxMAEMAgeFortyFiveFiftyNine & \pbMxMAEMAgeSixtySeventyFour & \pbMxMAEMAgeSeventyFiveEightyNine & \pbMxMAEMAgeNinetyHundred \\
pyBL M Bias & \pbMxBiasMAgeZero & \pbMxBiasMAgeOneFourteen & \pbMxBiasMAgeFifteenTwentyNine & \textbf{\pbMxBiasMAgeThirtyFortyFour} & \pbMxBiasMAgeFortyFiveFiftyNine & \pbMxBiasMAgeSixtySeventyFour & \pbMxBiasMAgeSeventyFiveEightyNine & \pbMxBiasMAgeNinetyHundred \\
\bottomrule
\end{tabular*}

\smallskip
\noindent{\footnotesize FF = flow-field; pyBL = pyBayesLife; F = female; M = male.}
\end{table}

% Table 8: sex differential by age band (inline with constants)
\begin{table}[htbp]
\caption{$\lx$-weighted MAE and bias of the sex differential
$\delta(x) = \log(m_x^M) - \log(m_x^F)$ by age band.}
\label{tab:sexdiff}
\begin{tabular*}{\textwidth}{@{\extracolsep{\fill}}lrrrr@{}}
\toprule
Age band & FF MAE & FF Bias & pyBL MAE & pyBL Bias \\
\midrule
0       & \textbf{\ffSdMAEAgeZero}              & \textbf{\ffSdBiasAgeZero}              & \pbSdMAEAgeZero              & \pbSdBiasAgeZero \\
1--14   & \textbf{\ffSdMAEAgeOneFourteen}       & \textbf{\ffSdBiasAgeOneFourteen}       & \pbSdMAEAgeOneFourteen       & \pbSdBiasAgeOneFourteen \\
15--29  & \textbf{\ffSdMAEAgeFifteenTwentyNine} & \textbf{\ffSdBiasAgeFifteenTwentyNine} & \pbSdMAEAgeFifteenTwentyNine & \pbSdBiasAgeFifteenTwentyNine \\
30--44  & \textbf{\ffSdMAEAgeThirtyFortyFour}   & \textbf{\ffSdBiasAgeThirtyFortyFour}   & \pbSdMAEAgeThirtyFortyFour   & \pbSdBiasAgeThirtyFortyFour \\
45--59  & \textbf{\ffSdMAEAgeFortyFiveFiftyNine} & \textbf{\ffSdBiasAgeFortyFiveFiftyNine} & \pbSdMAEAgeFortyFiveFiftyNine & \pbSdBiasAgeFortyFiveFiftyNine \\
60--74  & \textbf{\ffSdMAEAgeSixtySeventyFour}  & \ffSdBiasAgeSixtySeventyFour  & \pbSdMAEAgeSixtySeventyFour  & \textbf{\pbSdBiasAgeSixtySeventyFour} \\
75--89  & \textbf{\ffSdMAEAgeSeventyFiveEightyNine} & \ffSdBiasAgeSeventyFiveEightyNine & \pbSdMAEAgeSeventyFiveEightyNine & \textbf{\pbSdBiasAgeSeventyFiveEightyNine} \\
90--100 & \textbf{\ffSdMAEAgeNinetyHundred}     & \ffSdBiasAgeNinetyHundred     & \pbSdMAEAgeNinetyHundred     & \textbf{\pbSdBiasAgeNinetyHundred} \\
\bottomrule
\end{tabular*}

\smallskip
\noindent{\footnotesize FF = flow-field; pyBL = pyBayesLife.}
\end{table}

\begin{figure}[!htbp]
\centering
\includefigifexists[width=\textwidth]{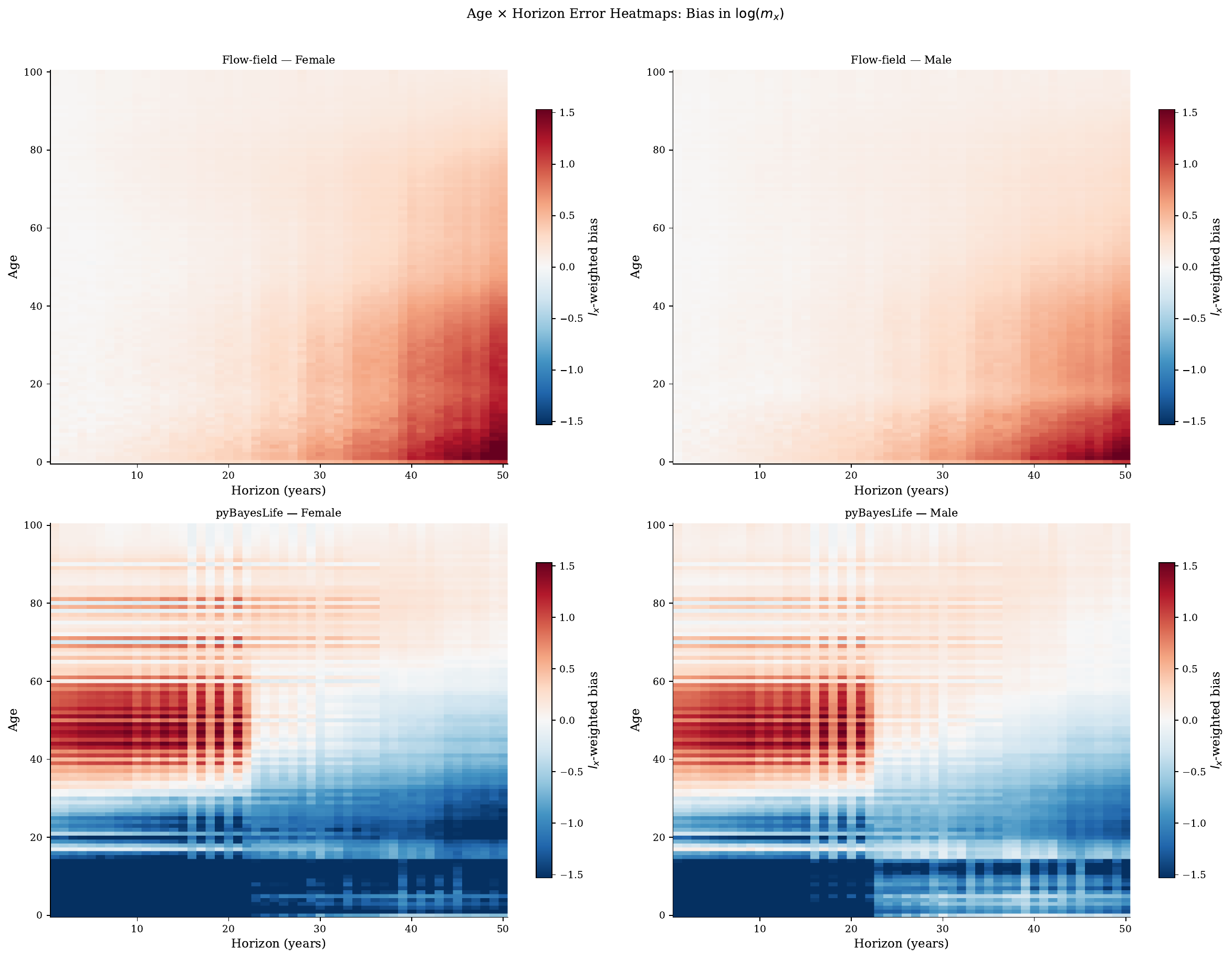}
\caption{Error heatmaps: $\lx$-weighted mean bias in $\log(\nmx)$ by
age (rows) and horizon (columns).  Top row: flow-field (left: female, right: male).  Bottom row: \pkg{pyBayesLife}.  The flow-field errors are uniformly small (near-white) with no systematic age$\times$horizon structure.  \pkg{pyBayesLife} shows strong structured bias reflecting the information bottleneck of reconstructing an age schedule from scalar $\ezero$.  The horizontal banding and qualitative shift at $h \approx 22$ in the \pkg{pyBayesLife} panels is a compositional artifact: each decade origin trains a separate Lee--Carter model with its own $b_x$ (age pattern of mortality change), and as the horizon increases, later-decade origins drop out of the test pool because they exhaust their available HMD validation data -- the decade-2010 origins disappear at $h = 13$, decade-2000 at $h = 22$ -- discretely shifting the average age pattern of the reconstruction.}
\label{fig:error-heatmaps}
\end{figure}

\paragraph{Error structure.}
\Cref{fig:error-heatmaps} displays the $\lx$-weighted mean bias in $\log(\nmx)$ as a function of age (rows) and forecast horizon (columns) for each method and sex.  The flow-field errors are uniformly small with no systematic age$\times$horizon structure -- the heatmaps are near-white throughout.  The \pkg{pyBayesLife} heatmaps reveal the structural signature of the $\ezero$-mediated reconstruction: strong systematic bias varying by age, reflecting the information bottleneck of recovering a 101-dimensional age pattern from a single scalar.

\begin{figure}[!htbp]
\centering
\includefigifexists[width=\textwidth]{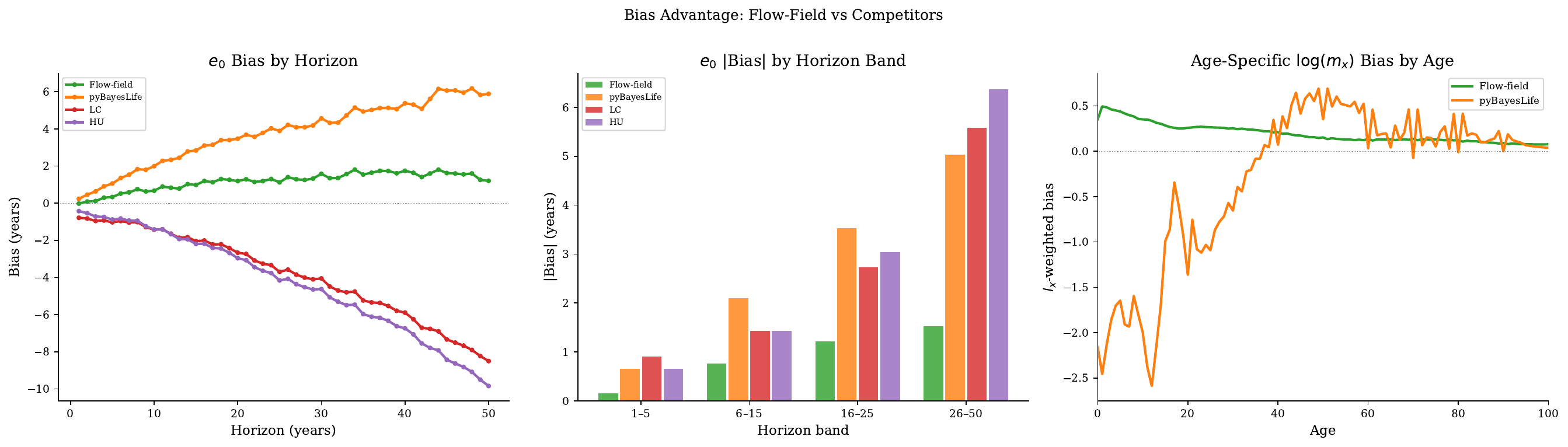}
\caption{Bias advantage across all methods.  Left: $\ezero$ bias by
horizon for all four methods on \commonN{} common test points.  The flow-field (green) stays near zero; Lee--Carter (red) and Hyndman--Ullah (purple) drift to $-3$ to $-8$~years; \pkg{pyBayesLife} (orange) drifts positive.  Centre: $|\text{bias}|$ by horizon band. Right: age-specific $\log(\nmx)$ bias by age -- the flow-field is near zero across ages while \pkg{pyBayesLife} shows strong age-dependent structure.}
\label{fig:bias-advantage}
\end{figure}

\Cref{fig:bias-advantage} summarises the bias contrast across all methods and metrics.  The left panel shows $\ezero$ bias by horizon for all four methods: the flow-field stays near zero at every horizon while Lee--Carter and Hyndman--Ullah accumulate large negative bias and \pkg{pyBayesLife} accumulates positive bias.  The right panel shows the age-specific bias profile: the flow-field's near-zero bias across ages contrasts with \pkg{pyBayesLife}'s strong age-dependent pattern.

% ══════════════════════════════════════════════════════════════════════════════
\section{Application to Non-HMD Populations}
\label{sec:external}
% ══════════════════════════════════════════════════════════════════════════════

The flow field is defined in $s_1$ space: the speed function $g^*(s_1)$, the trajectory functions $f_k^*(s_1)$, and the Tucker reconstruction from PCA scores.  A population outside the HMD need not be \emph{in} the decomposition -- it enters the flow field at its current mortality level and rides the canonical dynamics forward.

\textbf{Tier~1 ($\ezero$ only).}  Given a time series of $\ezero$ values, the system maps $\ezero$ to $s_1$ via a LOWESS-fitted canonical relationship $s_1(\ezero)$ estimated from the HMD training data, computes the country's recent $s_1$ velocity from forward differences, and forecasts using the hierarchical speed blend of \cref{eq:speed}.  PCA scores are set to the canonical $f_k^*(s_1)$ at each horizon -- the country is assumed to follow the average sex-age structure for its mortality level.  Output: full sex- and age-specific schedules via Tucker reconstruction.

\textbf{Tier~2 (age-specific rates).}  Given female and male age-specific mortality rates, the observed schedules are projected into Tucker space by solving $G_{ct} = S^{+} Z A^{+\top}$ via the pseudoinverses of the shared basis matrices.  This yields PCA scores -- including $s_1$ -- that may deviate from canonical.  The forecast navigates in $s_1$ space, and the score relaxation of \cref{eq:relax} preserves the structural deviations (PCs~2--5) in the near term and gradually converges toward the canonical trajectory.

\begin{figure}[!htbp]
\centering
\includegraphics[width=\textwidth]{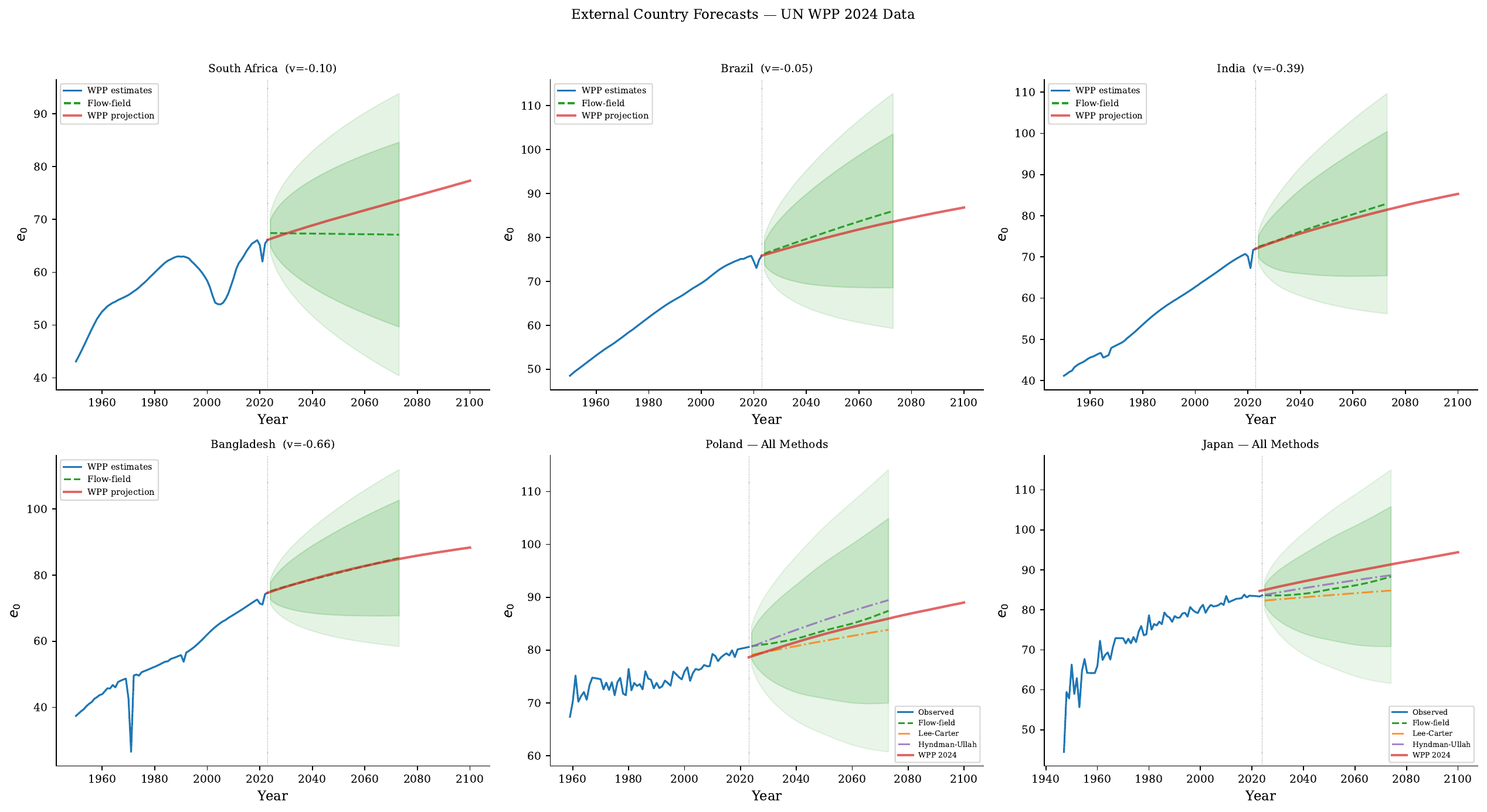}
\caption{External country forecasts and method comparison.  Top row:
South Africa, Brazil, India -- flow-field (green) vs WPP 2024 medium variant (red) using real WPP $\ezero$ estimates as input.  Bottom row: Bangladesh, Poland, Japan.  The Poland and Japan panels compare all methods: flow-field (green), Lee--Carter (orange), Hyndman--Ullah (purple), and WPP 2024 (red).  At 50-year horizons, Lee--Carter and Hyndman--Ullah diverge from the WPP and observed trajectory, while the flow-field tracks the WPP projection closely.  All flow-field panels include bias-corrected 80\% and 95\% prediction intervals.}
\label{fig:external}
\end{figure}

\Cref{fig:external} demonstrates the system on four non-HMD countries using UN WPP 2024 $\ezero$ estimates, with the WPP medium-variant projections shown for comparison.  The flow-field forecasts are broadly consistent with the UN projections -- which use the \citeauthor{RafteryChunnGerlandSevcikova2013} parametric double-logistic methodology -- providing independent validation that approaches, despite their very different architectures, converge on similar assessments of future mortality improvement.  Differences between the two forecasts are informative: where the flow-field is more optimistic (or pessimistic) than the WPP, it reflects the different information each method uses -- the flow-field relies on the country's own recent velocity and the HMD-wide canonical dynamics, while the WPP uses country-specific Bayesian posterior estimates of the double-logistic parameters.  \Cref{fig:external} shows the flow-field and WPP projections side by side to the 2070s -- the two systems agree closely for Brazil, India, and Bangladesh, while South Africa shows a larger gap because the HMD-wide speed function has little experience with the rapid gains possible during ART scale-up.

This capability rests on a strong assumption: that the non-HMD population's mortality dynamics are well-approximated by the HMD-wide flow field.  This assumption is most plausible for populations at mortality levels well-represented in the HMD ($\ezero \approx 50$--85) and least plausible for populations with distinctive mortality patterns driven by causes not well-represented in HMD populations (e.g.\ malaria, HIV/AIDS in the absence of treatment).  Even in these cases, the flow field provides a principled baseline forecast that can be adjusted with external information -- and the Tucker reconstruction guarantees structurally coherent schedules regardless of input quality.

% ══════════════════════════════════════════════════════════════════════════════
\section{Discussion}
\label{sec:discussion}
% ══════════════════════════════════════════════════════════════════════════════

The flow-field forecaster unifies two previously separate traditions in mortality modelling: the tensor decomposition approach to multi-population structure \citep{RussolilloGiordanoHaberman2011,DongHuangHaberman2020,ClarkMDMx2026} and the level-dependent forecasting approach \citep{RafteryChunnGerlandSevcikova2013}.  The unification is enabled by a conceptual reframing: rather than treating the Tucker temporal components as time series to be extrapolated, we treat the decomposition's score space as a \emph{dynamical system} in which the state evolves according to a flow field parameterised by mortality level.

In leave-country-out cross-validation with a 50-year horizon (\commonN{} test points), the system achieves an overall $\ezero$ MAE of \ffMAE~years -- comparable to Lee--Carter (\lcMAE), Hyndman--Ullah (\huMAE), and \pkg{pyBayesLife} (\pbMAE). Lee--Carter and Hyndman--Ullah are more accurate at short horizons ($h = 1$--10) but accumulate large systematic bias at long horizons (\lcBias{} and \huBias~years respectively); \pkg{pyBayesLife} accumulates positive bias (\pbBias~years); the flow-field's aggregate bias is the smallest at \ffBias~years.  The MAE advantage is concentrated at $h = 26$--50, precisely the regime that matters for the 50--75~year projections used in population planning and actuarial work.

The age-specific comparison against \pkg{pyBayesLife} -- our de novo reimplementation of the \citeauthor{RafteryChunnGerlandSevcikova2013} pipeline, trained exclusively on HMD data -- reveals that the flow-field's architectural advantage extends well beyond the scalar $\ezero$. On \mxCommonN{} age-specific test points evaluated against raw HMD mortality rates, the flow-field achieves an $\lx$-weighted $\log(\nmx)$ MAE of \ffMxMAElx, compared to \pbMxMAElx{} for \pkg{pyBayesLife} -- a \mxRatio-fold improvement that persists at every age band, every horizon, and for both sexes.  The flow-field also reproduces the observed sex differential far more accurately, because both sexes emerge from the same Tucker surface.  The performance gap is structural rather than implementational: collapsing the mortality surface to a scalar and then attempting to recover it discards information that cannot be reconstructed.

This performance arises from a system with a distinctive combination of properties that we summarise below.

% ──────────────────────────────────────────────────────────────────────────────
\subsection{Parsimony}
\label{sec:discussion:parsimony}
% ──────────────────────────────────────────────────────────────────────────────

The production system has effectively zero tuned parameters.  The per-component structural score relaxation rates ($\alpha_{s,k} \approx 0.95$--$0.98$) are measured from the empirical autocorrelation structure of deviations from canonical dynamics.  The era half-life ($\tau = 12$) is selected from a coarse grid, and the speed blend weight $w = 1.0$ has negligible influence (MAE varies by only a few hundredths of a year across its full range).  The forecast reduces to a nearly parameter-free flow integration: advance $s_1$ by the era-weighted canonical speed at each step, relax structural scores toward canonical at their empirical rates, reconstruct the full mortality schedule via Tucker, and compute $\ezero$ from the surface.  There is no ARIMA fitting, no state-space estimation, no drift computation, and no time-series machinery of any kind.

This is in sharp contrast to Lee--Carter (which requires SVD decomposition, drift estimation, and ARIMA modelling of $k_t$), Hyndman--Ullah (which adds functional data analysis and rank selection), and the Kalman-based MDMx forecaster \citep{ClarkMDMx2026} (which requires a full state-space model with hierarchical drift targets and observation noise).

% ──────────────────────────────────────────────────────────────────────────────
\subsection{Integrated framework}
\label{sec:discussion:integrated}
% ──────────────────────────────────────────────────────────────────────────────

The \citeauthor{RafteryChunnGerlandSevcikova2013} approach -- the methodology underlying the UN WPP -- forecasts $\ezero$ using a Bayesian double-logistic model and then maps it to age-specific rates using a separate model life table system -- the extended Coale--Demeny and UN regional model life tables \citep{CoaleDemeny1966,UnitedNations2022WPP}.  The two components are fitted independently, creating a seam between the $\ezero$ forecast and the age-pattern reconstruction -- a seam that can introduce inconsistency between the projected level and the projected age structure.

The flow-field system is fully integrated: the same Tucker PCA space serves as the forecasting coordinate system and the reconstruction basis.  The trajectory functions $f_k^*(s_1)$ constitute a continuous model life table system in Tucker coordinates, so the forecast \emph{is} the reconstruction -- there is no separate mapping step, no seam, and no possibility of inconsistency between the level forecast and the age-pattern reconstruction.

% ──────────────────────────────────────────────────────────────────────────────
\subsection{Structural sex-age coherence}
\label{sec:discussion:coherence}
% ──────────────────────────────────────────────────────────────────────────────

The Tucker decomposition factorises the four-dimensional sex--age--country--year tensor through shared basis matrices $\bS$ ($2 \times r_1$) and $\bA$ ($110 \times r_2$).  Because these bases are shared across all countries and years, any forecast produced by the system -- regardless of horizon, era weighting, or country -- is guaranteed to lie in the span of these bases.  This provides \emph{structural} sex-age coherence: the forecast mortality schedule at $h = 50$ has exactly the same structural properties (smooth age profiles, plausible sex differentials, monotonically increasing old-age mortality) as the observed schedules that trained the decomposition.  Implausible outcomes -- negative mortality rates, sex-crossovers in the wrong direction, wild age-pattern oscillations -- are impossible by construction.

\begin{figure}[!htbp]
\centering
\includegraphics[width=\textwidth]{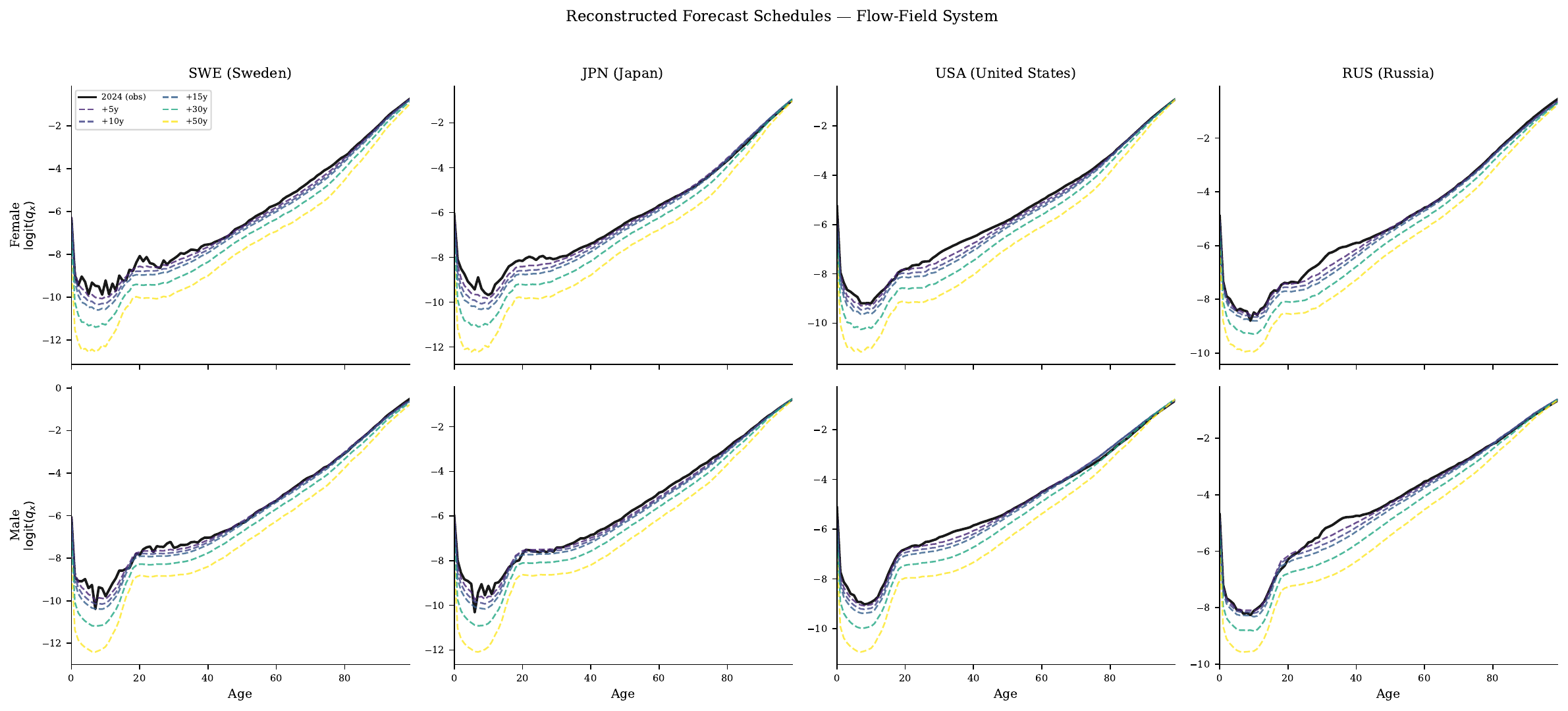}
\caption{Reconstructed forecast mortality schedules for Sweden,
Japan, USA, and Russia.  Black: last observed $\logit(\qx)$. Coloured dashed: forecast at 5-year horizons.  The Tucker reconstruction maintains smooth age profiles and coherent sex structure at all horizons.}
\label{fig:schedule}
\end{figure}

The score relaxation continues the country's current sex-age rotation dynamics into the forecast.  A country with a distinctive age pattern -- Russia's excess working-age male mortality, Japan's exceptional old-age female survival -- retains these features for decades into the forecast, converging \emph{gradually} toward the canonical pattern at empirically measured rates (half-lives 12--32~years).  The transition between observed and forecast mortality schedules is smooth by construction: the forecast starts at the country's actual last-observed Tucker scores and evolves continuously through score space, with no discontinuity at the forecast origin (\cref{fig:schedule}).

This is fundamentally different from the Lee--Carter and Hyndman--Ullah approaches, which extrapolate each temporal component independently and can produce implausible age-pattern crossovers and divergent sex differentials at long horizons.  The Tucker framework makes such pathologies structurally impossible.

% ──────────────────────────────────────────────────────────────────────────────
\subsection{Complete sex-specific mortality schedules}
\label{sec:discussion:schedules}
% ──────────────────────────────────────────────────────────────────────────────

The system produces complete single-year-of-age (0--109), sex-specific $q_x$ schedules at every forecast horizon -- not just $\ezero$ or abridged life tables.  These schedules are ready for direct input to population projection models, actuarial calculations, or health-burden estimation without further interpolation or graduation.  The Tucker reconstruction guarantees that each schedule is a plausible member of the family of mortality patterns observed in the HMD, with smooth age profiles and coherent sex structure.

% ──────────────────────────────────────────────────────────────────────────────
\subsection{Long-horizon accuracy and systematic bias}
\label{sec:discussion:longhorizon}
% ──────────────────────────────────────────────────────────────────────────────

The horizon profile (\cref{fig:benchmark-4method}) reveals a crossover: at short horizons ($h = 1$--10), Lee--Carter, Hyndman--Ullah, and \pkg{pyBayesLife} are more accurate because their jump-off adjustment and time-series modelling capture recent country-specific momentum effectively.  At $h = 26$--50, the flow-field error (MAE~\ffMAEVlongH~years) is substantially below Lee--Carter (\lcMAEVlongH), Hyndman--Ullah (\huMAEVlongH), and \pkg{pyBayesLife} (\pbMAEVlongH).

The bias contrast is starker and arguably more consequential than the MAE difference.  The flow-field's aggregate bias is \ffBias~years -- the smallest in magnitude among the four methods -- while Lee--Carter (\lcBias~years) and Hyndman--Ullah (\huBias~years) systematically underpredict, and \pkg{pyBayesLife} (\pbBias~years) systematically overpredicts, future life expectancy.  The mechanism is clear: Lee--Carter's random walk with drift and Hyndman--Ullah's ARIMA extrapolation both project temporal components linearly (or via low-order autoregressive models) into territories far beyond any historically observed values.  When the true rate of mortality improvement decelerates -- as it has at the mortality frontier -- the extrapolation overshoots the decline, producing forecasts that are systematically too pessimistic about future survival.  The flow-field avoids this because it navigates through a score space parameterised by mortality level: the canonical speed function is anchored by the cross-sectional experience of 47 countries at each level, and the trajectory cannot drift beyond the observed manifold.

For applied demography, this distinction matters.  Random forecast error (captured by MAE) averages out across populations and over time; systematic bias does not.  A pension system designed around life expectancy forecasts that are 3--4~years too low will be structurally underfunded.  Social security trust fund projections built on negatively biased mortality forecasts will overstate solvency.  Health system capacity planning that underestimates longevity will be perpetually behind demand.  The low bias of the flow-field system addresses what is, from a policy perspective, a persistent failure mode of existing mortality forecasting methods.

The era-weighted speed function (\cref{sec:architecture:era}) is essential to this result.  Without it, the canonical speed function averages over disparate eras, producing a substantial bias of its own.  With the truncated exponential kernel ($\tau = 12$, $W = 40$), the speed function at each forecast origin reflects contemporary dynamics.  Combined with $s_1$-space navigation (\cref{sec:architecture:navigation}), which eliminates the navigation/surface divergence, the aggregate bias is near zero.

\Cref{fig:bias-advantage} summarises the bias contrast: the flow-field stays near zero at every horizon and across ages, while the other three methods accumulate systematic bias of 3--8~years. For applied demography, the flow-field's low $\ezero$ bias and near-zero age-specific bias are distinctive and consequential properties of the system.

% ──────────────────────────────────────────────────────────────────────────────
\subsection{The cost of collapsing to a scalar}
\label{sec:discussion:collapse}
% ──────────────────────────────────────────────────────────────────────────────

The age-specific comparison in \cref{sec:agespecific} provides the most direct evidence for the flow-field's architectural advantage. The \mxRatio$\times$ performance gap in $\lx$-weighted $\log(\nmx)$ MAE is not a failure of the \citeauthor{RafteryChunnGerlandSevcikova2013} implementation -- our \pkg{pyBayesLife} reimplementation actually improves on the original through better MCMC parameterization, HMD-specific variance estimation, guaranteed positive rates, and machine-precision $\ezero$ recovery (\cref{sec:appendix:raftery} documents the specific issues that motivated each improvement).  Rather, the gap reflects a fundamental architectural limitation of the $\ezero$-mediated approach.

The mortality surface $\log(\nmx)$ for two sexes and $A$ single-year ages contains $2A$ degrees of freedom per country-year.  Projecting this to $\ezero$ reduces it to a single scalar, discarding all information about the age pattern, the sex differential in the age pattern, and country-specific deviations from average age patterns. The Lee--Carter reconstruction attempts to recover these $2A - 1$ lost dimensions from a single number, which is fundamentally underdetermined.  The Lee--Carter model assumes $\log(\nmx) = a_x + b_x \cdot k_t$, where $b_x$ (the age pattern of mortality change) is fixed from the training period.  Given a target $\ezero$, the model finds $k_t$ by bisection -- but the actual age pattern of future mortality change may differ from the training-period $b_x$, since the ages driving mortality improvement shift over time \citep{LiLeeGerland2013}.

When the target $\ezero$ is wrong by $\delta$ years, the bisection adjusts $k_t$ to compensate, distributing the adjustment across ages according to $b_x$.  At ages where $\nmx$ is very small (e.g.\ ages 1--14, where $\nmx \approx 10^{-4}$), a small absolute change in $\nmx$ corresponds to a large change in $\log(\nmx)$.  This amplification produces the \mxRatioAgeZero$\times$ gap at age~0 and \mxRatioAgeOneFourteen$\times$ gap at ages~1--14.  But the horizon results demonstrate that even when $\ezero$ is forecast correctly, the reconstruction still introduces substantial error: at $h = 1$--5, \pkg{pyBayesLife} forecasts $\ezero$ better than the flow-field (MAE \pbMAEShortH{} vs.\ \ffMAEShortH) yet produces age-specific errors \mxRatioShortH$\times$ larger.  The reconstruction itself -- the assumption that a fixed $b_x$ adequately describes how mortality change is distributed across ages -- is the binding constraint.

The sex differential comparison (\cref{fig:sex-diff-mx}) provides further evidence.  The flow-field reproduces the observed sex differential $\delta(x) = \log(m_x^M) - \log(m_x^F)$ far more accurately than \pkg{pyBayesLife}, because both sexes emerge from the same Tucker decomposition surface.  In the $\ezero$-mediated pipeline, the sex differential is determined by the gap model's scalar prediction of male $\ezero$, which is then independently rotated through the Lee--Carter age pattern -- a process that cannot capture the age-varying sex differential observed in the data. The error heatmaps (\cref{fig:error-heatmaps}) make the contrast vivid: the flow-field errors are uniformly small across all ages and horizons, while \pkg{pyBayesLife}'s errors show the structured pattern characteristic of the reconstruction bottleneck.

The flow-field avoids this entirely by forecasting $\logit(\nqx)$ at every age simultaneously, with the Tucker decomposition enforcing smooth, coherent age profiles.  It maintains three forms of coherence that the $\ezero$-mediated pipeline cannot: sex coherence through the shared sex mode of the Tucker decomposition rather than a separate gap model; age coherence through a low-rank subspace estimated from all observed mortality schedules rather than a one-dimensional $b_x$ subspace from a single country; and cross-country coherence through shared Tucker factor matrices rather than independently estimated Lee--Carter age patterns.

The \pkg{pyBayesLife} reimplementation itself is a contribution: a complete, R-free, WPP-free Python package that improves on the original in MCMC parameterization, variance estimation, Lee--Carter accuracy, and numerical robustness (\cref{sec:appendix:raftery} provides a complete accounting of the eleven issues addressed).  It serves as a transparent, reproducible benchmark and is available for use in any context where the \citeauthor{RafteryChunnGerlandSevcikova2013} pipeline is needed without R or WPP dependencies.

% ──────────────────────────────────────────────────────────────────────────────
\subsection{Applicability beyond the HMD}
\label{sec:discussion:external}
% ──────────────────────────────────────────────────────────────────────────────

The system forecasts any population for which an $\ezero$ time series is available (Tier~1) or for which age-specific mortality rates can be projected into Tucker space (Tier~2).  The external country demonstrations (\cref{fig:external}) show forecasts for South Africa, Brazil, India, and Bangladesh -- none of which are in the HMD -- alongside UN WPP projections.  The flow-field forecasts produce complete sex-specific, single-year-of-age mortality schedules guaranteed to be structurally coherent with the HMD experience -- a property that the WPP's $\ezero$-to-MLT pipeline does not guarantee.

% ──────────────────────────────────────────────────────────────────────────────
\subsection{Model life table system as byproduct}
\label{sec:discussion:mlt}
% ──────────────────────────────────────────────────────────────────────────────

The trajectory functions $f_k^*(s_1)$ define a continuous model life table system in Tucker coordinates.  Given any mortality level $s_1$ (or equivalently any $\ezero$, mapped to $s_1$ via the canonical relationship), the canonical structural scores $s_k = f_k^*(s_1)$ for $k = 2, \ldots, 5$ together with $s_1$ itself reconstruct the ``typical'' sex-specific mortality schedule at that mortality level via the Tucker basis matrices.  This system is a byproduct of the forecasting framework, but it is usable independently -- in the same spirit as the SVD-based model life table systems of \citet{Clark2019} -- for example, to generate model life tables for populations without age-specific data, or as a reference standard for evaluating the plausibility of observed schedules.

% ──────────────────────────────────────────────────────────────────────────────
\subsection{Suggestive trajectory behavior at very low mortality levels: a lifespan limit?}
\label{sec:discussion:mxplateau}
% ──────────────────────────────────────────────────────────────────────────────

Whether human life expectancy and maximum lifespan face fixed biological limits is one of the most contested questions in demography and biogerontology.

\citet{OeppenVaupel2002} showed that record national life expectancy has risen at a remarkably steady pace of roughly 2.5~years per decade since 1840, with no sign of deceleration -- every prediction of an upper limit has been broken within five years. \citet{Vaupel2010} reinforced this by documenting that human senescence has been \emph{postponed} by a decade: people reach old age in better health, and the rate of age-related deterioration has not itself accelerated.

The opposing camp argues that biological constraints impose a ceiling.  \citet{Fries1980} proposed a fixed maximum lifespan near 85~years around which morbidity would compress. \citet{OlshanskyCarnesCassel1990} challenged the extrapolation view directly, arguing that eliminating all causes of death would add only modest years because ageing itself -- the accumulating damage to cells and tissues -- is the binding constraint. \citet{DongMilhollandVijg2016} presented evidence that the maximum reported age at death plateaued in the mid-1990s around 115~years, suggesting a species-specific ceiling.  Most recently, \citet{Olshansky2024} examined 1990--2019 data from the ten longest-lived populations and concluded that life expectancy gains have decelerated sharply and radical extension is implausible without interventions that slow ageing itself.

Between these poles, \citet{BarbiLagonaVaupelWachter2018} found that Italian death rates plateau after age~105 rather than continuing to rise exponentially -- a pattern consistent with heterogeneous frailty rather than a hard wall, and one that leaves the theoretical maximum open-ended.  The current empirical picture thus shows clear deceleration in life expectancy gains at the population level, an unresolved question about whether a fixed maximum lifespan exists at the individual level, and broad agreement that further progress depends on whether medicine can slow the biological process of ageing rather than simply treating its consequences one disease at a time.

This work contributes a data point to this debate. \Cref{fig:e0-mapping} displays the deceleration of $\ezero$ as a function of mortality level (first PCA score) as mortality falls. The gentle roll-off starts when $\ezero$ is roughly 65, well within the HMD point cloud. \Cref{fig:speed-diagnostic} (leftmost panel) displays the smoothed forward differences (first derivative in time) in the level component (PCA 1) which explains the bulk of this deceleration. There was a period of large, approximately constant (steady velocity) negative decrements from about -25 to -8, followed by a brief slow-down (a hump) and then another roughly linear period of slowly decreasing negative decrements (slow deceleration) from about 4 to 17, followed by a change in the slope of the decrement (kink) to a final roughly linear period of more slowly decreasing decrements (slower deceleration). This maps out three eras of steady mortality decline separated by two transitions. The last two are decelerations, with the last being the slowest deceleration at the smallest levels of change. This suggests convergence, and the biological theory suggests the target is zero - no further decrement in mortality level.

% ──────────────────────────────────────────────────────────────────────────────
\subsection{Limitations and extensions}
\label{sec:discussion:limitations}
% ──────────────────────────────────────────────────────────────────────────────

The system's principal limitation is its weaker performance at short forecast horizons ($h = 1$--10), where Lee--Carter, Hyndman--Ullah, and \pkg{pyBayesLife} produce lower MAE.  The flow-field's era-weighted canonical speed function captures the average improvement pace at each mortality level but does not model country-specific short-term momentum -- the recent acceleration or deceleration that ARIMA and jump-off adjustments exploit.  A hybrid system that uses time-series short-horizon forecasts and transitions to flow-field dynamics at longer horizons is a natural extension.

The remaining small aggregate bias (\ffBias~years) reflects residual uncertainty in the canonical speed function at the mortality frontier, where HMD data is sparse.  A forecast from 1970 correctly weights the improvement pace of the 1950s--1970s, but the subsequent 50~years include both acceleration and deceleration that no fixed-kernel approach can foresee.

This is an empirical method that requires a large, diverse set of well-observed life tables to serve as training and calibration data. This creates two important limitations. First, the HMD training set is overwhelmingly European.  Populations with distinctive cause-of-death profiles (sub-Saharan Africa, tropical Asia) may not follow HMD-typical dynamics, and the system's applicability to such populations rests on the assumption that the Tucker flow field generalises across epidemiological contexts. Second, the range of mortality levels represented by the HMD does not include the super-low mortality that populations may obtain in the medium-to-distant future. Consequently, there is no data-driven trajectory path into or through those very low mortality regimes, and the current approach that extrapolates within-cloud behavior to edge points and beyond may not adequately reflects what will really happen. This is an open question for the theory of mortality change.
 
Natural extensions include: $\ezero$-dependent era weighting; adaptive kernels that update as the forecast evolves; a hybrid short/long-horizon system combining ARIMA and flow-field dynamics; conditioning on covariates such as GDP per capita or health expenditure; a fully Bayesian treatment in the spirit of \citet{RafteryChunnGerlandSevcikova2013}; and uncertainty quantification that accounts for parameter uncertainty in the flow field itself.

Future work will explore theory-driven trajectories into the very-low mortality space that has not yet been mapped by the HMD and how this work might contribute more to the debate about human lifespan.

% ══════════════════════════════════════════════════════════════════════════════
\section{Computational Environment and Acknowledgements}
\label{sec:computational}
% ══════════════════════════════════════════════════════════════════════════════

All computations were performed on an Apple MacBook Pro with an Apple M1~Max processor and 64\,GB unified memory, running macOS. The analysis pipeline is implemented in Python~3.14 within a Quarto notebook environment, managed by \texttt{uv} (package installer) and
\texttt{pyenv} (Python version management), with Positron as the IDE. Core dependencies include NumPy, SciPy, pandas, scikit-learn, statsmodels, matplotlib, and DuckDB.  The \pkg{pyBayesLife} modules additionally use JAX and NumPyro for MCMC estimation.  The full pipeline is contained in a single Quarto notebook ($\sim$6{,}000~lines) that produces all figures, tables, and cached objects. Document preparation uses \LaTeX\ via KOMA-Script (\texttt{scrartcl}) with Palatino/\texttt{mathpazo} typography.

The Lee--Carter and Hyndman--Ullah benchmarks are computed by the R \texttt{demography} package \citep{HyndmanUllah2007} via a subprocess bridge, using HMD graduated $m_x$ rates and person-year exposures to ensure the benchmarks employ the exact published algorithms rather than simplified reimplementations.

Mortality data are from the Human Mortality Database (\url{https://www.mortality.org}).  External country $\ezero$ estimates and projections are from the United Nations World Population Prospects 2024 \citep{UnitedNations2022WPP}.

An interactive Shiny web application demonstrating the life table generator, fitter, summary-indicator prediction, and mortality forecaster is deployed at
\url{https://samclark.shinyapps.io/mdmx/}.

The complete source code is available from the author.

Claude (Anthropic, Claude Opus 4.6) served as a research assistant throughout the development of this project.  Its contributions included writing and debugging Python code for the computational pipeline, drafting and editing \LaTeX\ manuscript text, performing literature searches, conducting numerical cross-checks between the Quarto output and manuscript claims, and iterating on architectural decisions through interactive empirical experimentation.  All substantive scientific decisions -- including defining and framing the questions; designing the analytical approach; choosing the specific methods; optimizing and fine-tuning each method; validating and interpreting results; and organizing and creating the manuscript -- were made by the author.  The AI assistant's outputs were reviewed, verified, and revised by the author before incorporation.

% ══════════════════════════════════════════════════════════════════════════════
\clearpage
\section{Notation}
\label{sec:notation}
% ══════════════════════════════════════════════════════════════════════════════

We follow the notation of \citet{ClarkMDMx2026} throughout. Tensors of order three or higher are calligraphic uppercase ($\M$, $\G$), matrices are bold uppercase ($\bS$, $\bA$, $\bC$, $\bT$), vectors are bold lowercase ($\bm{s}$, $\bm{v}$), and scalars are italic ($\alpha$, $h$).  \Cref{tab:ff-notation} collects the symbols introduced in this paper; see the main MDMx manuscript for the full notation table.

{%
\small
\setlength\LTleft{0pt}
\setlength\LTright{0pt}
\begin{longtable}{@{}l@{\hspace{1em}}l@{\hspace{1em}}p{0.58\textwidth}@{}}
\caption{Principal notation for the flow-field forecaster.}
\label{tab:ff-notation} \\
\toprule
Symbol & Dim. & Meaning \\
\midrule
\endfirsthead
\multicolumn{3}{@{}l}{\small\itshape\tablename~\thetable, continued} \\[4pt]
\toprule
Symbol & Dim. & Meaning \\
\midrule
\endhead
\midrule
\multicolumn{3}{r@{}}{\small\itshape continued on next page} \\
\endfoot
\bottomrule
\endlastfoot
\multicolumn{3}{@{}l}{\textit{Tucker decomposition (from MDMx)}} \\
$\M$ & $S \times A \times C \times T$ & mortality tensor ($\logit(\qx)$; $S{=}2$, $A{=}110$, $C{=}48$, $T{=}274$) \\ $\bS, \bA, \bC, \bT$ & varies & factor matrices for sex, age, country, year \\ $\G$ & $r_1 \times r_2 \times r_3 \times r_4$ & core tensor; ranks $(2, 42, 46, 100)$ \\ $G_{ct}$ & $r_1 \times r_2$ & effective core matrix for country $c$, year $t$ \\ $\hat{M}_{:,:}(h)$ & $S \times A$ & reconstructed mortality schedule at horizon $h$ \\
\addlinespace
\multicolumn{3}{@{}l}{\textit{PCA of the effective core}} \\
$V$ & $N \times r_1 r_2$ & PCA loading matrix (first $N{=}5$ components of $\vect(G_{ct})$; rows are components, columns are features) \\ $\bar{g}$ & $1 \times r_1 r_2$ & PCA centering vector: $\overline{\vect(G_{ct})}$ \\ $s_k$ & -- & $k$-th PCA score ($k = 1, \ldots, 5$) \\ $\bm{s}_{c,t}$ & $1 \times N$ & PCA score vector for country $c$, year $t$ \\
\addlinespace
\multicolumn{3}{@{}l}{\textit{Flow-field functions (in $s_1$ space)}} \\
$g^*(s_1)$ & -- & speed function: $\mathrm{d}s_1/\mathrm{d}t$ as a function of $s_1$ \\ $g^*_\tau(s_1)$ & -- & era-weighted speed function (LOWESS with half-life $\tau$) \\ $f_k^*(s_1)$ & -- & trajectory function: canonical $s_k$ at level $s_1$ ($k = 2, \ldots, 5$) \\ $v_{s_1,\text{country}}$ & -- & country's trailing-mean $s_1$ velocity at origin \\
\addlinespace
\multicolumn{3}{@{}l}{\textit{Forecasting parameters}} \\
$w$ & -- & speed blend weight ($w{=}1$: pure canonical; $w{=}0$: pure country) \\ $w_{\text{era}}(t)$ & -- & era weighting kernel: truncated exponential with half-life $\tau$ and window $W$ \\ $\alpha_v$ & -- & speed relaxation rate (empirical) \\ $\alpha_{s,k}$ & -- & score relaxation rate for component $k$ ($k = 2, \ldots, 5$; empirical; half-lives 12--32~yr) \\ $s_k^{\text{actual}}$ & -- & country's observed score for component $k$ at the forecast origin \\ $\tau$ & -- & era half-life (truncated exponential kernel for speed function weighting) \\ $\tau_{\text{blend}}$ & -- & jump-off residual decay half-life ($\tau_{\text{blend}} = 2$~yr) \\ $W$ & -- & era hard window (data older than $W$ years is discarded; $W{=}40$) \\ $h$ & -- & forecast horizon (years ahead) \\ $s_1^*$ & -- & tail extension transition point in $s_1$ space (corresponds to $\ezero \approx 78$) \\ $w_{\text{lin}}$ & -- & smoothstep blending weight for tail extension: $t^2(3-2t)$ \\
\addlinespace
\multicolumn{3}{@{}l}{\textit{Prediction intervals}} \\
$\sigma_1$ & -- & one-step-ahead standard deviation of $\ezero$ forecast error \\ $\kappa$ & -- & PI calibration factor ($\text{SD}$ of CV z-scores) \\ $\sigma(h)$ & -- & horizon-dependent PI width: $\kappa \cdot \sigma_1 \cdot \sqrt{h}$ \\ $b(h)$ & -- & bias correction function (LOWESS of CV errors vs.\ horizon) \\ $\Delta_0$ & $S \times A$ & jump-off residual: full Tucker reconstruction minus 5-PC approximation at the forecast origin \\ $t_k$ & -- & LOWESS slope of trajectory function $f_k^*$ at $s_1^*$ (joint tangent component) \\
\addlinespace
\multicolumn{3}{@{}l}{\textit{Empirical convergence}} \\
$\beta(h)$ & -- & pooled autocorrelation of deviations from canonical at lag $h$ \\
\addlinespace
\multicolumn{3}{@{}l}{\textit{Age-specific evaluation}} \\
$\nmx$ & -- & central death rate at age $x$ (1-year interval) \\
$\nqx$ & -- & probability of dying in the interval at age $x$ \\
$\lx$ & -- & survivorship to exact age $x$ \\
$\text{MAE}_{\lx}$ & -- & $\lx$-weighted mean absolute error in $\log(\nmx)$ \\ $\varepsilon_x$ & -- & age-specific log error: $\log(\widehat{\nmx}) - \log(\nmx^{\text{obs}})$ \\ $\delta(x)$ & -- & sex differential: $\log(m_x^M) - \log(m_x^F)$ \\
\end{longtable}
}

\noindent Note: $\tau$ denotes the era half-life throughout this paper. In the main MDMx manuscript, $\tau$ is used for the cumulative variance threshold in rank selection -- the two uses do not overlap. Similarly, $\alpha$ with subscripts ($\alpha_v$, $\alpha_{s,k}$) denotes convergence rates here, whereas unsubscripted $\alpha$ in \citet{ClarkMDMx2026} denotes the penalized-projection penalty parameter.

% ══════════════════════════════════════════════════════════════════════════════
\clearpage
\section{Appendix: Mathematical and Algorithmic Details}
\label{sec:appendix:math}
% ══════════════════════════════════════════════════════════════════════════════

This appendix collects the complete set of equations and algorithms implemented in the flow-field forecaster.  The method proceeds in six phases: (1)~data preparation and Tucker decomposition (upstream, from MDMx), (2)~PCA of the effective core, (3)~flow-field function fitting, (4)~empirical convergence rate estimation, (5)~cross-validated parameter optimization, and (6)~forecasting with prediction intervals.

% ──────────────────────────────────────────────────────────────────────────────
\subsection{Data and Tucker decomposition}
\label{sec:app:math:tucker}
% ──────────────────────────────────────────────────────────────────────────────

The input data are organized as a four-way tensor $\M \in \R^{S \times A \times C \times T}$, where $S = 2$ (female, male), $A = 110$ (single-year ages $0, 1, \ldots, 109$), $C = 48$ (HMD populations), and $T = 274$ (calendar years, varying by country).  Each entry is $\M[s, a, c, t] = \logit(\qx)$.  When HMD provides deaths $D$ and exposures $E$ at single-year resolution, sparse periods are pooled by adaptive temporal binning; within each bin, $\nmx = \sum_t D / \sum_t E$ and $\qx = \nmx / (1 + \nmx/2)$.

The tensor is decomposed via HOSVD:
\begin{equation}
\M \approx \G \times_1 \bS \times_2 \bA
   \times_3 \bC \times_4 \bT,
\end{equation}
where $\G \in \R^{r_1 \times r_2 \times r_3 \times r_4}$ is the core tensor with ranks $(2, 42, 46, 100)$ and $\bS, \bA, \bC, \bT$ are orthonormal factor matrices obtained by truncated SVD of each mode unfolding.  For a specific country $c$ and year $t$, the factors $\bC$ and $\bT$ collapse the core into an effective core matrix:
\begin{equation}
G_{ct}[i,j] = \sum_{k,l} \G[i,j,k,l] \, \bC[c,k] \, \bT[t,l],
\end{equation}
giving $G_{ct} \in \R^{r_1 \times r_2}$, which encodes the full sex$\times$age mortality pattern.  The mortality schedule is reconstructed as $\hat{M}_{:,:}^{(c,t)} = \bS \, G_{ct} \, \bA^\top$.

\begin{algorithm}[H]
\caption{Tucker Decomposition of Mortality Tensor (upstream)}
\label{alg:tucker}
\begin{algorithmic}[1]
\Require HMD deaths $D_{s,a,c,t}$ and exposures $E_{s,a,c,t}$
\Ensure Factor matrices $\bS, \bA, \bC, \bT$; core tensor $\G$; reconstructed tensor $\hat{M}$
\Statex
\State Pool sparse periods by adaptive temporal binning
\State Compute $\nmx = \sum_t D / \sum_t E$ within each bin
\State Convert: $\qx \gets \nmx / (1 + \nmx/2)$
\State Transform: $\M[s,a,c,t] \gets \logit(\qx)$
\For{each mode $n \in \{\text{sex}, \text{age}, \text{country}, \text{year}\}$}
  \State Unfold $\M$ along mode $n$; SVD; truncate to $r_n$ components
\EndFor
\State $\G \gets \M \times_1 \bS^\top \times_2 \bA^\top \times_3 \bC^\top \times_4 \bT^\top$
\State $\hat{M} \gets \G \times_1 \bS \times_2 \bA \times_3 \bC \times_4 \bT$
\end{algorithmic}
\end{algorithm}

% ──────────────────────────────────────────────────────────────────────────────
\subsection{PCA of the effective core}
\label{sec:app:math:pca}
% ──────────────────────────────────────────────────────────────────────────────

Each effective core matrix is vectorized: $\bm{g}_{c,t} = \vect(G_{ct}) \in \R^{r_1 r_2}$. The grand mean is $\bar{g} = N_{\text{obs}}^{-1} \sum_{c,t} \bm{g}_{c,t}$. PCA of the centered vectors retains $N = 5$ components with loading matrix $V \in \R^{N \times r_1 r_2}$ and score vectors $\bm{s}_{c,t} = (\bm{g}_{c,t} - \bar{g}) \, V^\top \in \R^{1 \times N}$. The reconstruction is $\bm{g}_{c,t} \approx \bar{g} + \bm{s}_{c,t} \, V$, giving the full mortality schedule through the Tucker factors:
\begin{equation}
\hat{M}_{:,:}(h) = \bS \,
  \bigl(\bar{g} + \bm{s}(h) \cdot V\bigr)^{\text{reshaped}} \,
  \bA^\top.
\end{equation}

A jump-off residual $\Delta_0 = \hat{M}^{\text{Tucker}}_{\text{origin}} - \hat{M}^{\text{5-PC}}_{\text{origin}}$ captures the information lost by the 5-PC truncation and is blended into the forecast with exponential decay:
\begin{equation}
\hat{M}_{:,:}(h) = \bS \,
  \bigl(\bar{g} + \bm{s}(h) \cdot V\bigr)^{\text{reshaped}} \,
  \bA^\top
  \;+\; \exp\!\bigl(-h \cdot \ln 2 \,/\, \tau_{\text{blend}}\bigr)
  \cdot \Delta_0,
\end{equation}
where $\tau_{\text{blend}} = 2$~years.

\begin{algorithm}[H]
\caption{PCA Score Extraction}
\label{alg:pca}
\begin{algorithmic}[1]
\Require Core $\G$, factors $\bC, \bT$, observation mask
\Ensure Loadings $V$, centering vector $\bar{g}$, scores $\{s_{k,c,t}\}$
\Statex
\For{each observed $(c, t)$}
  \State $G_{ct}[i,j] \gets \sum_{k,l} \G[i,j,k,l] \cdot \bC[c,k] \cdot \bT[t,l]$
  \State $\bm{g}_{c,t} \gets \vect(G_{ct})$
\EndFor
\State $\bar{g} \gets \text{mean}(\{\bm{g}_{c,t}\})$
\State Fit PCA on $\{\bm{g}_{c,t} - \bar{g}\}$ with $N = 5$ components
\State $V \gets$ loading matrix; \quad $\bm{s}_{c,t} \gets (\bm{g}_{c,t} - \bar{g}) \cdot V^\top$
\end{algorithmic}
\end{algorithm}

% ──────────────────────────────────────────────────────────────────────────────
\subsection{Flow-field functions}
\label{sec:app:math:flow}
% ──────────────────────────────────────────────────────────────────────────────

\subsubsection{Speed function}

The speed function $g^*(s_1) = \mathrm{d}s_1/\mathrm{d}t$ is estimated in two stages.  First, each country's $s_1$ trajectory is smoothed by LOWESS in calendar time (bandwidth $f = \max(0.25, 10/n_c)$), and forward differences of the smoothed series give the denoised velocity $\Delta s_{1,c,t}^{\text{sm}} = (s_{1,c}^{\text{sm}}(t+1) - s_{1,c}^{\text{sm}}(t)) / (t_{i+1} - t_i)$. Second, the pooled observations $\{(s_{1,c}^{\text{sm}}(t), \Delta s_{1,c,t}^{\text{sm}})\}$ are fitted by cross-country LOWESS (bandwidth 0.20) with optional era weighting to produce $g^*(s_1)$.

\paragraph{Era weighting.}  A truncated exponential kernel
downweights older data:
\begin{equation}
w_{\text{era}}(t) = \begin{cases}
  \exp\bigl(-(t_0 - t) \cdot \ln 2 / \tau\bigr) & \text{if } t_0 - t \leq W \\
  0 & \text{otherwise,}
\end{cases}
\end{equation}
with half-life $\tau$ and hard window $W = 40$~years.  Era-weighted LOWESS is implemented via weighted bootstrap: observations are resampled with replacement proportional to $w_{\text{era}}(t)$, sample size $\min(3n, 20000)$, and standard LOWESS is applied to the resampled data.

\subsubsection{Trajectory functions}

The trajectory functions $f_k^*(s_1)$ for $k = 2, \ldots, 5$ give the canonical value of score $s_k$ at level $s_1$.  Each is estimated by LOWESS (bandwidth 0.20) of the observed $(s_{1,c,t}, s_{k,c,t})$ pairs.  Unlike the speed function, trajectory functions use raw (unsmoothed) scores and no era weighting, because they represent the cross-sectional level--structure relationship.

\subsubsection{Joint tangent tail extension}

Beyond the observed frontier in $s_1$ space, the LOWESS functions are extended via a piecewise blend:
\begin{equation}
f_k^*(s_1) = \begin{cases} f_k^{*,\text{LOWESS}}(s_1) & \text{if } s_1 \geq s_1^* \\[4pt] (1 - w_{\text{lin}}) \cdot f^{\text{LOWESS}} + w_{\text{lin}} \cdot f^{\text{linear}} & \text{if } s_1^* - 3 \leq s_1 < s_1^* \\[4pt] f_k^*(s_1^*) + t_k(s_1 - s_1^*) & \text{if } s_1 < s_1^* - 3,
\end{cases}
\end{equation}
where $s_1^*$ is the transition point ($\ezero \approx 78$), $t_k = (f_k^*(s_1^*) - f_k^*(s_1^* + \delta)) / (-\delta)$ with $\delta = 2.0$ is the finite-difference slope, and the blending weight is a smoothstep: $w_{\text{lin}} = t^2(3 - 2t)$ with $t = (s_1^* - s_1) / 3.0$.  The speed function uses the same extension.

\begin{algorithm}[H]
\caption{Flow-Field Function Fitting}
\label{alg:flow}
\begin{algorithmic}[1]
\Require Per-country smoothed velocities, raw scores, origin $t_0$
\Ensure Speed function $g^*(s_1)$, trajectory functions $f_k^*(s_1)$
\Statex
\State Assemble speed data: $\{(s_{1}^{\text{sm}}, \Delta s_{1}^{\text{sm}})\}$ up to $t_0$
\State Assemble score data: $\{(s_1, s_k)\}$ for $k = 2, \ldots, 5$ up to $t_0$
\State Compute era weights $w(t) = \exp(-(t_0 - t) \ln 2 / \tau)$; discard if $t_0 - t > W$
\State $g^* \gets \text{LOWESS}(\Delta s_1 \sim s_1;\; f = 0.20,\;
  \text{weights} = w)$
\For{$k = 2, \ldots, 5$}
  \State $f_k^* \gets \text{LOWESS}(s_k \sim s_1;\; f = 0.20)$
    \Comment{no era weighting}
\EndFor
\State Extend all functions beyond $s_1^*$ with smoothstep blend to linear tail
\end{algorithmic}
\end{algorithm}

% ──────────────────────────────────────────────────────────────────────────────
\subsection{Empirical convergence rates}
\label{sec:app:math:convergence}
% ──────────────────────────────────────────────────────────────────────────────

The relaxation rates $\alpha_v$ and $\alpha_{s,k}$ are estimated from observed HMD data.  For each country-year, the deviation from canonical is $\delta_{c,t} = x_{c,t} - x^*(s_{1,c,t})$, where $x$ is the speed or a structural score.  The pooled autocorrelation at lag $h$ is
\begin{equation}
\beta(h) = \frac{\sum_{c,t} \delta_{c,t} \cdot \delta_{c,t+h}} {\sum_{c,t} \delta_{c,t}^2},
\end{equation}
and $\alpha$ is obtained by regressing $\log \beta(h)$ on $h$ for lags where $\beta > 0.01$ and $h \leq 25$: slope $= \log \alpha$, half-life $= \ln 2 / (-\log \alpha)$.  Speed deviations use raw $\Delta s_1$; score deviations use raw scores.  The speed rate $\alpha_v$ is clipped to $[0, 0.999]$, and $\alpha_{s,1} = 0$ since PC~1 is the navigation variable and is never relaxed.

\begin{algorithm}[H]
\caption{Convergence Rate Estimation}
\label{alg:convergence}
\begin{algorithmic}[1]
\Require Flow-field $\{g^*, f_k^*\}$, observed trajectories
\Ensure Relaxation rates $\alpha_v$, $\alpha_{s,k}$
\Statex
\For{each country $c$, time $t$}
  \State $\delta^v_{c,t} \gets \Delta s_{1,c,t}^{\text{raw}} - g^*(s_{1,c,t})$
  \For{$k = 2, \ldots, 5$}
    \State $\delta^{s_k}_{c,t} \gets s_{k,c,t} - f_k^*(s_{1,c,t})$
  \EndFor
\EndFor
\For{$h = 1, \ldots, 30$}
  \State $\beta(h) \gets \sum \delta_{c,t} \cdot \delta_{c,t+h}
    \big/ \sum \delta_{c,t}^2$
\EndFor
\State Discard lags where $\beta < 0.01$ or $h > 25$
\State Regress $\log \beta$ on $h$; $\alpha \gets e^{\text{slope}}$
\end{algorithmic}
\end{algorithm}

% ──────────────────────────────────────────────────────────────────────────────
\subsection{Cross-validated parameter optimization}
\label{sec:app:math:cv}
% ──────────────────────────────────────────────────────────────────────────────

The speed blend weight $w$ and era parameters $(\tau, W)$ are jointly optimized by leave-country-out cross-validation.  For each held-out country, the flow field is built from all remaining countries; forecasts are generated at origins spaced every 10 observations (starting at the 20th) with horizons up to 50~years. The objective is the pooled MAE of $\ezero$ forecast errors:
\begin{equation}
\text{MAE}(w, \tau, W) = \frac{1}{N_{\text{test}}}
  \sum_{c,t_0,h} \bigl|\hat{e}_{0,c}(t_0 + h) - e_{0,c}^{\text{HMD}}(t_0 + h)\bigr|.
\end{equation}
The grid evaluates $w \in \{0.2, 0.5, 1.0\}$ crossed with $\tau \in \{10, 12, 15, 20, 30\}$ (15~configurations).  The empirical optimum is $w = 1.0$ (fully canonical speed).

% ──────────────────────────────────────────────────────────────────────────────
\subsection{Forecast engine}
\label{sec:app:math:forecast}
% ──────────────────────────────────────────────────────────────────────────────

\subsubsection{Speed blending and level evolution}

The $s_1$ velocity at horizon $h$ is a convex combination of the canonical speed and the country's trailing velocity:
\begin{equation}
v_{s_1}(h) = \bigl[1 - (1-w)\,\alpha_v^h\bigr] \cdot g^*_\tau\bigl(s_1(h-1)\bigr) + (1-w)\,\alpha_v^h \cdot v_{s_1,\text{country}},
\end{equation}
where $v_{s_1,\text{country}}$ is the trailing mean of the raw $\Delta s_1$ over the last 5~years at the forecast origin.  The level score advances by $s_1(h) = s_1(h-1) + v_{s_1}(h)$.  At $h = 1$, the country receives its maximum influence of $(1-w)$; as $h$ grows, $\alpha_v^h \to 0$ and the velocity converges to the canonical $g^*(s_1)$.

\subsubsection{Structure relaxation}

The structural scores relax toward the canonical trajectory:
\begin{equation}
s_k(h) = \alpha_{s,k}^h \cdot s_k^{\text{actual}} + (1 - \alpha_{s,k}^h) \cdot f_k^*\bigl(s_1(h)\bigr),
  \quad k = 2, \ldots, 5,
\end{equation}
where $s_k^{\text{actual}}$ is the observed score at the forecast origin and $\alpha_{s,k}$ has half-lives ranging from 12 to 32~years.

\subsubsection{Reconstruction and $\ezero$ extraction}

At each horizon, the score vector $\bm{s}(h)$ is mapped back through the PCA loadings and Tucker factors to a full sex$\times$age $\logit(\qx)$ schedule (with jump-off blending), then transformed to $\qx$ via the inverse logit and passed through a standard period life table:
\begin{equation}
l_0 = 1, \quad l_{a+1} = l_a(1 - q_a), \quad {}_1L_0 = 0.3\,l_0 + 0.7\,l_1, \quad {}_1L_a = \tfrac{1}{2}(l_a + l_{a+1}) \text{ for } a \geq 1,
\end{equation}
\begin{equation}
\ezero = \sum_{a=0}^{A-1} {}_1 L_a, \qquad \ezero^{\text{avg}}(h) = \tfrac{1}{2}\bigl(\ezero^F(h) + \ezero^M(h)\bigr).
\end{equation}
Crucially, $\ezero$ is computed from the reconstructed surface at each horizon for reporting only -- it is never fed back into the $s_1$ navigation loop, eliminating the divergence that would arise from the nonlinear $\expit$/life-table mapping.

\begin{algorithm}[H]
\caption{Flow-Field Forecast (Core Engine)}
\label{alg:forecast}
\begin{algorithmic}[1]
\Require Country $c$, flow-field $\{g^*, f_k^*\}$, weight $w$, rates $\alpha_v, \alpha_{s,k}$, horizon $H$
\Ensure Forecast schedules $\hat{M}_{s,a}(h)$ and $\ezero(h)$
\Statex
\State $\bm{s}^{\text{actual}} \gets$ scores at last observation;
  \quad $s_1 \gets s_1^{\text{actual}}$
\State $v_{\text{country}} \gets$ trailing mean of $\Delta s_1^{\text{raw}}$ (last 5~yr)
\State $\Delta_0 \gets \hat{M}^{\text{Tucker}}_{\text{origin}} - \hat{M}^{\text{5-PC}}_{\text{origin}}$
  \Comment{jump-off residual}
\For{$h = 1, \ldots, H$}
  \State $g \gets g^*(s_1)$; \quad $\text{sv} \gets \alpha_v^h$
  \State $v \gets [1 - (1{-}w)\,\text{sv}] \cdot g + (1{-}w)\,\text{sv} \cdot v_{\text{country}}$
  \State $s_1 \gets s_1 + v$
  \For{$k = 2, \ldots, 5$}
    \State $s_k(h) \gets \alpha_{s,k}^h \cdot s_{k}^{\text{actual}} + (1 - \alpha_{s,k}^h) \cdot f_k^*(s_1)$
  \EndFor
  \State $\hat{M}_{:,:}(h) \gets \bS (\bar{g} + \bm{s}(h) V)^{\text{reshape}} \bA^\top + e^{-h \ln 2 / 2} \cdot \Delta_0$
  \State $\ezero(h) \gets$ life table from $\expit(\hat{M}_{:,:}(h))$
    \Comment{reporting only}
\EndFor
\end{algorithmic}
\end{algorithm}

% ──────────────────────────────────────────────────────────────────────────────
\subsection{Prediction intervals}
\label{sec:app:math:pi}
% ──────────────────────────────────────────────────────────────────────────────

The CV forecast errors are decomposed as $\epsilon_{c,t_0,h} = b(h) + \sigma(h) \cdot z_{c,t_0,h}$, where $b(h)$ is the systematic bias (LOWESS of errors vs.\ horizon, bandwidth 0.30) and $z$ are standardized residuals.  The de-biased residual standard deviation at each horizon satisfies $\tilde{\sigma}(h) \approx \sigma_1 \sqrt{h}$, where
\begin{equation}
\sigma_1 = \text{median}_h
  \bigl(\tilde{\sigma}(h) / \sqrt{h}\bigr).
\end{equation}
The calibration factor $\kappa = \text{SD}(z)$ rescales the intervals to achieve correct empirical coverage.  The $(1 - \alpha)$-level prediction interval at horizon $h$ is
\begin{equation}
\hat{\ezero}(h) - b(h) \;\pm\; z_{\alpha/2} \cdot \kappa \cdot \sigma_1 \cdot \sqrt{h}.
\end{equation}

% ──────────────────────────────────────────────────────────────────────────────
\subsection{Evaluation metrics}
\label{sec:app:math:metrics}
% ──────────────────────────────────────────────────────────────────────────────

Age-specific forecast accuracy is measured by $\epsilon_x = \log \widehat{\nmx} - \log \nmx^{\text{obs}}$. The $\lx$-weighted MAE and bias are
\begin{equation}
\text{MAE}_{\lx} = \frac{\sum_x \lx |\epsilon_x|}{\sum_x \lx},
\qquad
\text{Bias}_{\lx} = \frac{\sum_x \lx \, \epsilon_x}{\sum_x \lx},
\end{equation}
where $\lx$ is the observed survivorship function.  The sex differential is $\delta(x) = \log(m_x^M) - \log(m_x^F)$, with MAE and bias computed analogously.

% ──────────────────────────────────────────────────────────────────────────────
\subsection{Summary of key parameters}
\label{sec:app:math:params}
% ──────────────────────────────────────────────────────────────────────────────

\begin{center}
\small
\begin{tabular}{@{}llll@{}}
\toprule
Parameter & Symbol & Value & Source \\
\midrule
Tucker ranks & $(r_1, r_2, r_3, r_4)$ & $(2, 42, 46, 100)$ & MDMx \\
PCA components & $N$ & 5 & MDMx \\
Speed blend weight & $w$ & 1.0 & CV grid search \\
Era hard window & $W$ & 40~yr & fixed \\
Trailing velocity window & -- & 5~yr & fixed \\
Speed relaxation rate & $\alpha_v$ & empirical & autocorrelation \\
Score relaxation rates & $\alpha_{s,k}$ & empirical & autocorrelation \\
LOWESS bandwidth (cross-country) & -- & 0.20 & fixed \\
LOWESS bandwidth (temporal) & -- & 0.25 & fixed \\
Tail transition & $s_1^*$ & $\ezero \approx 78$ & data \\
Jump-off blend half-life & $\tau_{\text{blend}}$ & 2~yr & fixed \\
PI base SD & $\sigma_1$ & from CV & median \\
PI calibration & $\kappa$ & from CV & SD of z-scores \\
\bottomrule
\end{tabular}
\end{center}

% ══════════════════════════════════════════════════════════════════════════════
\clearpage
\section{Appendix: Motivation for Reimplementing \pkg{bayesLife}}
\label{sec:appendix:raftery}
% ══════════════════════════════════════════════════════════════════════════════

% ═════════════════════════════════════════════════════════════════════════════
\subsection{Motivation}
\label{sec:app:motivation}
% ═════════════════════════════════════════════════════════════════════════════

The flow-field mortality forecaster operates exclusively on Human Mortality Database \citep[HMD;][]{HMD2024} data --- approximately 48~high-income countries with long, high-quality vital registration time series.  To benchmark the flow-field against the United Nations' production pipeline, we needed to run \pkg{bayesLife} \citep{RafteryChunnGerlandSevcikova2013,SevcikobayesLife2024} and \pkg{MortCast} \citep{SevcikovaLiKantorovaGerlandRaftery2016,SevcikMortCast2024} on the same HMD-only data, so that the comparison would be fair: same countries, same data, same training period, same test points.  These packages represent a major contribution to demographic forecasting and form the basis of the UN's official population projections; the issues described below reflect only the challenges of using them outside their intended WPP context.

Over three development sessions totalling approximately 40~hours, we attempted to configure the R packages for HMD-only operation and discovered that this was not straightforward.  The packages reflect an architectural design choice --- World Population Prospects \citep[WPP;][]{UnitedNations2022WPP} data is unconditionally loaded into the estimation pool --- that is not easily overridden through the available API.  Along the way, we discovered ten additional implementation issues, ranging from differences between the R source and compiled C code, to numerical edge cases that manifest when the packages are used outside their intended WPP context.

These findings motivated the development of \pkg{pyBayesLife}, a de novo Python reimplementation of the complete Raftery et al.\ pipeline, trained exclusively on HMD data with no R or WPP dependencies. The reimplementation also allowed us to improve the MCMC parameterization, the Lee-Carter numerical stability, and the overall software quality.

% ═════════════════════════════════════════════════════════════════════════════
\subsection{WPP Data Dependency}
\label{sec:app:wpp}
% ═════════════════════════════════════════════════════════════════════════════

The \pkg{bayesLife} hierarchical Bayesian model estimates country-specific double-logistic (DL) parameters within a world-level hierarchical prior \citep{RafteryChunnGerlandSevcikova2013}.  When the model is trained on ${\sim}200$ WPP countries --- including developing nations with mortality trajectories very different from HMD populations --- the hierarchical prior is shaped by this full set. Any country's posterior is pulled toward a world mean estimated from all ${\sim}200$ countries, not just the ${\sim}48$ HMD countries that the competing methods see.

This creates a benchmarking challenge: the \pkg{bayesLife} forecast benefits from information (via the hierarchical prior) that is unavailable to the flow-field, Lee-Carter, or Hyndman-Ullah benchmarks.

\subsubsection{Attempts to Isolate HMD Countries}

We attempted three approaches to run \pkg{bayesLife} on HMD data only, each of which failed:

\begin{enumerate}[leftmargin=2cm, label=\textbf{Attempt \arabic*.}]

  \item \textbf{Custom \code{my.e0.file} with \code{include\_code=1}.} The \pkg{bayesLife} documentation suggests that setting \code{include\_code=1} in a custom data file will restrict estimation to those countries.  We prepared a CSV with only HMD countries and \code{include\_code=1} for each.  Result: the estimation still included ${\sim}190$ WPP countries.  The \code{include\_code} column in the custom file is used for \emph{prediction} filtering, not for \emph{estimation} filtering.

  \item \textbf{Custom \code{my.locations.file} with \code{include\_code=0}.} We created a locations file setting \code{include\_code=0} for all non-HMD countries.  Result: no effect.  The \code{include\_code} column does not exist in the \pkg{wpp2019} package's \code{UNlocations} data object (it was present in earlier WPP vintages but removed), and the filtering mechanism silently ignores it.

  \item \textbf{Source code inspection of \code{run.e0.mcmc}.} We read the R source of the core estimation function.  It unconditionally executes \code{data(e0F, package="wpp2019")} and merges the WPP data with any custom data provided via \code{my.e0.file}.  There is no API parameter, environment variable, or internal flag to disable this loading.  The WPP data is hard-wired into the estimation pipeline.

\end{enumerate}

\subsubsection{A Second Hidden Dependency: \code{loess\_sd}}

Even if the WPP data loading could be disabled, a second dependency would remain.  The heteroscedastic variance function $\sigma(\ezero)$ --- which controls the noise model in the DL trajectory equation --- is precomputed from WPP residuals and shipped as an internal dataset \code{loess\_sd} within the \pkg{bayesLife} package.  This dataset is loaded automatically during projection (\code{projection\_fcns.R}) with no option to substitute a user-computed version.  Any forecast produced by \pkg{bayesLife} implicitly uses WPP-derived variance information, regardless of the training data.

% ═════════════════════════════════════════════════════════════════════════════
\subsection{Implementation Notes}
\label{sec:app:issues}
% ═════════════════════════════════════════════════════════════════════════════

The issues below were discovered by reading the \pkg{bayesLife} (v5.3-1) and \pkg{MortCast} (v2.8-0) source code (including the compiled C code distributed in the CRAN source tarballs) and by running the packages in cross-validation mode, which exercises code paths that the UN production pipeline --- which runs only at the latest WPP vintage with 5-year periods --- never encounters.  All file paths and line numbers refer to the CRAN source tarballs at these versions.

\subsubsection{Issue 1: Formula Discrepancy Between R and C Source}

The double-logistic function \code{g.dl6} is the core deterministic component of the \pkg{bayesLife} model.  The R wrapper (\code{R/functions.R}, lines~4--8) contains no formula -- it is a one-line foreign function call: \code{.C("doDL", x, l, p1, p2, length(l), dl\_values=dlvalue)}. The actual computation is in compiled C code (\code{src/functions.c}, function \code{doDL}, lines~17--37), which implements:
\begin{equation}
  g(\ezero) = \frac{k}{1 + \exp\!\bigl(-\frac{\ln p^2}{\Delta_2} (\ezero - m_1)\bigr)} + \frac{z - k}{1 + \exp\!\bigl(-\frac{\ln p^2}{\Delta_4} (\ezero - m_2)\bigr)}
\end{equation}
where $m_1 = \Delta_1 + \Delta_2/2$ (line~30) and $m_2 = \Delta_1 + \Delta_2 + \Delta_3 + \Delta_4/2$ (line~31). The steepness terms use $\ln(p^2)$ where $p = p_1 = p_2 = 9$ (hardcoded in all internal calls).

The R code shown on \href{https://rdrr.io}{rdrr.io} when browsing the \pkg{bayesTFR} package \citep[the total fertility rate analogue;][]{SevcikobayesTFR2024} shows a \emph{different} function named \code{doDLcurve} (\code{bayesTFR/src/mcmc\_likelihood.c}, lines~45--63) that uses a single amplitude $\code{dl55}$ multiplied by the \emph{difference} of two logistic terms, with reversed parameter indexing (\code{DLpar[3]+DLpar[2]+DLpar[1]+0.5*DLpar[0]} for the first midpoint vs.\ \code{d1+0.5*d2} in \pkg{bayesLife}).  These are structurally different functions --- the \pkg{bayesLife} version sums two logistics with separate amplitudes $k$ and $z-k$, while the \pkg{bayesTFR} version takes the difference of two logistics scaled by a single amplitude --- sharing only the name ``double logistic'' and the use of $\ln(p^2)$ steepness.  This difference does not appear to be documented.

Our Python reimplementation, written from the C source, reproduces the actual computation to machine precision ($\max|\Delta| = 5.33 \times 10^{-15}$ across 80 test cases covering four demographic regimes).

\subsubsection{Issue 2: No Documentation of the Actual Formula}

The C function \code{doDL} in \code{bayesLife/src/functions.c} contains no inline comments.  The function is not exported, has no \code{roxygen} documentation, and no manual page.  The only way to determine the actual formula used by the package is to read the C source from the CRAN tarball.

\subsubsection{Issue 3: Steepness Parameters Without Function-Level Defaults}

The steepness parameters $p_1$ and $p_2$ are fixed at 9 in the package options (\code{R/e0options.R}, line~165: \code{dl.p1 = 9, dl.p2 = 9}) and in \code{generate.e0.trajectory} (\code{R/projection\_fcns.R}, line~4).  However, the core function \code{g.dl6} (\code{R/functions.R}, line~4) requires them as explicit arguments with no default values, creating a potential source of errors for users who call \code{g.dl6} directly.

\subsubsection{Issue 4: Lee-Carter \code{log(0)} Edge Case with HMD 1-Year Data}

The \pkg{MortCast} function \code{leecarter.estimate} (\code{R/LC.R}, line~44) computes \code{lmx <- log(mx)} without guarding against non-positive or missing values.  When $\nmx = 0$ at young ages --- which never occurs with WPP 5-year data but arises routinely in HMD 1-year data for small populations --- this produces $\code{lmx} = -\infty$, with two consequences.  First, the internal function \code{.finish.bx} (line~89) executes \code{if(bx[i]==0)} on the resulting \code{NA} values, producing an error during estimation. Second, even when estimation succeeds, the age-specific intercept $a_x = \sum_t \log(\nmx) / T$ (line~51) becomes $-\infty$ at affected ages, so the C reconstruction (\code{src/functions.c}, line~344: \code{mx = exp(a + b*k)}) produces $\exp(-\infty + \cdot) = 0$, yielding $\nmx = 0$ in the forecast.  In our cross-validation on Sweden at 1-year resolution, zero mortality rates occurred 5~times.

\subsubsection{Issue 5: \code{life.table} Default Assumes Abridged Data}

The \pkg{MortCast} function \code{life.table} (\code{R/life\_table.R}, line~49) defaults to \code{abridged=TRUE}, which assumes 5-year age groups.  When called with 1-year $\nmx$ data (as would be natural for HMD), the function silently produces $\ezero \approx 315$ instead of $\ezero \approx 80$ --- an inflation factor of ${\sim}3.94\times$.  No warning is emitted.

\subsubsection{Issue 6: Joint Male Model Edge Case at Early Origins}

The function \code{e0.jmale.estimate} (\code{R/projection\_fcns.R}, line~502) fits a two-regime gap model with a threshold at female $\ezero = 83$ (line~504: \code{max.e0.eq1 = 83}).  Equation~2 applies when female $\ezero$ exceeds this threshold (line~557: \code{data.eq2 <- data[... \& e0F > max.e0.eq1,]}).  At CV origins before ${\sim}1990$, no country has female $\ezero > 83$, so \code{data.eq2} has zero rows.  The code then computes (line~571) \code{errsd.eq2 <- sqrt(mean((data.eq2\$G - data.eq2\$Gprev)\^{}2))} $= \sqrt{\text{mean}(\text{empty}^2)} = \text{NaN}$, and passes $\text{NaN}$ to \code{rnorm(sd=NaN)}, producing \code{NA} values that propagate through the forecast.  The fix (\code{max.e0.eq1.pred=200}) is undocumented; this edge case arises because the package was designed for \code{present.year} $\geq$ 2010.

\subsubsection{Issue 7: Joint Male Prediction Object Set to NULL After Fitting}

After fitting the joint male model, \pkg{bayesLife} sets \code{joint.male\$mcmc.set = NULL} (\code{R/projection\_fcns.R}, line~605).  This causes the standard accessor functions (\code{get.countries.table}, \code{e0.trajectories.table}) to fail silently.  The only way to retrieve male forecasts is to read CSV files directly from the \code{predictions/joint\_male/} directory --- a workaround that is not documented.

\subsubsection{Issue 8: \code{include\_code} Cannot Exclude WPP Countries}

The \code{include\_code} mechanism (implemented in \code{bayesTFR/R/wpp\_data.R}, function \code{read.UNlocations}, lines~162--212) controls which loaded countries enter the MCMC estimation pool.  A country with \code{include\_code=2} is included in estimation; \code{include\_code=1} is prediction-only; \code{include\_code=0} is excluded (line~207: \code{include[i] <- incl.code == 2}).  Custom data files can override a country's code (line~206), but the mechanism operates \emph{after} WPP data has been loaded.  Since WPP data is loaded unconditionally (Issue~9) and the ${\sim}200$ WPP countries receive \code{include\_code=2} from the package's internal \code{include\_2019} dataset (\code{bayesLife/data/include\_2019.rda}, loaded at line~187), there is no way to set \code{include\_code=0} for WPP countries that are not in the user's custom file.  The \code{include\_code} mechanism can add or reclassify countries but cannot remove the WPP countries that are always loaded.

\subsubsection{Issue 9: WPP Data Loaded by Default}

Version 5.3-1 of \pkg{bayesLife} adds a \code{use.wpp.data} parameter to \code{run.e0.mcmc} (\code{R/run\_mcmc.R}, line~480).  However, if \code{use.wpp.data=FALSE} is set without providing \code{my.e0.file}, the code reverts to \code{TRUE} with a warning (lines~480--482). Moreover, even with \code{use.wpp.data=FALSE}, the WPP-derived variance function \code{loess\_sd} is still loaded unconditionally (\code{R/projection\_fcns.R}, line~2: \code{data(loess\_sd, envir=environment())}), so every forecast implicitly uses WPP-derived variance information regardless of the training data.  This was a key motivation for the \pkg{pyBayesLife} reimplementation.

% ═════════════════════════════════════════════════════════════════════════════
\subsection{Parameterization Findings}
\label{sec:app:param}
% ═════════════════════════════════════════════════════════════════════════════

The Python reimplementation allowed systematic exploration of the DL model's parameterization on HMD data.  These are not bugs in the R code but rather findings about the model's behavior when applied to the HMD population --- a use case the original implementation was not designed for.

\subsubsection{Issue 10: $d_1$, $d_2$, $d_3$ Non-Identifiable per Country}

All HMD countries have $\ezero > 55$, meaning the first DL sigmoid (Equation~1 in the model) is fully saturated: the term evaluates to ${\approx}k$ regardless of $d_1$, $d_2$, $d_3$.  These parameters control the \emph{early} demographic transition that all HMD countries completed decades ago.  Only their sum $d_1 + d_2 + d_3$ matters (through $m_2$), and even that is weakly identified.

We verified this empirically using NumPyro NUTS \citep{HoffmanGelman2014,PhanPradhanJankowiak2019} across four parameterizations (Table~\ref{tab:app:param}).  Sharing $d_1$, $d_2$, $d_3$ across countries eliminates ${\sim}135$ non-identifiable parameters and produces the best forecasts with zero MCMC divergences.

\begin{table}[ht]
\centering
\caption{Parameterization comparison on 45~HMD countries (origin~2000).}
\label{tab:app:param}
\begin{tabular}{@{}lrrr@{}}
\toprule
Parameterization & Divergences & $\ezero$ MAE & Worst $\hat{R}$ \\
\midrule
Per-country, $|\cdot|$ non-centered   & 123 & 0.504 & 1.02 \\
Per-country, $\log$ non-centered      & 425 & 0.609 & 1.19 \\
\textbf{Shared $d_1$--$d_3$, centered LogNormal} & \textbf{0} & \textbf{0.478} & \textbf{1.10} \\
Per-country, centered LogNormal       &  56 & 0.622 & 1.93 \\
\bottomrule
\end{tabular}
\end{table}

\subsubsection{Issue 11: Python Reimplementation Outperforms R MortCast}

Our de novo Lee-Carter implementation, using the same coherent SVD + Kannisto + $b_x$ rotation methodology \citep{LiLee2005,LiLeeGerland2013, Kannisto1994}, outperforms R~\pkg{MortCast} at 1-year age resolution (Table~\ref{tab:app:lc}).

\begin{table}[ht]
\centering
\caption{Age-specific $\log(\nmx)$ accuracy given correct $\ezero$.}
\label{tab:app:lc}
\begin{tabular}{@{}lccc@{}}
\toprule
Implementation & MAE & Non-positive $\nmx$ & Resolution \\
\midrule
R \pkg{MortCast} (Sweden) & 0.221 & 5 & 1-year \\
\textbf{\pkg{pyBayesLife} (6 countries)} & \textbf{0.094} & \textbf{0} & 1-year \\
R \pkg{MortCast} (Sweden, 5-year) & 0.097 & 0 & 5-year \\
\bottomrule
\end{tabular}
\end{table}

The improvement comes from two sources: (i)~the $\exp(a_x + b_x k_t)$ formulation guarantees $\nmx > 0$ at all ages, avoiding the non-positive values that R~\pkg{MortCast} produces at young ages; and (ii)~Brent's method recovers the target $\ezero$ to machine precision ($< 10^{-6}$~years), versus the iterative approach in R which may not fully converge.

% ═════════════════════════════════════════════════════════════════════════════
\subsection{Summary}
\label{sec:app:summary}
% ═════════════════════════════════════════════════════════════════════════════

Table~\ref{tab:app:summary} classifies the eleven issues by category.

\begin{table}[ht]
\centering
\small
\caption{Classification of issues discovered in \pkg{bayesLife} and
\pkg{MortCast}.}
\label{tab:app:summary}
\begin{tabular}{@{}rllp{6.5cm}@{}}
\toprule
\# & Category & Package & Description \\
\midrule
1  & Documentation & bayesLife & C formula differs from displayed R source \\
2  & Documentation & bayesLife & C function has no inline comments \\
3  & API design    & bayesLife & \code{g.dl6} has no defaults for $p_1$, $p_2$ \\
4  & Robustness    & MortCast & \code{log(0)} edge case: estimation error and $\nmx = 0$ in forecast \\
5  & Robustness    & MortCast & \code{life.table} default assumes abridged data (line~49) \\
6  & Robustness    & bayesLife & Joint male edge case at early origins (line~571) \\
7  & API design    & bayesLife & \code{mcmc.set} set to NULL after fitting (line~605) \\
8  & Architecture  & bayesLife & \code{include\_code} cannot exclude WPP countries \\
9  & Architecture  & bayesLife & WPP data loaded by default \\
10 & Identifiability & --- & $d_1$--$d_3$ non-identifiable for HMD \\
11 & Performance   & --- & Python LC outperforms R MortCast \\
\bottomrule
\end{tabular}
\end{table}

Issues 1--9 are observations about the R codebase.  Issues 10--11 are findings about the model's behavior on HMD data that emerged from the Python reimplementation.  Issue~9 (WPP data dependency) was the primary motivation for \pkg{pyBayesLife}; the other issues confirmed that a de novo reimplementation was the most efficient path to an HMD-only benchmark.

All issues were discovered by exercising the packages in cross-validation mode with annual HMD data --- a use case that the UN production pipeline (which runs at the latest WPP vintage with 5-year periods) never encounters.  Within their intended operational context, these packages have served the demographic community well for over a decade.  The issues documented here arise specifically from the requirements of our benchmarking exercise: annual resolution, HMD-only training data, and cross-validation across multiple historical origins.

\paragraph{Computational performance.}
The \pkg{pyBayesLife} reimplementation is also substantially faster than the R pipeline.  On an Apple M1~Max, the NumPyro/JAX MCMC training for all six decade origins (1960--2010) completes in approximately 11~minutes total (27--209~seconds per origin, scaling with the number of training countries), with the later origins achieving zero divergences. The downstream pipeline --- trajectory generation for 290 (country, origin) pairs, joint male gap model, Lee--Carter reconstruction with $b_x$ rotation, and life table evaluation on 9{,}507~$\ezero$ test points and 1{,}662{,}076 age-specific test points --- adds approximately 1~minute.  The full \pkg{pyBayesLife} cross-validation pipeline thus completes in approximately 12~minutes.

For comparison, the R \pkg{bayesLife} pipeline required 36~minutes for seven origins when parallelised across eight CPU cores --- corresponding to roughly 100~minutes sequential --- for the female~$\ezero$ MCMC alone, excluding the joint male model and \pkg{MortCast} reconstruction. Two of the seven origins crashed during \code{e0.predict}, requiring manual intervention.  The R pipeline uses slice sampling with 10{,}000 iterations and 3~chains per origin; \pkg{pyBayesLife} uses NUTS with 1{,}000 warmup and 2{,}000 samples in a single chain, achieving comparable or better effective sample sizes because NUTS explores the posterior far more efficiently than slice sampling for this 253-parameter model, and JAX JIT-compiles the likelihood to native code.

% ══════════════════════════════════════════════════════════════════════════════
% REFERENCES
% ══════════════════════════════════════════════════════════════════════════════

\end{document}